\documentclass[aps,pra,twocolumn,showpacs,notitlepage,superscriptaddress,letterpaper,floatfix]{revtex4-2}
\usepackage[utf8]{inputenc}
\usepackage{amsmath,amssymb}
\usepackage{graphicx}
\usepackage{hyperref}
\usepackage{cleveref}
\usepackage{todonotes}

\begin{document}
\title{A quantum electromechanical interface for long-lived phonons}
\date{\today}
\author{Alkim Bozkurt}
\affiliation{The Gordon and Betty Moore Laboratory of Engineering, California Institute of Technology, Pasadena, California 91125}
\affiliation{Institute for Quantum Information and Matter, California Institute of Technology, Pasadena, California 91125}
\author{Han Zhao}
\affiliation{The Gordon and Betty Moore Laboratory of Engineering, California Institute of Technology, Pasadena, California 91125}
\affiliation{Institute for Quantum Information and Matter, California Institute of Technology, Pasadena, California 91125}

\author{Chaitali Joshi}
\affiliation{The Gordon and Betty Moore Laboratory of Engineering, California Institute of Technology, Pasadena, California 91125}
\affiliation{Institute for Quantum Information and Matter, California Institute of Technology, Pasadena, California 91125}

\author{Henry G. LeDuc}
\affiliation{Jet Propulsion Laboratory, California Institute of Technology, Pasadena, California 91109}
\author{Peter K. Day}
\affiliation{Jet Propulsion Laboratory, California Institute of Technology, Pasadena, California 91109}

\author{Mohammad Mirhosseini}
\email{mohmir@caltech.edu}
\homepage{http://qubit.caltech.edu}
\affiliation{The Gordon and Betty Moore Laboratory of Engineering, California Institute of Technology, Pasadena, California 91125}
\affiliation{Institute for Quantum Information and Matter, California Institute of Technology, Pasadena, California 91125}
\begin{abstract}
   Controlling long-lived mechanical oscillators in the quantum regime holds promises for quantum information processing. Here, we present an electromechanical system capable of operating in the GHz-frequency band in a silicon-on-insulator platform. Relying on a novel driving scheme based on an electrostatic field and high-impedance microwave cavities based on TiN superinductors, we are able to demonstrate a parametrically-enhanced electromechanical coupling of ${g/2 \pi} = 1.1$ MHz, sufficient to enter the strong-coupling regime with a cooperativity of $\mathcal{C} = 1200$. The absence of piezoelectric materials in our platform leads to long mechanical lifetimes, finding intrinsic values up to $\tau_\text{d} = 265~ \mu$s ($Q = 8.4 \times {10}^6$ at $\omega_\mathrm{m}/2\pi = 5$ GHz) measured at low-phonon numbers and millikelvin temperatures. Despite the strong parametric drives, we find the cavity-mechanics system in the quantum ground state by performing sideband thermometry measurements. Simultaneously achieving ground-state operation, long mechanical lifetimes, and strong coupling sets the stage for employing silicon electromechanical resonators as memory elements and transducers in hybrid quantum systems, and as a tool for probing the origins of acoustic loss in the quantum regime.
\end{abstract}
\maketitle

\begin{figure*}[!t]
\centering\includegraphics[width=14cm]{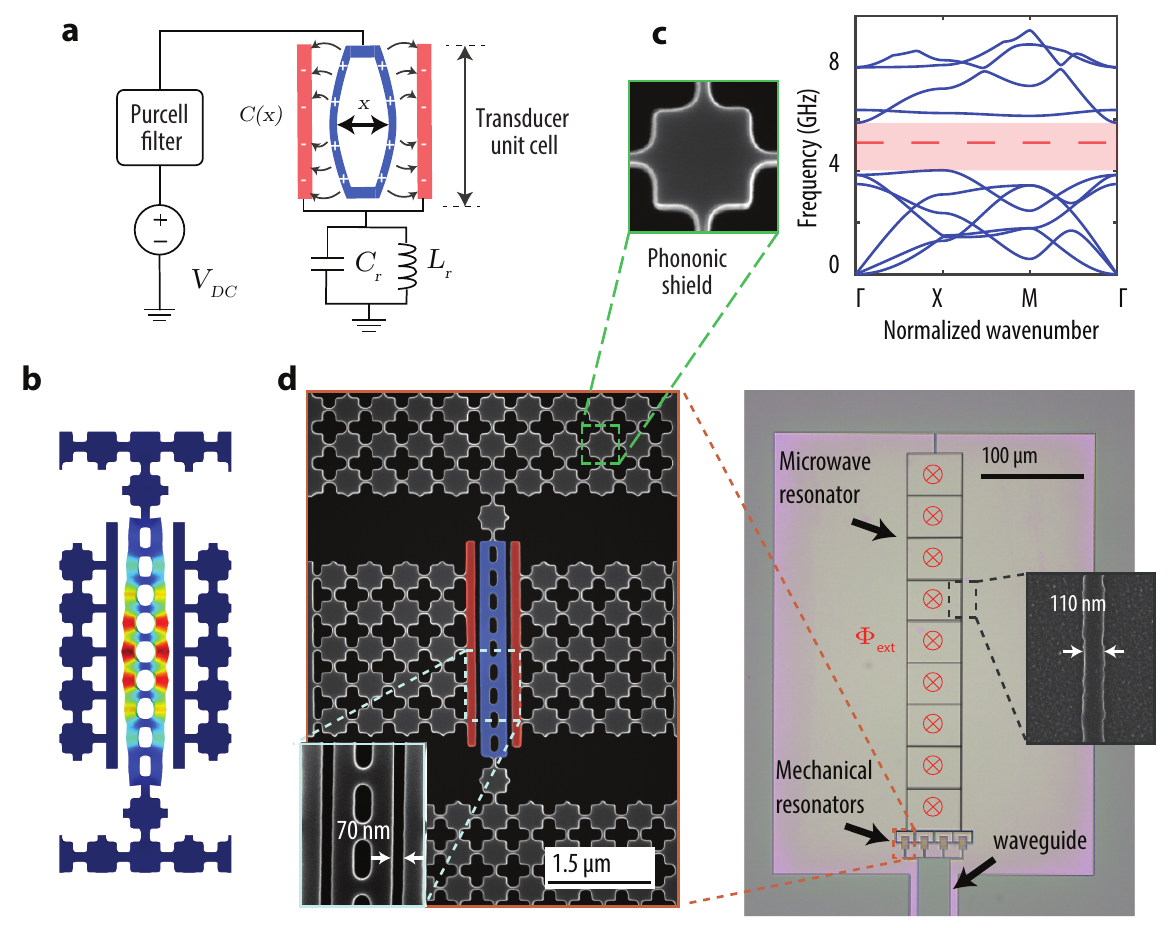}
\caption{{\bf Electrostatic transduction.} (a) The circuit diagram of an electrostatic transducer coupled to a microwave cavity (shown as an LC resonator). The Purcell filter is used to deliver the static biasing voltage to the transducer without incurring radiative loss to the microwave cavity. (b) Simulated displacement profile of the confined mechanical resonance with the largest electromechanical coupling. (c) Schematics of the unit cell and the band diagram of a phononic shield. (d) Left: The scanning electron microscope (SEM) image of the fabricated transducer clamped by the phononic shields. The inner and outer electrodes are shown in false color. Right: The optical microscope image of the fabricated microwave resonator. Inset shows the TiN nanowires.}
    \label{fig:dev}
\end{figure*}


Phonons, the quanta of energy stored in vibrations in solids, promise unique opportunities for storing and communicating quantum information. The intrinsic mechanisms for phonon dissipation get suppressed at low temperatures \cite{mcguigan1978}, leading to extremely low acoustic loss in single crystalline materials \cite{renninger2018,beccari2022b}. Additionally, the inability of sound waves to propagate in vacuum makes it possible to trap phonons in wavelength-scale dimensions via geometric structuring, leading to near-complete suppression of environment-induced decay \cite{maccabe2020b}. Finally, phonons interact with solid-state qubits and the electromagnetic waves across a broad spectrum, making them near-universal intermediaries for cross-platform information transfer \cite{10.1038/s41567-020-0797-9}. Motivated by these properties, pioneering work in the past two decades has enabled sensitive measurement and control of mechanical oscillators in the quantum regime via optical and electrical interfaces, making them viable candidates for quantum sensors, memories, and transducers \cite{SafaviNaeini:2019ci,10.1364/optica.425414}.

While optomechanical experiments have been successful in measuring phonons with millisecond-to-second lifetimes \cite{maccabe2020b,wallucks2020}, 
accessing long-lived mechanical resonances with electrical circuits has been more challenging. In the gigahertz frequency range, where the spectral proximity to superconducting qubits holds the most promise for quantum technologies, piezoelectricity is the predominant mechanism for converting microwave photons to phonons. Piezoelectric devices have been used with remarkable success in coupling mechanical modes to superconducting qubits \cite{10.1038/s41586-018-0719-5, 10.1038/s41567-022-01591-2,wollack2022}. However, their need for hybrid material integration, sophisticated fabrication process, and reliance on lossy poly-crystalline materials \cite{wollack2021b} has limited the state-of-the-art experiments to sub-microseconds mechanical lifetimes in devices with compact geometries \cite{wollack2022,10.1038/s41586-020-3038-6}. This evidently large gap between the mechanical lifetimes accessible to optical and electrical interfaces motivates pursuing less invasive forms of electromechanical interaction. Creating better electrical interfaces for long-lived phonons holds the potential for revolutionizing our current quantum toolbox by pairing the superior coherence of acoustics with the massive nonlinearity of Josephson junction circuits \cite{chu2020}.




Here, we realize electromechanical coupling between microwave photons in a superconducting circuit and long-lived phonons in a 5-GHz crystalline silicon oscillator. To achieve this, we rely on \emph{electrostatic transduction}, where we use a static electric field, as opposed to conventionally used radio-frequency drives, to realize a parametrically-enhanced interaction in a microwave cavity with a motion-dependent capacitor. The absence of alternating currents from the driving field in this scheme eliminates conductive loss, allowing us to achieve large parametrically-enhanced coupling rates without causing heating in the system. To further enhance electromechanical interactions, we rely on frequency-tunable high-impedance microwave resonators made from TiN superinductors. Relying on these innovations, we are able to demonstrate electromechanical interaction in the strong coupling regime, enabling the coherent exchange of microwave photons and phonon at a cooperativity of $\mathcal{C}\approx1200$. We measure mechanical lifetimes in the few-phonon regime, demonstrating quality factors in excess of 8 million (at 5 GHz) in our best devices. To the best of our knowledge, this is the highest value measured via electrical interfaces in this frequency band. Crucially, we observe no parasitic heating for a large range of electrostatic biasing fields in our system, allowing us to operate in the quantum ground state as verified by calibrated sideband thermometry measurements in a dilution refrigerator. The combination of long lifetimes, strong interaction, and the compact geometry of our platform promises future experiments in employing mechanical modes as microwave-frequency quantum memories, improved microwave-optical transducers, and new measurement capabilities for exploring the origins of acoustic loss crystalline materials.

\section*{A phononic crystal electrostatic transducer}
The operating principles of our experiment can be understood by considering a capacitor with mechanically moving electrodes connected to an external DC voltage source. In this setting, the mechanical vibrations of the capacitor electrodes create a time-dependent dipole oscillating at mechanical resonance frequency. Connecting this charged moving capacitor (i.e. the \emph{transducer}) to an electromagnetic cavity (see \cref{fig:dev}a) leads to an interaction between the voltage operator of the photons in the cavity $(\hat{V}/V_\mathrm{zpf} = i(\hat{a}-\hat{a}^\dagger))$ and the quantized mechanical displacement operator $(\hat{x}/x_\mathrm{zpf} = (\hat{b}+\hat{b}^\dagger))$. This interaction can be described by the Hamiltonian (see \cref{APP.1})
\begin{equation}
\label{eq:1}
    \hat{H}_\mathrm{int} = i \left(x_\mathrm{zpf}\partial_x C\right) V_\mathrm{zpf} V_\mathrm{DC}(\hat{a}\hat{b}^\dagger - \hat{a}^\dagger\hat{b}) = i\hbar g_\mathrm{em}(\hat{a}\hat{b}^\dagger - \hat{a}^\dagger\hat{b})
\end{equation}
Here, $x_\mathrm{zpf}$ and $V_\mathrm{zpf}$ represent the zero-point motion and voltage of the phonon and photon fields, respectively. The coupling rate is a function of the geometry (through $\partial_x C$) and the applied bias voltage $V_{\text{DC}} $, and arises as a result of the change in the stored electrostatic energy as a function of mechanical motion. The DC voltage in this process can be understood as a `pump' in a parametric process \cite{teufel2011b}. Unlike the conventional parametric electromechanics, however, the pump is solely comprised of electric fields at zero frequency and is not accompanied by alternating currents. As we will see, this distinction is crucial in our experiment because it increases the net coupling rate at large voltages without being limited by the dissipation in the superconducting cavity \cite{peterson2019a,kalaee2019}.

Despite its conceptual simplicity, electrostatic transduction is challenging to realize at GHz frequencies \cite{rouxinol2016a,vanlaer2018}. Getting substantial coupling requires increasing the motion-dependent capacitance and the zero-point displacement. This combination has been previously achieved in low-mass, narrow-gap suspended capacitors, which support MHz-frequency mechanical resonances \cite{teufel2011a}. However,the frequency scaling of acoustic loss in metals (speculated to be caused at grain boundaries \cite{mason1947,zeng2010,wollack2021b}) makes these structures unsuitable for GHz frequencies. Additionally, the short wavelength of GHz-frequency phonons leads to increased acoustic radiative loss to the surrounding environment, making it challenging to localize high-$Q$ resonances.

Our solution is to utilize planar nano-structured devices made from crystalline silicon membranes. Adding a thin layer of metal on top of the membrane allows us to form a capacitor in this platform while relying on phononic crystals to engineer localized resonances. Our transducer consists of a phononic crystal resonator made of a periodic array of multiple unit cells (see \cref{fig:dev}b), patterned on the inner electrode of a vacuum-gap capacitor. The in-plane movement of the `breathing' mechanical mode in this structure leads to the modulation of the capacitance. We have maximized the rate of change of this motion-dependent capacitance by fabricating capacitors with narrow gaps in the range of 65-70 nm. Additionally, we have maximized the capacitance by increasing the number of phononic crystal unit cells to the limit set by the onset of disorder effects, which lead to mode breakup as observed in finite-element simulations (see \cref{APP.3}). A key benefit of this planar geometry is the possibility of creating phononic crystals with a wide band gap for all phonon polarizations \cite{maccabe2020b}. We use these `phononic shields' for clamping the transducer to its surrounding membrane (see \cref{fig:dev}c, d). 

The size mismatch is a central challenge in coupling the presented phononic crystal transducer to a microwave circuit. Set by the small wavelength of phonons ($\sim 1~ \mu$m at the target frequency of 5 GHz), the small size of the transducer translates to a motion-dependent capacitance that is much smaller than the typical capacitance of a microwave cavity. This mismatch leads to a poor electric energy density overlap (see \cref{APP.1}), which dilutes the electromechanical interaction. Formally, this effect is captured by a linear dependence of the electromechanical coupling to the zero-point voltage of microwave photons ($V_\mathrm{zpf}$, see \cref{eq:1}). Recent progress in developing circuits for error-protected superconducting qubits has led to established techniques \cite{shearrow2018a, 10.1103/PhysRevLett.122.010504, 10.1038/s41586-020-2687-9} for magnifying zero-point voltage of microwave photons in high-impedance resonators. In our experiment, we achieve this by galvanically connecting the transducer to a microwave resonator (\cref{fig:dev}d) consisting of a 110-nm wide nanowire formed from thin-film titanium nitride (thickness $\sim 15$ nm). The inertia of charge carriers in this disordered superconductor leads to a large inductance, which is enhanced by patterning structures with small cross sections \cite{zmuidzinas2012, shearrow2018a}. Beyond magnifying the electromechanical interactions, kinetic inductance is tunable via external magnetic fields, which allows us to control the resonance frequency of the microwave cavity in-situ \cite{xu2019a}.


\section*{Characterization and measurements}
\begin{figure}[t]
\centering\includegraphics[width=\linewidth]{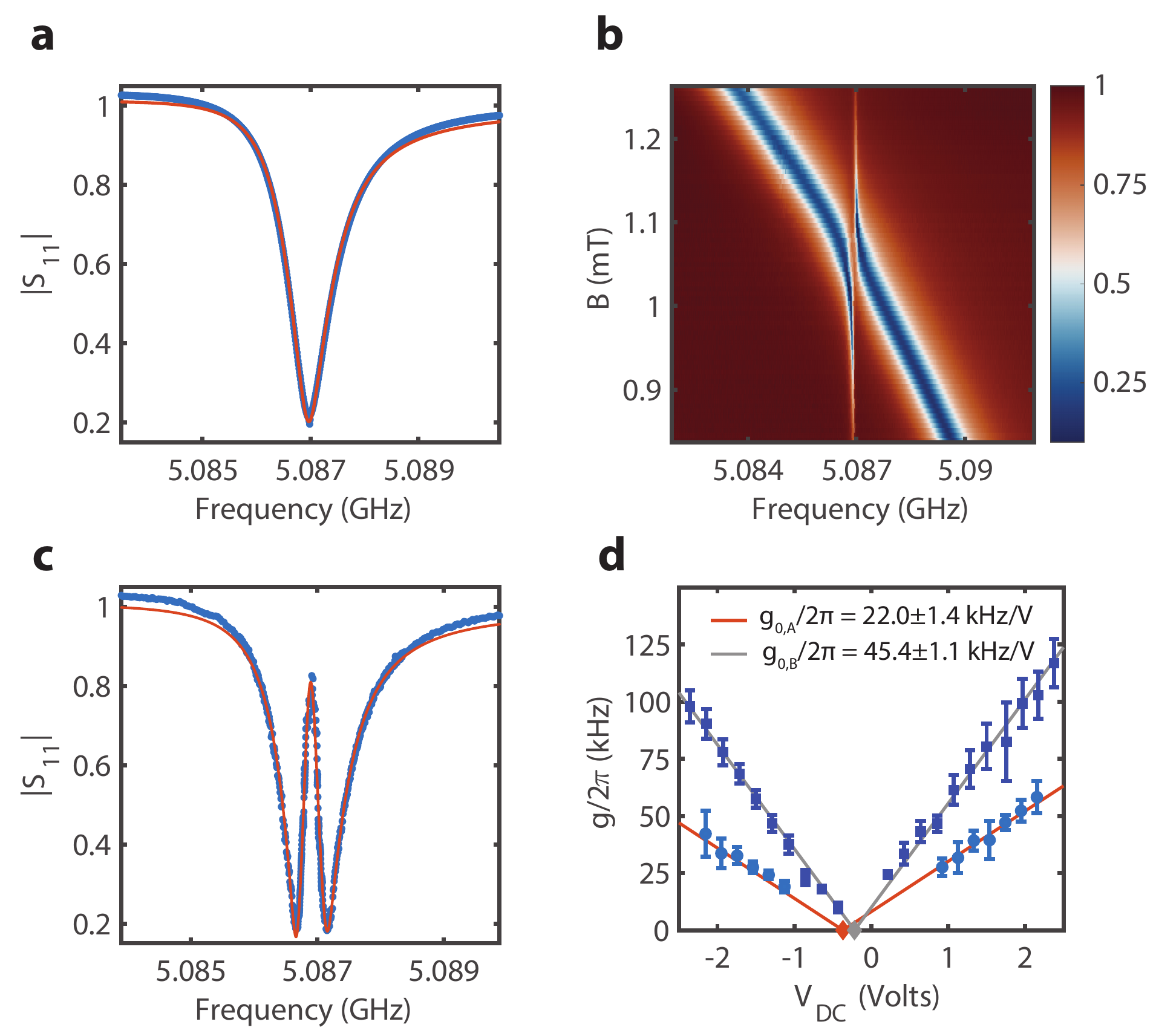}
\caption{\textbf{Electromechanically-induced transparency.} (a) The reflection spectrum of a TiN microwave cavity with the DC voltage set to zero and with the cavity detuned from the mechanical resonance. From the theoretical fit (red line) we find  $\kappa_i/2\pi = 520\,\text{kHz}$ and $\kappa_e/2\pi = 800\,\text{kHz}$.(b) The reflection spectrum of the cavity biased at $V_\text{DC} = 10 ~\text{V}$ as a function of the external magnetic field showing a signature of the mechanical mode at 5.087 GHz. (c) A cross-section of the data at (b), showing the reflection spectrum of the resonant microwave-mechanics system. The red lines depicts a theory fits to the EIT lineshape. (d) The extracted coupling rates from EIT traces at different voltages, plotted for two devices. The zero-coupling voltage offset voltages for device A (B) is -0.36V (-0.22V). The data in parts (a-c) are taken with device A.}
    \label{fig:EIT}
\end{figure}


We have characterized fabricated electromechanical resonators in a dilution refrigerator with a base temperature of 20 mK (see fabrication details at \cref{APP.methods}). A coplanar waveguide is connected to the device for simultaneously applying DC voltages to our mechanical capacitor and probing our microwave resonator in reflection via its coupling to the waveguide. In the absence of electromechanical coupling, we can measure the bare microwave cavity response using a vector network analyzer (VNA, see \cref{fig:EIT}a). To locate the mechanical resonance, we apply a DC voltage and continuously tune the frequency of the microwave resonator via an external magnetic field (see \cref{APP.2}). The electromechanical interaction leads to a large reflection at the point where the microwave and mechanical frequencies cross, in a phenomenon known as the  electromechanically-induced transparency (EIT) (\cref{fig:EIT}b) \cite{teufel2011a}. We extract the electromechanical coupling rate using a fit to the theory expressions for the EIT response (see supplementary \cref{APP.1}, \cref{fig:EIT}c), and plot it as a function of the  applied voltage in \cref{fig:EIT}d. As evident, the coupling rate is found to be a linear function of the voltage bias with a slope ($g_\text{0,B}/2\pi = 45.4 \pm 1.1 \,\text{kHz/V}$) closely matching the results from numerical modeling (see \cref{APP.3}). In addition, we present measurement results from a second device with an identical geometry, but with a small coupling $g_\text{0,A}/2\pi = 22.0 \pm 1.4 \,\text{kHz/V}$, which is likely impacted by mode-breakup due to fabrication disorder (see \cref{APP.3}). An interesting feature in the data is the small non-zero coupling at the zero-voltage bias, where zero coupling is achieved at a negative offset voltage. This feature is found to be persistent in a shorted capacitor geometry and in the absence of a voltage source, which may indicate the presence of trapped charges in the transducer \cite{fleetwood1993}.

\section*{Coherence Properties}
\begin{figure*}[]
\centering\includegraphics[width=16.5cm]{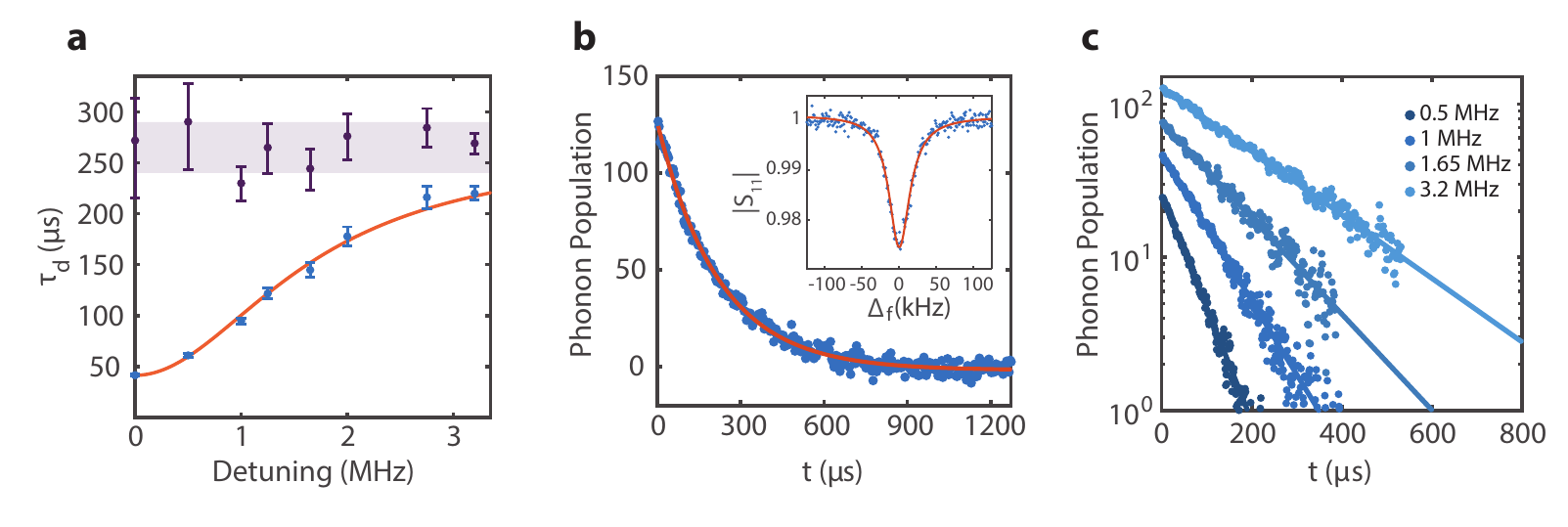}
\caption{\textbf{Mechanical lifetime measurements}.  (a) The total lifetime ($\tau_\text{d}$) extracted from ringdown measurements for different mechanics-microwave detunings $\Delta/2\pi$ (blue points). The red line shows a theory fit, finding $\tau_\text{d,i} = 265 \pm 25~\mu s$. This range is indicated by the purple shaded region. The purple data points show $\tau_\text{d,i}$ found at each detuning by subtracting the contribution from the electromechanical readout. (b) Free energy decay (ringdown) data for the maximum detuning of 3.2 MHz. The fit finds total lifetime of $220 \pm 6~\mu s$. Inset: Mechanical spectrum measured at a probe power corresponding to a maximum of 4 phonons in the mechanical resonator. The theory fit gives a linewidth of 33 kHz, corresponding to $\tau_\text{c} \approx 5~\mu s$. (c) Log-scale ringdown plots as a function of the phonon number inside the mechanical resonator. All data is from device A with a DC voltage of 1.2 V.}
    \label{fig:coh}
\end{figure*}



We next investigate the coherence properties of our devices by exciting the mechanical resonator with a pulse and registering its free decay via electromechanical readout (see \cref{APP.5}). In this measurement, the electromechanical readout rate is a function of the detuning between the mechanical and microwave resonances $\Gamma_\text{em} = g^2\kappa/(\Delta^2+(\kappa/2)^2)$. At any given detuning, the total decay rate of the mechanics is given by $\Gamma = \Gamma_i + \Gamma_\text{em}$, where $\Gamma_i$ is the intrinsic decay rate. In order to precisely measure $\Gamma_i$, we perform multiple measurements where we gradually increase $\Delta$, leading to a gradual reduction in $\Gamma_\text{em}$ until the total decay becomes dominated by the intrinsic part. Fitting the total decay rate expression as a function of $\Delta$, we extract an intrinsic $\tau_\text{d}$ lifetime of $265 \pm 25~\mu s$  (\cref{fig:coh}a), corresponding to a Q factor of $8\times10^6$.  At the largest detuning we achieve, we can reach the regime where the intrinsic decay dominates the dynamics and obtain a total energy decay lifetime $\tau_\text{d}$ of $220 \pm 6~\mu s$ (\cref{fig:coh}b), which strongly supports the large intrinsic lifetime inferred from the fits. We note that the measured mechanical lifetime is remarkably large for a device with compact geometry, corresponding to a quality factor that is more than two orders of magnitude large than the state-of-the-art piezoelectric devices \cite{wollack2022}.

Apart from the energy relaxation lifetime, the coherence time of our mechanics bears significance for future quantum applications. We find the coherence time as the reciprocal of the linewidth extracted from fitting the EIT response in the large detuning regime ($\Delta >> \Gamma_i$). Using this method, we find  $\tau_\text{c} \approx ~5~\mu s$ (\cref{fig:coh}b), a value that is substantially shorter than the lifetime. This discrepancy between the decay and coherence times hints at the presence of frequency jitter and has been previously observed in optomechanical experiments with silicon phononic crystal resonators. While the exact nature of this dephasing remains unknown, it has been hypothesized to be caused
by interaction with the two-level system (TLS) defects \cite{maccabe2020b,wallucks2020}. To better understand possible dephasing and decay sources, we carry out ringdown measurements for different numbers of phonons in the mechanical resonator. We change the intra-cavity phonon number in accordance with the readout efficiency $\Gamma_\text{em}/\Gamma$, keeping the output detection powers nearly constant at the lowest levels detectable by our amplifiers to ensure we can reach the low-phonon regime while maintaining feasible measurement times. Interestingly, we observe exponential decay with no sign of power dependence down to the single-phonon level (see \cref{fig:coh}c). Similarly, we find the coherence time extracted from EIT measurements to be insensitive to phonon numbers in the range of 4-500 (see \cref{APP.5}). Although we do not find signatures of saturable loss (a common signature of a TLS bath) in this measurement setting, we find that the coherence and decay times change as a function of the applied DC voltage, providing direct evidence for the presence of TLS defects. This behavior is explored in detail further in the manuscript.

\section*{Probing the limits of parametric-enhancement}
\begin{figure}[t]
\centering\includegraphics[width=\linewidth]{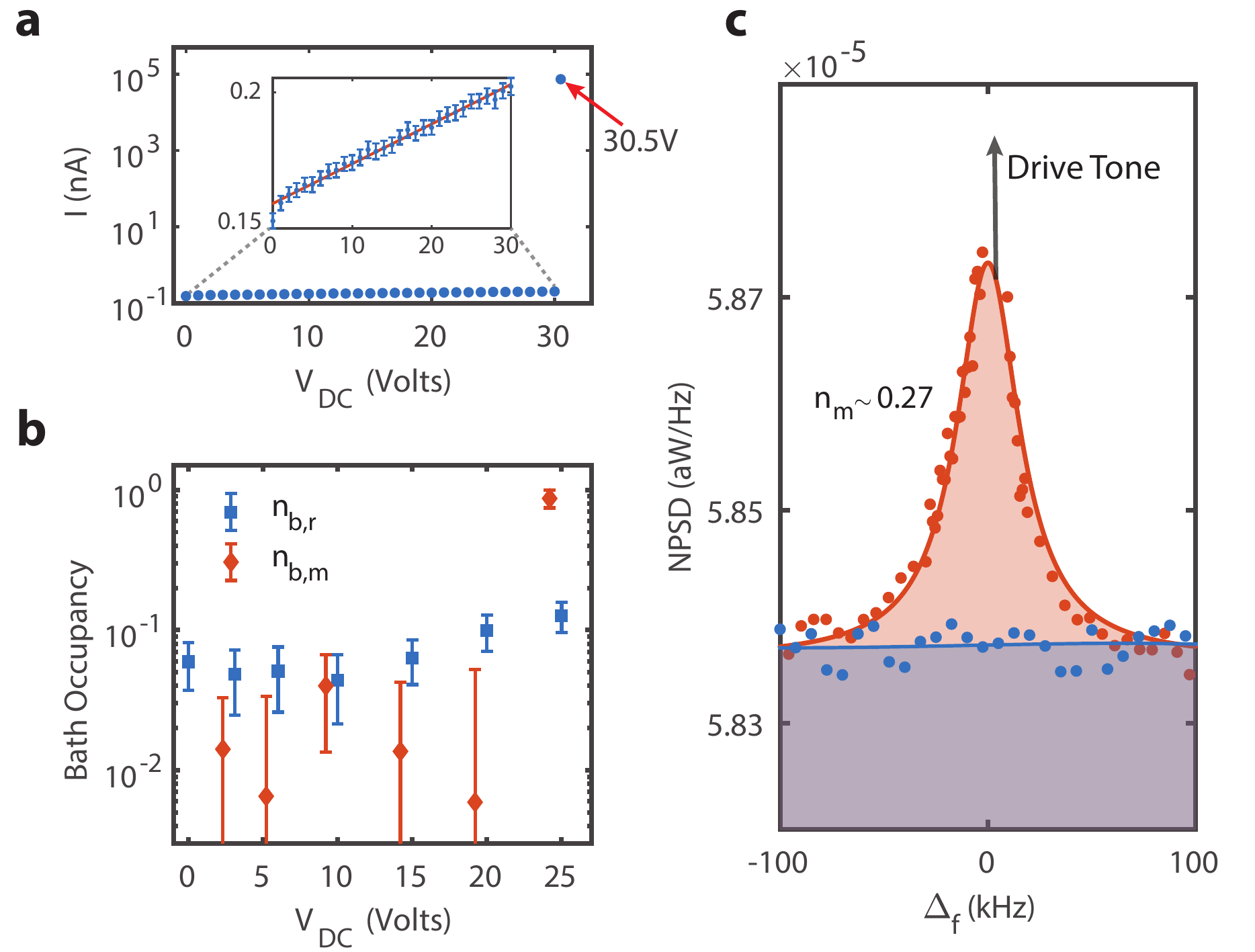}
\caption{\textbf{Probing the limits of parametric-enhancement.} (a) The leakage current through the transducer as a function of the applied voltage, measured with a picoammeter. Errorbars are too small to be visible. Inset: A zoomed-in view of the data for the range 0-30 V, with a line fit to a $500 ~\text{G}\Omega$ resistance. (b) The extracted mechanics ($n_\text{b,m}$, red points) and microwave bath ($n_\text{b,r}$, blue points) occupancy as a function of the applied voltage. We have added a small horizontal offset between the two data sets for greater visibility. (c) The driven response of the mechanics at $3.1$ V (red points) detected by a spectrum analyzer. The solid line shows a fit to a Lorentzian response. The mechanical resonator is populated with $0.27$ phonons due to the inelastic scattering of the drive tone. The detected emission from the mechanical resonator at 15 V in the absence of the drive (blue points). The background level is set by the added noise from our amplifier chain. All data is from device A. }
    \label{fig:ther}
\end{figure}


A naturally emerging question is about the maximum achievable rate of the electromechanical interactions in our platform. This rate is set by the magnitude of the DC voltage that can be applied before the onset of any spurious heating or instabilities in the system. 

We do not expect to see any significant leakage current passing through the devices because of the freezing of the charge carriers at the low measurement temperatures. However, applying large voltages to the narrow-gap capacitors in our devices leads to strong electric fields which may lead to ionization and dielectric breakdown \cite{gutierrez-d.2001}. We measure the leakage current through the transducer structures as a function of applied voltage. As evident in \cref{fig:ther}a, we see a leakage current (with a characteristic resistance of $\sim500 ~\text{G}\Omega$) that is found to be dominantly caused by the cables in our measurement, bounding the actual leakage through the device to below the measurement sensitivity of our measurement. When increasing the voltage beyond 30 V, we observe an abrupt spike in the leakage current that was initially attributed to dielectric breakdown. However, imaging of our devices post warm-up at the room-temperature indicates that the sudden leakage is most likely due to the onset of pull-in instabilities, resulting in the shut-down of the capacitor gap (see \cref{APP.4}). Repeating this experiment on four identical test devices, we find the onset of this instability in a consistent range ($29$-$31$ V).

To probe the signatures of any potential pump-induced heating (previously observed in similar devices under radio-frequency drives \cite{kalaee2019}), we perform side-band thermometry \cite{teufel2011b}. The thermometry measurement process for the mechanical resonator is visualized in \cref{fig:ther}c. We drive the system with a weak tone and measure the incoherent emission from the inelastic scattering of the drive to locate the mechanical resonance (see \cref{APP.4}) \cite{zhang2014}. A subsequent measurement with no drives shows a negligibly small emission, which is calibrated in experiments with long averages to extract the resonator's occupation. We have found the system of mechanics-microwave to be in the quantum ground state for the entire range of applied voltages in our device (0-25 V). Despite observing no heating, we note that the presence of a strong electromechanical back-action cooling (for the mechanical mode) and radiative cooling through the on-chip waveguide (for the microwave cavity) may mask a weak heating process. To find a more sensitive trace of any potential heating, we perform thermometry in a regime where the microwave cavity is detuned far away from the mechanical resonator to reduce the effect of electromechanical back-action cooling and deduce the temperature of the phenomenological intrinsic baths that the mechanical and microwave resonators interact with (see \cref{APP.4} for details).  \Cref{fig:ther}b shows the measured microwave ($n_\text{b,r}$) and mechanical bath occupancy ($n_\text{b,m}$) as a function of the applied DC bias. As evident, the baths remain in the ground state for a large voltage range, with the exception of the mechanical bath at 25 V, where the occupancy rises to $0.86\pm0.08$ phonons (attempts to do measurements at higher voltages were unsuccessful due to the onset of the pull-in instability). The slightly higher occupation of the microwave bath across all voltages (including 0 V) is likely caused by the absence of IR-shielding in our measurement setup, which leads to the generation of quasiparticles, and can raise the microwave bath temperature in devices with a large kinetic inductance \cite{grunhaupt2018,serniak2018}. Regardless, despite the observation of these subtle features testifying to the accuracy of our measurement technique, we do not observe any indication of significant pump-induced heating in the mechanical resonator or the microwave cavity for a wide range of voltages, which is a promising feature of our electrostatic driving scheme.

\section*{The Strong-Coupling Regime}
\begin{figure}[t]
\centering\includegraphics[width=\linewidth]{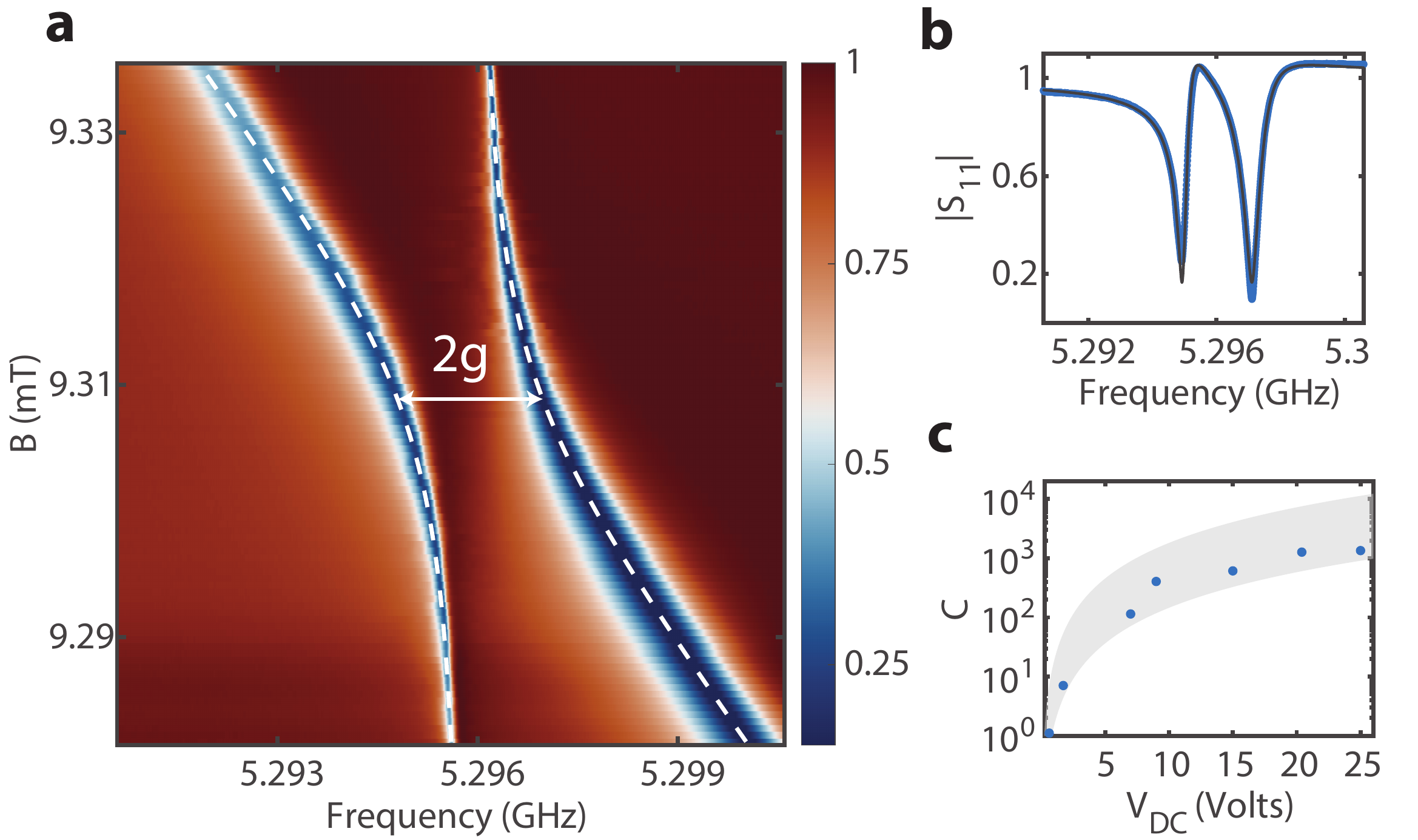}
\caption{\textbf{Demonstration of strong-coupling regime.} (a) Measured reflection spectrum showing avoided crossing between the mechanical mode biased at $V_\mathrm{DC} = 25 $V and the microwave cavity tuned with a magnetic field. The theory fit (dashed lines) gives a parametrically-enhanced coupling rate of $g/2\pi = 1.08$ MHz. (b) A cross-section of the data in (a) showing the splitting between the two hybridized modes on resonance. The black solid line is a theory fit which includes the Fano effect accounting for the observed asymmetry. (c) The electromechanical cooperativity measured at different voltage values. The intrinsic decay rates are obtained by ringdown measurements. The data points in all parts are taken with device B.
The grey shaded region shows the range of cooperativity values calculated using the maximum values of couplings and lifetimes measured in different devices.  }
    \label{fig:str}
\end{figure}
 
 Having discovered the safe range of voltages we can apply to our devices, we investigate the maximum electromechanical coupling rate within reach. For this purpose we use the device B, which has larger $g_0$. Setting the external voltage to 25 V, we sweep our microwave frequency by changing the external magnetic field, which gives rise to an avoided crossing feature between the microwave and mechanics as seen in \cref{fig:str}a. Fitting the frequencies of the two hybridized modes, we extract a parametrically-enhanced electromechanical coupling rate of $g = 1.08\,\text{MHz}$. This value corresponds to $g_0 = 42.7 \,\text{kHz/V}$ which matches the values we have calculated at low voltages, indicating that the parametric enhancement scales linearly with voltage in the entire measurement range. Achieving the strong-coupling regime is manifested clearly in the measurements of the reflection spectrum at resonance, where we see a pair of hybridized modes with linewidths of $(\kappa+\gamma)/2 = 692$ kHz, satisfying $2g > \kappa+\gamma$ (\cref{fig:str}b). Achieving the strong-coupling regime in our system is significant as it allows the coherent exchange of phonons and microwave photons, a pre-requisite for utilizing electromechanical systems in a range of quantum applications.  Another key figure of merit in coupling mechanical modes to qubits and microwave resonators is cooperativity, defined as the ratio of the electromechanical readout rate to the intrinsic mechanical decay rate $\mathcal{C} = 4g^2/(\kappa_i \Gamma_i)$. We use the measured electromechanical coupling rates and the mechanical intrinsic decay rates (from the ringdown measurements) to find the cooperativity as a function of bias voltage (see \cref{fig:str}b). For the maximum voltage value of 25 volts, we find $\Gamma_i/2\pi = 4.8~ \text{kHz}$ ($\tau_\text{d}$ = $33~ \mu s$), corresponding to a cooperativity of 1270 (see \cref{APP.4}). As a guide, we also mark the values of cooperativity assuming the maximum coupling rates and lifetimes measured across different devices on the same plot, finding estimates that exceed ${10}^4$ at 25 V. 


\section*{Interaction with Two-level Systems}
\begin{figure*}[!t]
\centering\includegraphics[width=14cm]{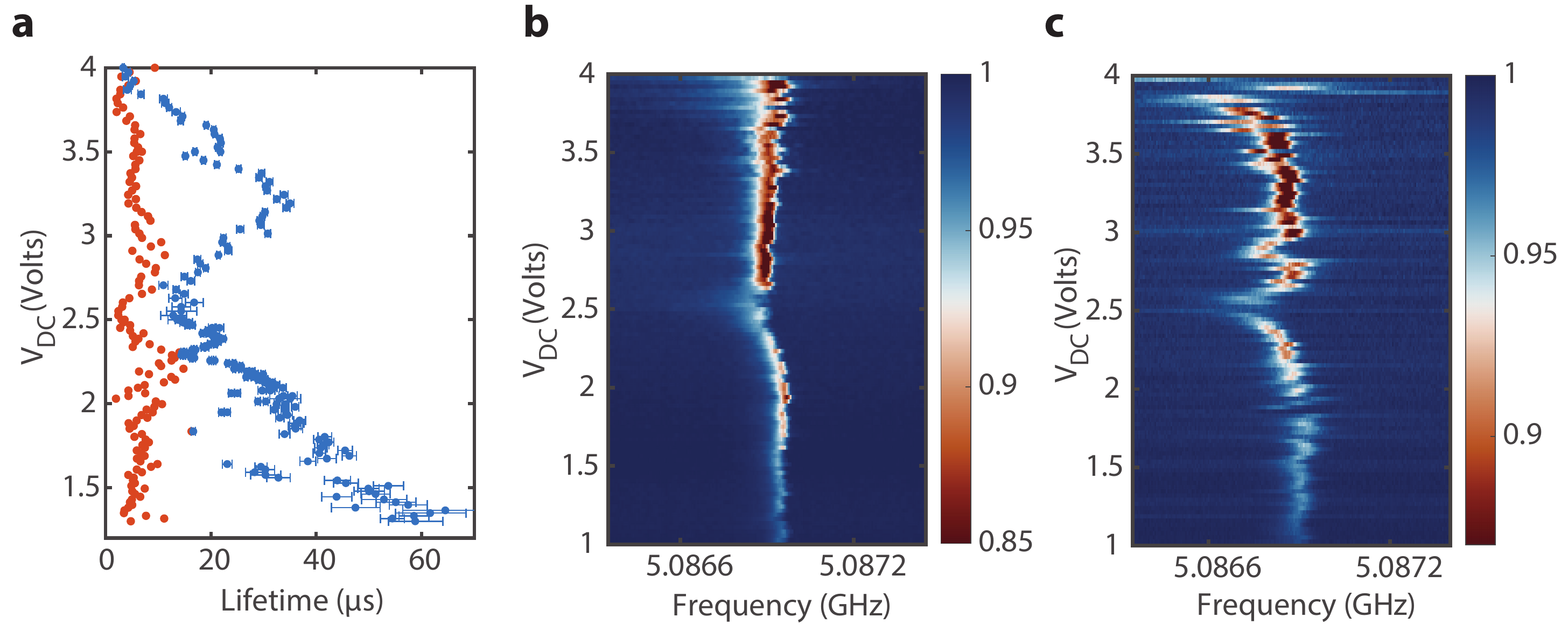}
\caption{\textbf{Interaction of the mechanical resonator with the TLS bath}.  (a) The measured decay time ($\tau_\text{d}$, blue data points) and coherence time ($\tau_\text{c}$, red data points) measured as a function of bias voltage. The probe power level is set to -39 dbm. (b) The mechanical reflection spectrum at a probe power of -40 dBm. (c) The measurement in part (b) repeated with a probe at -52 dBm power. Probe powers at $2.5$ V are reported as a reference. For each voltage point, the power level is adjusted to retain the intra-cavity phonon number in the range 500-3000. All data is from device A.}
    \label{fig:TLS}
\end{figure*}

As noted earlier, we suspect that the mechanical dephasing in our measurements can be attributed to coupling to TLS, which is previously shown to be the dominant loss mechanism for acoustic resonators with substantial surface participation at millikelvin temperatures \cite{maccabe2020b,wollack2021b}. Modeled phenomenologically as two nearly-degenerate energy conﬁgurations of electrons in amorphous materials, a TLS manifests as a resonant defect with both electrical and acoustic susceptibilities \cite{Phillips:1987ge,muller2019}. Using previous theory work as a guide \cite{ramos2013} and extracting the spatial distribution of the strain field from FEM numerical modeling, we estimate the spectral density ($3/$ GHz) and the coupling rate ($13$ MHz) of individual TLS defects to the nanomechanical resonators in our system. The large coupling rate and the small density suggest a departure from the continuum TLS-bath picture observed in the past work \cite{10.1103/physrevb.93.041411, 10.1038/s41534-020-00348-0}, and offers the possibility of observing mechanics-TLS interactions at the individual defect level.

To make this observation, we take advantage of the TLS frequency tuning via the Stark shift \cite{lisenfeld2019} from the electrostatic bias in our system (estimated to create TLS frequency shifts at a rate of $20$ GHz$/$V, see \cref{APP.5}). \Cref{fig:TLS}a shows the measured decay and coherence times as a function of voltage, manifesting strong modifications indicative of interaction with TLS defects. Further, we make a measurement of the mechanical spectrum as a function of voltage, observing abrupt frequency shifts (in form of avoided crossings) commensurate with the changes in the coherence and decay times. Finally, by comparing measurement results at two probe powers, we see a saturation behavior in the vicinity of the voltage values where the mechanical mode is heavily affected by TLS (see \cref{fig:TLS}b,c). A more detailed measurement of mechanical linewidth as a function of phonon number at these points provides a good fit to the widely used TLS model (see \cref{APP.5}). We note that in our device we cannot fully saturate the signatures of the TLS before the onset of acoustic Kerr nonlinearities at large phonon numbers \cite{bachtold2022}.


\section*{Conclusions and Outlook}
In conclusion, we present an integrated cavity electromechanical system capable of achieving MHz-level coupling rates at a mechanical frequency of several GHz. Using this system, we demonstrate achieving the strong coupling regime, with a cooperativity exceeding 1200. Relying on an electrostatic driving field, we are able to obtain a large parametric enhancement of the interaction with negligible parasitic heating, leading to operation in the quantum ground state. Device fabrication is performed using a TiN-on-SOI material system, which is compatible with superconducting qubits and optomechanical crystals. Additionally, by relying on thin films and single-crystalline silicon, we are able to show mechanical quality factors above 8 million, corresponding to two orders of magnitude improvement over piezoelectric devices in similar geometries \cite{10.1038/s41586-020-3038-6,wollack2022}. Finally, we note the material-agnostic nature of the underlying process in our experiment, which holds the potential for adoption in platforms hosting spin qubits \cite{10.1103/physrevx.9.031022}.

Looking ahead, we envision several avenues for further improvement of the presented devices. The electromechanical coupling rates can be readily increased by multiple folds upon integration of electrostatic transducers with microwave cavities with ultra-high impedance \cite{10.1103/physrevapplied.14.044055,10.1038/s41586-020-2687-9}, reaching full parity with piezo-electric platforms. Additionally, while we observe record-long lifetimes in devices with electrical connectivity, our measurements remain much shorter than the second-long results from optomechanical experiments in silicon structures with no metallic components \cite{maccabe2020b,wallucks2020}. This observation motivates a systematic study of the sources of residual acoustic loss, including the role of metallic components, fabrication disorder in the acoustic shields, and two-level-system defects. A better understanding of the loss mechanisms along with the implementation of proper mitigation techniques is expected to lead to longer mechanical lifetimes. With moderate improvements, achieving the millisecond regime is expected to be within reach in near future, with the potential to deliver transformative impacts on mechanics-based microwave-optical interconnects \cite{10.1364/optica.425414}, error-protected bosonic qubits \cite{10.1103/prxquantum.3.010329}, and quantum memories \cite{10.1103/physrevlett.123.250501,Pechal:tm}.

 \begin{acknowledgments}
We thank Oskar Painter and Mahmud Kalaee for fruitful discussions, which led to the conception of this work. This work was supported by the startup funds from the EAS division at Caltech, National Research Foundation (grant No. 2137776), and a KNI-Wheatley scholarship. C.J. gratefully acknowledges support the IQIM/AWS Postdoctoral Fellowship. M.M gratefully acknowledges support from the Q-NEXT.
\end{acknowledgments}

\appendix
\section{Methods}
\label{APP.methods}
\subsection{Fabrication}

A 4-inch silicon on insulator (SOI) wafer is utilized at the beginning of the fabrication process, and is covered with $\sim 15$ nm TiN films via sputtering \cite{leduc2010}. Following sputtering, the wafer is diced into $1\times1 \,\text{cm}^2$ dies. For the following steps the patterning of the structures is achieved by electron beam lithography. (i) Deposition of niobium markers followed by lift-off. (ii) Inductively coupled plasma - reactive ion etching (ICP-RIE) of TiN and Si with $\text{SF}_6/\text{Ar}$ and $\text{SF}_6/\text{C}_4\text{F}_8$ chemistry. These etching steps are used to define the capacitor vacuum gap, phononic crystal nanobeam, phononic shields and the release holes throughout the metalized and bare sections of the Si membrane. The devices are released with Hydrogen Fluoride (HF) post-fabrication.

\subsection{Measurement Setup}
The chip is wirebonded to a PCB, which is placed into a copper box and then mounted to the mixing stage of the dilution refrigerator at $\sim 15\,\text{mK}$. The box has a coil on top for magnetic field tuning the microwave resonators. The coil is obtained by hand winding a superconducting wire around a cylindrical extrusion.

The device is measured in reflection with the aid of a cryogenic circulator. A bias tee is placed between the chip and the circulator to enable DC biasing of the transducer and readout of the microwave cavity via the same CPW. The RF input line consists of multiple cascaded attenuators with a total attenuation of 74 dB. A tunable attenuator is further added to the input line to control the input power in a programmable manner. The DC input has no attenuation  and is directly attached to the bias tee (low frequency transmission band up to 500 MHz). At the output, we have an amplifier chain that consists of a HEMT amplifier thermalized to the 4K stage and a room temperature amplifier, with a total gain of $\sim 65\,\text{dB}$.

The external DC voltage for the electromechanical interaction and the current for the tuning coil are applied via a multi-channel programmable low-noise DC source. Since, significant amount of current is required to tune the microwave resonators, the normal metal parts in the coil wiring leads to spurious heating of the mixing stage. The coherent response of the mechanics-cavity system is probed via a vector network analyzer (VNA) in reflection. For thermometry and investigations of the full driven response of the system, the microwave emission is detected by a spectrum analyzer. For time-resolved measurements, we use Quantum Machines OPX+ module.  This tool enables the generation of pulses with an arbitrary waveform generator (AWG), heterodyne detection, demodulation of signals via a digitizer, and the processing of detected signals with an FPGA. 

\subsection{Device Parameters}
As part of our experiment we have fabricated and extensively characterized two devices. The parameters obtained from the measurements are described in \cref{param_table}. Based on these parameters, we select different devices to exhibit specific features of our platform. Due to its larger electromechanical coupling strength, device B can enter the strong coupling regime and attain large cooperativities above 1000. On the other hand, device A has the largest energy decay lifetime we have observed in our platform and is well suited to demonstrate the substantial lifetimes that can be achieved with our devices. Furthermore, as it requires less frequency tuning (due to a smaller mechanics-cavity detuning at zero magnetic field), we are able to measure it at a lower  temperature (20mK), where the heating due to the current on the coil is minimized, which makes it more suitable to perform sensitive sideband thermometry measurements and investigate TLS physics without thermal saturation of TLS.
\begin{table*}[!tb]
\caption{Summary of device parameters. The referenced maximum decay lifetime and the coherence time are at the few-phonon level.}
\begin{center}
\begin{tabular}{c || c c c c c c c c}
\label{param_table}
Device & $\omega_m/2\pi$ (GHz) &$\omega_r^\text{max}/2\pi$ (GHz)& $\kappa_i/2\pi$ (kHz) & $\kappa_e/2\pi$ (kHz) &  $g_0/2\pi$ (kHz/V) & $\tau_\text{d}^\text{max} \,(\mu\text{s})$ & $\tau_\text{c}^\text{max}\,(\mu\text{s})$ & T$_\text{MXC}$ (mK)  \\
\hline
A & 5.087 & 5.096 & 520 & 800  & 22.0 & 265 & 8 & 20 \\
B & 5.296 & 5.483 & 775 & 490 & 45.4 & 77 & 2 & 90 \\
\hline\hline
\end{tabular}
\end{center}
\end{table*}
\section{Electrostatic Interaction}
\label{APP.1}
\subsection{Hamiltonian Derivation}

The interaction term in the Hamiltonian for a capacitor with mechanically moving electrodes ($C_{\text{m}}$) is given as
\begin{equation}
    H_\text{int} = \frac{\hat{q}^2}{2C_{\text{m}}} \left({\frac{1}{C_{\text{m}}}\frac{\partial C_{\text{m}}}{\partial x}\hat{x}}\right).
\end{equation}

The displacement operator can be written as $\hat{x} = x_{\text{zpf}}(\hat{b} + \hat{b}^\dagger)$. We have both electrostatic charge due to external voltage source and RF charge associated with the microwave resonance on top of the capacitor. This leads to a charge operator which can be written as $\hat{q} = iQ_{\text{zpf}}(\hat{a} - \hat{a}^\dagger) + Q_{\text{DC}}$. Inserting these operators into the Hamiltonian, we obtain
\begin{equation}
    H_\text{int} = \frac{1}{2} \frac{\partial C_{\text{m}}}{\partial x} x_{\text{zpf}} \left[i\frac{Q_{\text{zpf}}}{C_{\text{m}}}(\hat{a} - \hat{a}^\dagger) + \frac{Q_\text{DC}}{C_{\text{m}}}\right] ^2(\hat{b} + \hat{b}^\dagger).
\end{equation}
Noting that $Q_\text{zpf}/C_{\text{m}} = V_{\text{zpf}}$, $Q_{\text{DC}}/C_{\text{m}} = V_{\text{DC}}$ and expanding the terms
\begin{multline}
    H_\text{int} = \frac{1}{2} \frac{\partial C_{\text{m}}}{\partial x} x_{\text{zpf}} (\hat{b} + \hat{b}^\dagger)\times\\
    \left[-V_{\text{zpf}}^2(\hat{a} - \hat{a}^\dagger)^2 + V_{\text{DC}}^2 + 2iV_{\text{zpf}}V_{\text{DC}}(\hat{a} - \hat{a}^\dagger)\right]. 
    \label{eq:fullHamiltonian}
\end{multline}
Keeping only the interaction terms between the DC voltage and the RF fields and carrying out the rotating wave approximation for our case of mechanics in resonance with microwave gives us
\begin{equation}
    H_\text{int} = i \frac{\partial C_{\text{m}}}{\partial x} x_{\text{zpf}} V_{\text{zpf}}V_\text{DC}(\hat{a}\hat{b}^\dagger - \hat{a}^\dagger\hat{b}) = i\hbar g_\text{em} (\hat{a}\hat{b}^\dagger - \hat{a}^\dagger\hat{b}).
\end{equation}
This Hamiltonian has the form of an artificial piezoelectric response with an interaction strength 
\begin{equation}
    \hbar g_\text{em} = \frac{\partial C_{\text{m}}}{\partial x} x_{\text{zpf}} V_{\text{zpf}}V_{\text{DC}}.
\end{equation}
This constitutes a parametric interaction where the interaction strength scales linearly with the applied external voltage. 

Upon obtaining the interaction term, we can write the full Hamiltonian of our system as 
\begin{equation}
    H/\hbar = \omega_r\hat{a}^\dagger\hat{a} + \omega_m\hat{b}^\dagger\hat{b} +ig(\hat{a}\hat{b}^\dagger - \hat{a}^\dagger\hat{b}),
    \label{eq:Ham_full}
\end{equation}
where we have dropped the subscript from $g_\text{em}$ for brevity. 
In probing our cavity-mechanics system, we use a microwave tone at frequency $\omega_d$. We can write the Langevin equations for \cref{eq:Ham_full} as
\begin{align}
    \dot{\hat{a}} &= -(i\omega_r + \kappa/2)\,\hat{a} - g\hat{b} - \sqrt{\kappa_e} \hat{a}_\text{in} \\
      \dot{\hat{b}} &= -(i\omega_m + \gamma/2)\,\hat{b} + g\hat{a}.
\end{align}
where we have neglected the thermal fluctuations entering the microwave and mechanics, as we are focused on the coherent response of our system.

Taking the Fourier transform of the Langevin equations, we obtain
\begin{align}
    (i\Delta + \kappa/2) \,\hat{a}(\omega) &= -  g\hat{b}(\omega) - \sqrt{\kappa_e} \,\hat{a}_\text{in}(\omega) \label{eq:input_mw}\\
      (i\delta + \gamma/2) \,\hat{b}(\omega) &=  g\hat{a}(\omega), \label{eq:input_mech}
\end{align}
where $\Delta = \omega_r-\omega$ and $\delta = \omega_m-\omega$. Substituting \cref{eq:input_mech} into \cref{eq:input_mw}, we can express the microwave cavity operator as
\begin{equation} 
    \hat{a}(\omega) = -  \frac{\sqrt{\kappa_e}\,\hat{a}_\text{in}(\omega)}{i\Delta + \kappa/2 + \frac{g^2}{i\delta + \gamma/2}}.
    \label{eq:cavity_operator}
\end{equation}
Using input-output theory, we can define the output operator as $\hat{a}_\text{in}(\omega) = \hat{a}_\text{out}(\omega) + \sqrt{\kappa_e} \hat{a}(\omega)$. Using \cref{eq:cavity_operator}, we obtain the familiar electromechanically induced transparency (EIT) expression
\begin{equation}
    \frac{\hat{a}_\text{out}(\omega)}{\hat{a}_\text{in}(\omega)}= 1 -  \frac{\kappa_e}{i\Delta + \kappa/2 + \frac{g^2}{i\delta + \gamma/2}}.
\end{equation}
We utilize this EIT expression to fit the reflection traces we obtain of the cavity-mechanics system.
\subsection{Coupling Perturbation Theory}

Within the framework of cavity electromechanics, we can consider our interaction to be caused by radiation pressure. More precisely, the stored electrical energy in our system changes with the mechanical displacement via the modulation of the capacitance. Looking at the term in the Hamiltonian that leads to electrostatic interaction in \cref{eq:fullHamiltonian}, we can see that the change in the cross electrical energy is the origin of this interaction and thus can be used to capture the change in the capacitance. It is possible to express this change in the energy via a perturbative integral, similar to the moving boundary integrals for electromechanical systems \cite{pitanti2015}. In this perturbative approach, we assume that the displacement of the material boundaries does not change the electric field but alters the local permittivity due to leading to a electromechanical coupling rate 
\begin{multline}
    \hbar g_{em} = x_{\text{zpf}} \oint dA \, \left(\mathbf{Q(r)}\cdot \hat n) (\Delta\epsilon\mathbf{E^\parallel(r)}_{\text{DC}}\mathbf{E^\parallel(r)}_{\text{RF}} \right. \\
    \left. - \Delta\epsilon^{-1}\mathbf{D^\perp(r)}_{\text{DC}}\mathbf{D^\perp(r)}_{\text{RF}}\right).
\end{multline}
Here, $x_{\text{zpf}} = \sqrt{\hbar/2m_{\text{eff}}\omega_m}$ is the zero point fluctuations of displacement and $m_{\text{eff}}$ is the effective mass of the acoustic resonator.  $\mathbf{Q(r)}$ is the normalized displacement where $\text{max}[\mathbf{Q(r)}] = 1$. $\Delta\epsilon = \epsilon_1 - \epsilon_2$ and  $\Delta\epsilon^{-1} = 1/\epsilon_1 - 1/\epsilon_2$ are the electrical permittivty contrast between the two materials that are on the boundary covered by the surface integral. $\mathbf{E^\parallel(r)}_{\text{DC}}$ ($\mathbf{E^\parallel(r)}_{\text{RF}}$) is the parallel electric field component obtained from electrostatic simulations of the capacitor with $V_{\text{DC}}$ ($V_{\text{zpf}}$) applied to the capacitors. Likewise, $\mathbf{D^\perp(r)}_{\text{DC}}$ ($\mathbf{D^\perp(r)}_{\text{RF}}$) is the perpendicular displacement field obtained from the same simulation. In this expression the voltage dependence of the coupling is directly embedded in the capacitor voltages used in the simulations. One can alternatively solve for a given voltage (such as 1V) and then scale the fields appropriately based on $V_{\text{DC}}$ and $V_{\text{zpf}}$. For instance, $g_0$ can be simply obtained by setting $V_{\text{DC}} = 1$V. 

It has been previously observed that the electrical response of the capacitor at DC and RF frequencies may differ from one another \cite{vanlaer2018}. Generally, the charge carriers freeze off at cryogenic temperatures, giving rise to massive resistivity values for silicon \cite{gutierrez-d.2001}. However, in capacitor structures under DC voltages, band bending leads to the formation of a narrow space charge region which effectively screens out the field at the bulk of the silicon \cite{beckers2019}. This leads to a DC response which can approximately be modelled by modelling silicon as a perfect conductor. On the other hand, our microwave field which oscillates at a frequency above the RC cutoff cannot be screened and silicon behaves like a perfect insulator for these fields.

Taking this into account, only fields perpendicular to the boundaries will exist for the DC field and the integral for the electromechanical coupling can be simplified to. 
\begin{equation}
    \hbar g_{em} \approx x_{\text{zpf}} \oint dA\, \left(\mathbf{Q(r)}\cdot \hat n) (\epsilon_0^{-1} \mathbf{D^\perp(r)}_{\text{DC}}\mathbf{D^\perp(r)}_{\text{RF}}\right).
\end{equation}
For our devices with $g_0/2\pi = 45.4 ~\text{kHz/V}$ this approach gives us a very accurate simulation result of $46~\text{kHz/V}$. If we were to assume that silicon was an insulator at all frequencies and the field distributions were identical, we would underestimate our coupling strength and obtain $32\text{kHz/V}$.

\subsection{Equivalent Circuit Model}
\begin{figure}[h]
    \centering
    \includegraphics[width = 7.5cm]{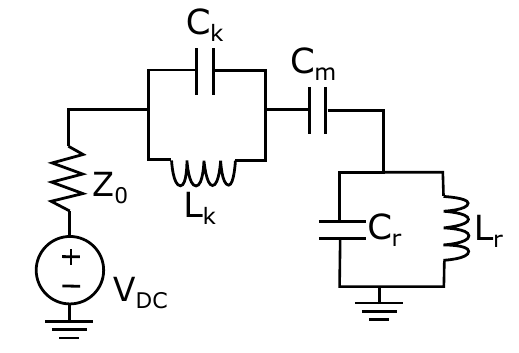}
    \caption{Equivalent electrical circuit of our system. }
    \label{fig:eq_circuit}
\end{figure}
The calculation of the electromechanical coupling strength via finite element method simulations is sufficient to completely describe the behavior of our system which consists of two coupled resonators. However, obtaining an electrical equivalent circuit for the mechanical resonator is crucial for making the analysis of complex electromechanical circuits more tractable. For this purpose, we model our mechanical resonator by a parallel LC resonator ($C_{\text{k}}$,$L_{\text{k}}$) in series with a capacitor $C_{\text{m}}$. This model can be seen in \cref{fig:eq_circuit}, where the coupling to the lumped element microwave resonator ($C_{\text{r}}$,$L_{\text{r}}$) is capacitive via $C_{\text{m}}$.

The equivalent mechanical capacitance can be expressed as 
\begin{equation}
    C_{\text{k}} = \frac{2C_{\text{m}}^2\omega_m}{\hbar \,\left(\partial_x C_\text{m}\, x_\text{zpf}\,V_\text{DC}\right)^2} - C_{\text{m}}.
\end{equation}
The equivalent mechanical inductance can be obtained following the calculation of $C_\text{k}$ by noting that $\omega_m = [L_\text{k}(C_\text{k}+C_\text{m})]^{-1/2}$. This circuit model gives us the correct expression for the electromechanical coupling strength, which is attained by capacitive coupling between the two circuit modes where 
\begin{equation}
    g_\text{em} =  \frac{C_{\text{m}}}{\sqrt{(C_{\text{k}}+C_{\text{m}})(C_{\text{r}}+C_{\text{m}})}}\sqrt{\omega_r\omega_m}.
\end{equation}

In this capacitive coupling picture, the value of $C_{\text{r}}$ primarily sets the zero point fluctuations of voltage for the microwave resonator since we have a small electromechanical participation ratio ($\eta = C_{\text{m}}/(C_{\text{m}}+C_{\text{r}})\ll 1$). The small participation ratio of our capacitor is caused by the small physical dimensions of our electromechanical capacitor which is commensurate with $1\mu$m transverse acoustic wavelength at 5 GHz. This leads to the electrical energy on top of the electromechanical capacitor to be substantially diluted compared to the total electrical energy stored on the microwave resonator. 

We express the circuit parameters for device B in Table \ref{table:param}. We note that the equivalent circuit model parameters for the mechanical resonator are dependent on the external voltage bias $V_{\text{DC}}$. This scaling is noted on the table, where $L_\text{k}$ and $C_\text{k}$ are provided for 1V. 
\begin{table}[h]
\caption{Equivalent circuit parameters for device B. The mechanical parameters $C_{\text{k}}$ and $L_{\text{k}}$ are dependent on the applied external voltage. The given values are for 1 V and the dependence on $V_{\text{DC}}$ is specified.}
\label{table:param}
\begin{center}
\begin{tabular}{c c}
Parameter & Value \\
\hline
$C_{\text{m}}$ & 0.2~ fF \\
$C_{\text{r}}$ &  12.1 ~fF\\
$L_{\text{r}} $    &  75.8 ~nH \\
$\eta$ &  1.5~\% \\
$C_{\text{k}}$ & 43.5~ nH $(1 \text{V}/V_{\text{DC}})^2$ \\ 
$L_{\text{k}}$ & 21.1~ fF $(V_{\text{DC}}/1 \text{V})^2$ \\
$f_{r,m}$ & 5.26 ~GHz \\
$g_0/2\pi$ & 45.4~ kHz/\text{V} \\
\hline\hline
\end{tabular}
\end{center}
\end{table}

In the circuit model, one can see that the mechanical resonator is capacitively coupled to the CPW which has an impedance $Z_0 = 50~\Omega$. This leads to some direct external decay to the CPW. However, even at the maximum voltage we can apply of 25 V, due to our massive equivalent capacitance this readout is at the level of 5 Hz, which is negligible and further emphasizes the need for a microwave cavity in order to enhance electromechanical readout of mechanics. 
\section{Microwave Devices}
\label{APP.2}
\subsection{Microwave Resonator}
\begin{figure}[t!]
    \centering
    \includegraphics[width = 6cm]{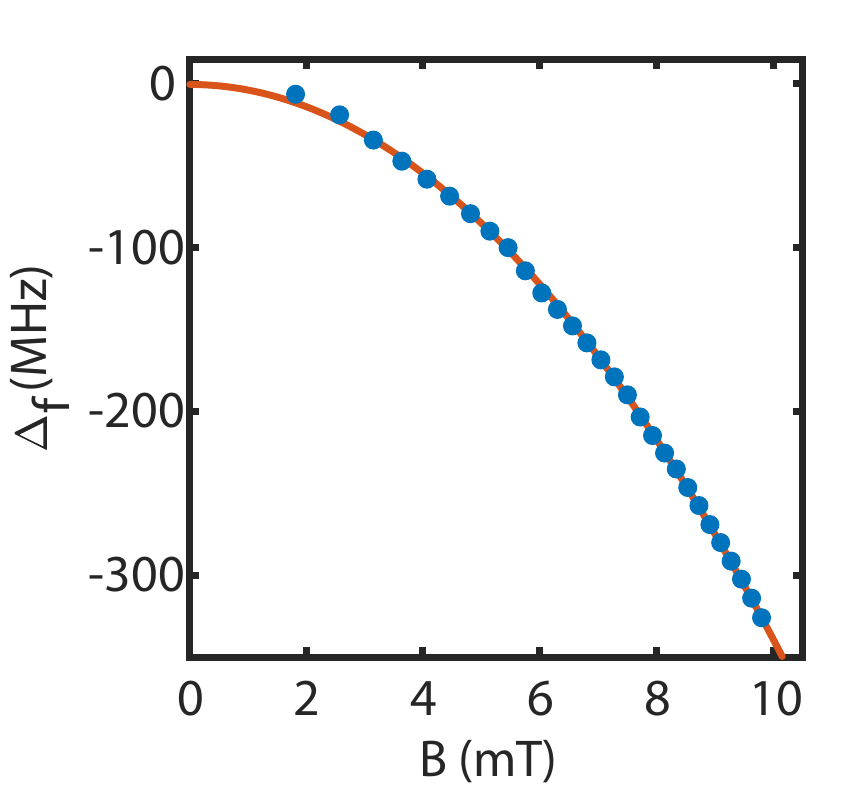}
    \caption{Change in the microwave resonance frequency based on the applied external perpendicular magnetic field. The solid line is a parabolic theory fit.}
    \label{fig:tuning}
\end{figure}

The $\lambda/4$ tunable microwave resonators in our work are formed by ICP-RIE etching of TiN films ($t\approx 15$ nm) with a sheet kinetic inductance of $40 \text{pH}/ \square$, which are sputtered on high resistivity ($>3 \text{k}\Omega$) SOI substrates (device layer thickness 220 nm). The high kinetic inductance TiN films are chosen for two main reasons: (i) obtaining a large impedance ($Z = \sqrt{L/C}$) resonator in order to enhance the electromechanical interaction ($g_{\text{em}} \propto \sqrt{Z}$) and (ii) attaining a high degree of tunability via an external magnetic field to bring the microwave into resonance with mechanics. These goals are satisfied by forming a $\lambda/4$ resonator, with the inductive component realized by a nanowire. The total kinetic in this structure is a function of film properties and geometry
\begin{equation}
    L_{\text{k}} = L_\square\frac{8}{\pi^2} \frac{l}{w},
    \label{eq:kinetic_ind}
\end{equation}
where $L_\square$ is the sheet inductance, $l$ is the nanowire length and $w$ is the nanowire width. In order to maximize the impedance, we etch our nanowires to be as narrow as possible, which in our case corresponds to a width of approximately 110 nm. Attempts to reduce wire width below this number led to reduced repeatability and a large disorder in resonator frequency. 

A current passing through a TiN nanowire modifies the kinetic inductance in a nonlinear fashion
\begin{equation}
    L_{\text{k}}(I) = L_{\text{k}}(0) \left[ 1 + \left(\frac{I}{I_*}\right)^2 \right].
\end{equation}
where $I_*$ is the critical current of the nanowire. Patterning of the nanowires to form closed loops and application of an external perpendicular magnetic field, provides a `wireless' means of modifying the kinetic inductance via the screening current induced through the loops \cite{xu2019a}. This is the mechanism we us to tune the resonator's frequency.

With these principles in mind, we utilize a ladder-like topology for our microwave resonator as seen in \cref{{fig:dev}}. Finite-element method (FEM) simulations indicate that our kinetic inductance excellently matches \cref{eq:kinetic_ind} and we have negligible geometric inductance, leading to a kinetic inductance participation ratio of approximately unity ($>98\%$). Furthermore, we calculate the impedance of our resonator to be 2.5 k$\Omega$.  The calculated lumped element equivalent circuit parameters for the microwave resonator of device B can be seen in Table \ref{table:param}. Despite having a substantial impedance, due to our mechanical capacitance being very small, we have a low electrical energy participation ratio for the mechanical capacitor ($\eta = C_{\text{m}}/(C_{\text{m}}+C_{\text{r}})$) of 1.5\%. This small participation ratio permits us to galvanically connect multiple electromechanical capacitors to our microwave resonator.

Following the fabrication of our devices, we find the measured resonance frequencies in good agreement with the device modeling, to within a random offset of approximately 300 MHz, which is attributed to fabrication disorder. We test the tunability of our devices by applying external magnetic fields via currents passing through a superconducting coil mounted on top of the sample box. For small tuning compared to the frequency, we expect the frequency shift to be given by 
\begin{equation}
    \Delta_f = -k B^2,
\end{equation}
where $k$ is a device dependent proportionality constant and $B$ is the external perpendicular magnetic field amplitude \cite{xu2019a}. This parabolic dependence is verified for device B on \cref{fig:tuning}. We can tune our device by about 6\% of their resonance frequency without causing any substantial degradation of the microwave intrinsic quality factor beyond $\kappa_i/2\pi \sim$ 500 kHz. 

\subsection{Purcell Filter}

\begin{figure}[t!]
    \centering
    \includegraphics[width = \columnwidth]{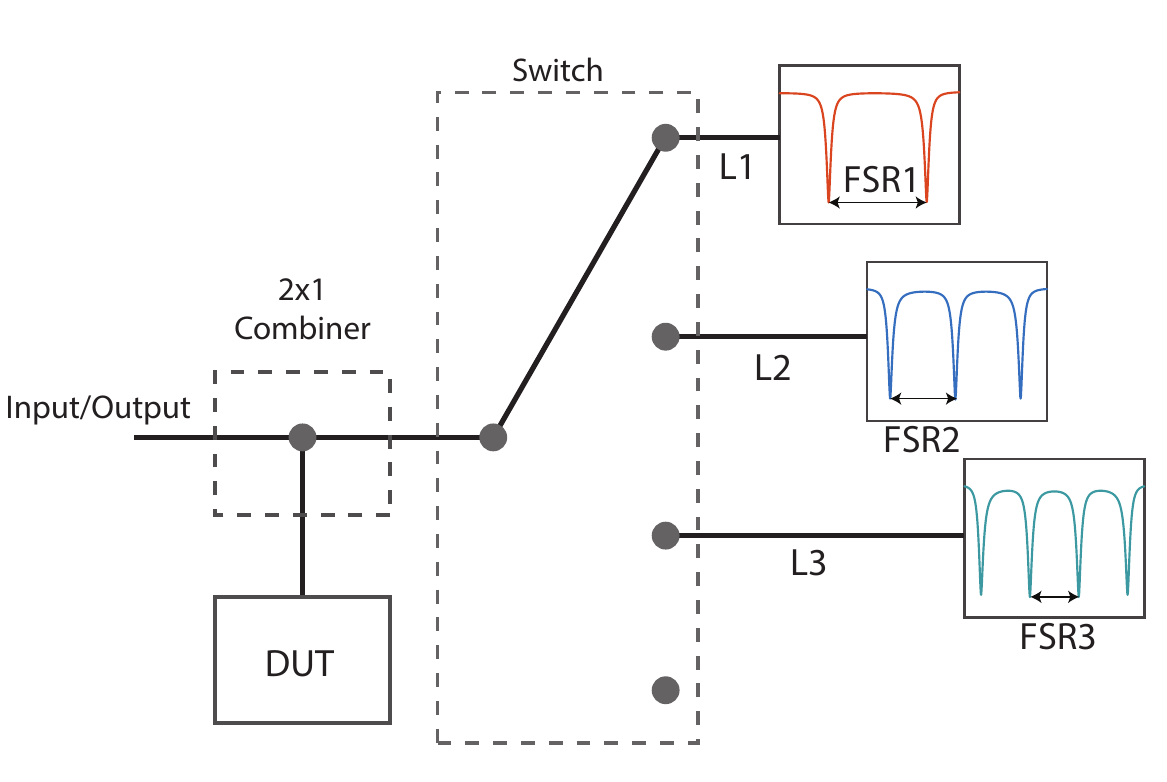}
    \caption{Circuit diagram of the Purcell filter configuration. The input/output line is attached to a circulator (not shown in the figure) used for investigation of the devices in reflection mode. The final line of the switch is not connected to a transmission line. The transmission lines have different lengths (L1,L2,L3). Their response is depicted as Lorentzian resonances separated by a given FSR that differs for each cable. DUT: Device under test.}
    \label{fig:sup_filter}
\end{figure}

\begin{figure*}[t]
    \centering\includegraphics[width = 0.9\textwidth]{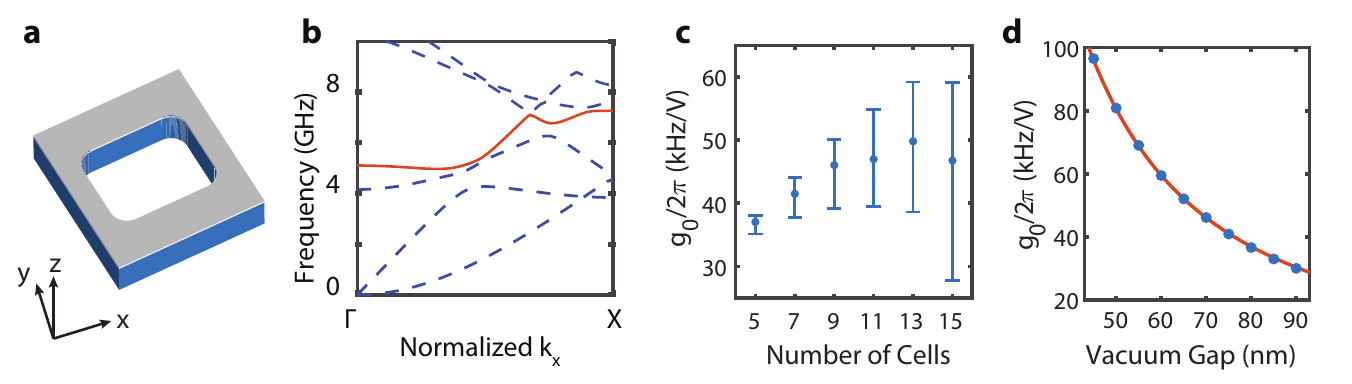}
    \caption{(a) Unit cell of the inner electrode of the electrostatic transducer. The beam is oriented along the x-direction. (b) Unit cell acoustic band diagram. (c) Disorder simulations of coupling strength vs number of unit cells of the inner electrode for $\sigma = 2$nm disorder and 55 realizations for each data point. The simulations are for a 70-nm vacuum gap.(d) Average coupling strength vs vacuum gap size for 9 unit cells. The simulated average values are in the presence of fabrication disorder. The fit is $\propto \text{gap}^{-1.66}$. }
    \label{fig:sup_mech_EMC}
\end{figure*}

In our design, we prioritize redundancy in terms of the number of mechanical resonators by attaching four electromechanical capacitors to a single microwave resonator. This is done to mitigate the impact of frequency shifts caused by fabrication disorder of the microwave resonators, where the different mechanical resonators spans roughly 500 MHz (the disorder in the mechanical frequencies is much smaller than the microwave disorder). However, this approach leads to an increased capacitance between our microwave resonator and the CPW, which gives rise to undesirably large external microwave decay. In order to increase the cooperativity of our system and reach the strong-coupling regime, this decay channel has to be suppressed. 

We utilize an off-chip Purcell filter for this purpose as depicted schematically in \cref{fig:sup_filter}. We place the filter in parallel to our device, where the notch filter reduces the external coupling strength of the microwave resonator. For this purpose, we make use of home-built microwave filters, realized as open transmission lines with a multi-pole spectrum. To get approximate frequency matching between our devices and the filter, we use a mechanical switch that enables us to choose among multiple filters with different free spectral ranges (in a span of 200 MHz to 500 MHz). In cases where we get perfect frequency matching between a device and a filter resonance, we can reduce the external decay rate $\kappa_e/2\pi$ to sub-200 kHz levels. Furthermore, we have a line in our switch which is not connected to a transmission line, permitting us to effectively turn our filter off. 

\section{Mechanics Design}
\label{APP.3}

The mechanical behavior of our structures is investigated via FEM simulations on COMSOL. The inner electrode of the electrostatic transducer is patterned to form a phononic crystal cavity consisting of multiple identical unit cells, with the unit cell depicted in \cref{fig:sup_mech_EMC}a. These unit cells are completely metallized with a 15-nm layer of TiN on top of a 220-nm Si membrane (the anisotropic elasticity tensor is used for silicon as in \cite{chan2012}).  We are primarily interested in the $\Gamma$ point of the breathing mode shown by the red band on the unit cell band diagram on \cref{fig:sup_mech_EMC}b. The breathing mode is selected because it has a displacement profile that significantly modulates the electromechanical capacitance. The $\Gamma$ mode is further necessary to ensure phase matching of the electromechanical interaction. Once multiple unit cells are attached to form a 1-D chain and are clamped by phononic shields that have a full bandgap, they hybridize to form a supermode that spans all the unit cells, which is our mechanical mode of interest. 

Ideally, we would like to have as many unit cells as possible in forming our central electrode, since the electromechanical interaction strength scales as $\propto \,\sqrt{N_\text{cells}}$. In practice, however, fabrication disorder leads to mode break-up, limiting the number of cells that can be utilized. We investigate this phenomenon via disorder simulations, which constitutes a crucial tool in investigating the disorder limited properties of periodically patterned acoustic and optical structures \cite{lethomas2011,minkov2013,maccabe2019}. The disorder can be modeled as random changes in the center positions and the width of the negative patterns, which are etched to form the actual structure. These random changes are represented as realizations of independent Gaussian random variables with zero mean and standard deviation $\sigma$. The precise algorithm is as follows: (i) the center of the etched filleted rectangles are varied by $\sigma$ in each direction (ii) the height and width of the rectangles are varied by $2\sigma$. Since each simulation represents a given disorder realization, to accurately extract the statistics we simulate multiple instances for a given number of unit cells. We investigate the impact of disorder on the coupling strength as shown in \cref{fig:sup_mech_EMC}c. In this situation, the changes in the mode profile and potential disorder induced mode breakup is captured in variations in the coupling strength, which is the experimentally relevant parameter to us. We can see that up until 9 unit cells, $g_0$ grows with the number of cells. However, beyond this number the growth rate diminishes and the disorder-induced variations increase substantially. At 9 unit cells and a 70-nm gap, we have a coupling strength of $46^{+4}_{-7}\, \text{kHz/V}$ which is in excellent agreement with our experimental results. Finally, we investigate the dependence of the coupling strength on the vacuum gap as shown in \cref{fig:sup_mech_EMC}d. The fit indicates that the coupling strength is a very strong function of the gap and the scaling is more rapid than that of a parallel plate capacitor. 

\begin{figure}[t!]
    \centering
    \includegraphics[width = \columnwidth]{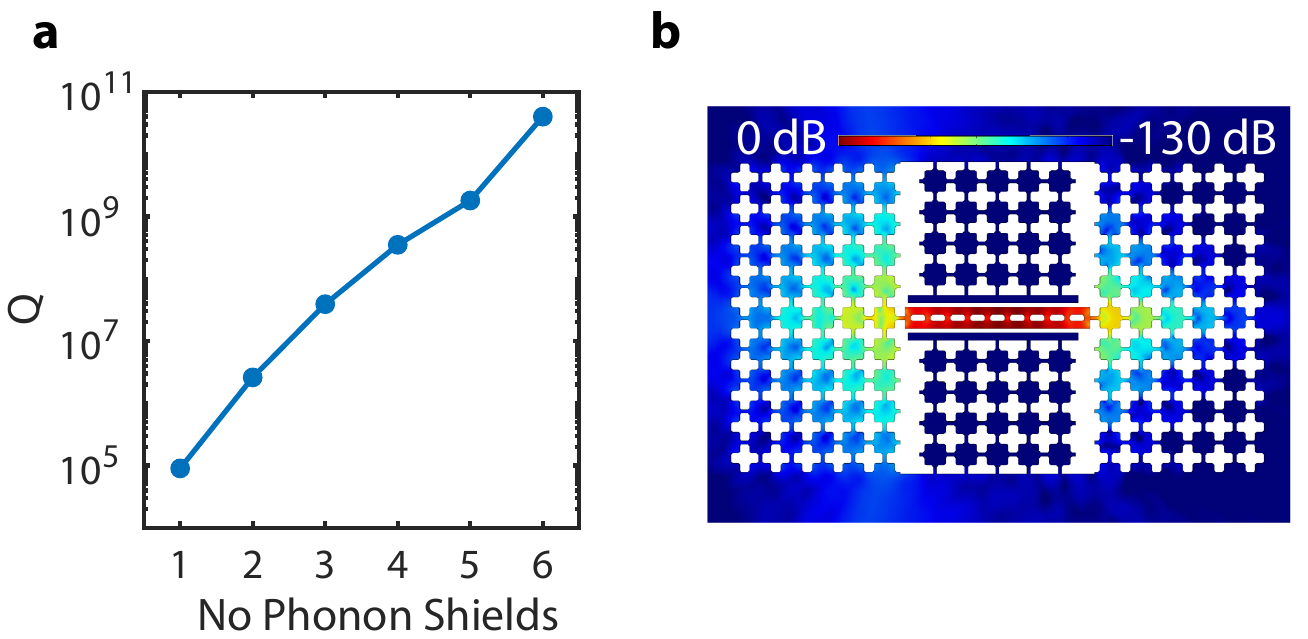}
    \caption{(a) Radiation limited acoustic quality factor vs number of phonon shield periods in the absence of disorder. (b) Logarithmic plot of mechanical energy density for 5 shield periods.}
    \label{fig:sup_shield}
\end{figure}

Since acoustic waves cannot propagate in vacuum, the only radiative leakage pathway is through the silicon substrate. We minimize this loss by clamping our acoustic resonator with a phononic shield that has a complete bandgap. We investigate the effectiveness of the phononic shields in suppressing radiation loss via finite-element simulations, where the continuum of modes in the substrate is modeled by a perfectly matched layer. The simulations provide us with the radiation limited mechanical $Q$ in the absence of fabrication disorder. As depicted in \cref{fig:sup_shield}a the mechanical quality factor roughly nearly  exponentially with the number of phononic shields. We have used 5 shield periods in our final design, for which we plot the acoustic energy distribution in \cref{fig:sup_shield}b. 
\section{Thermometry}
\label{APP.4}
\subsection{Theory}
We perform sideband thermometry by measuring the emission from the microwave and mechanical resonators with a spectrum analyzer. Our Hamiltonian is identical to that of a red-detuned optomechanical system with a large sideband resolution ($\omega_m/\kappa \gg 1$). Hence, in order to extract the thermal occupancy values for the microwave and mechanical thermal baths, we use the expression for the noise power spectral density following previous electromechanics work \cite{teufel2011b,fink2016}. The power spectral density following the amplifier chain is 
\begin{multline}
    S(\omega)= \hbar\omega 10^{\mathcal{G}/10} \left(n_\text{add} + \frac{1}{2}\right. + \\
    n_\text{wg}\left|\left(1-\frac{\kappa_e\chi_r}{1+g^2\chi_m\chi_r}\right)\right|^2 + n_\text{b,r}\frac{\kappa_e\kappa_i|\chi_r|^2}{|1+g^2\chi_m\chi_r|^2} \\ \left.+ n_\text{b,m}\frac{\kappa_e\gamma_i g^2|\chi_r|^2|\chi_m|^2}{|1+g^2\chi_m\chi_r|^2}\right)
    \label{eq:therm_full}.
\end{multline}

Here, $\mathcal{G}$ is the amplifier gain in dB and $n_\text{add}$ is the noise added by our amplifiers, which is dominated by the HEMT in our case. In this picture, the microwave resonator interacts with an intrinsic bath of occupancy $n_\text{b,r}$ with rate $\kappa_i$ and the waveguide having an occupancy of $n_{\text{wg}}$ with rate $\kappa_e$. 
 Similarly, the mechanical resonator is coupled to an intrinsic bath having an occupancy of $n_\text{b,m}$ with rate $\gamma_i$. The electromechanical interaction with strength $g$ will further lead to Purcell decay into the microwave for the mechanics, giving rise to electromechanical back-action.  The PSD expression also includes the the bare electrical and mechanical susceptibilities which are given as
\begin{align}
    \chi_r^{-1}(\omega) &= \kappa/2 -i(\omega-\omega_r) \\
        \chi_m^{-1}(\omega) &= \gamma_i/2 -i(\omega-\omega_m),
\end{align}
where $\omega_r$ ($\omega_m$) is the microwave (mechanics) resonance frequency. 
 In the weak coupling regime, we can express the mechanics and microwave resonator thermal occupancies as
\begin{align}
    n_r &= \frac{\kappa_in_\text{b,r}+\kappa_en_\text{wg}}{\kappa_e+\kappa_i}, \\
    n_m &= \frac{ n_\text{b,m}+Cn_r}{1+C}.
\end{align}
Here, $C = \frac{g^2\kappa\gamma_i^{-1}}{\Delta^2 + (\kappa/2)^2}$ is the effective cooperativity when the mechanical and microwave resonators are detuned by $\Delta$. In the absence of any detuning and at large bias voltages, the cooperativity becomes large, leading to substantial electromechanical cooling, which makes it challenging to unambiguously find the mechanical bath occupancy ($n_\text{b,m}$). Therefore, we operate in a low-cooperativity regime (large $\Delta$), which permits precise extraction of thermal bath occupancies. Extraction of the mechanical intrinsic bath occupancy is particularly important for quantum memory applications of our mechanical resonators, as there won't be permanent electromechanical back-action cooling in this scenario and the thermal decoherence rate depends on this bath occupancy.

To facilitate the analysis, we simplify the emission due to the mechanics intrinsic bath in \cref{eq:therm_full} as 
\begin{equation}
    S_{m}(\delta) \sim 4\tilde{n}_\text{m} \frac{\kappa_e}{\kappa} \frac{\Gamma_\text{em}}{\gamma}\frac{(\gamma/2)^2}{\delta^2 + (\gamma/2)^2}
    \label{eq:mech_emission}.
\end{equation}
Here, $\delta = \omega - \tilde\omega_m$ is the detuning between the emission frequency and mechanical resonance frequency, with $\tilde\omega_m$ being the mechanical frequency shifted by the optical spring effect. $\gamma = \gamma_i + \Gamma_\text{em}$ is the total mechanical linewidth and $\tilde{n}_\text{m} = n_\text{b,m} \gamma_i/\gamma$ is the thermal occupancy of the mechanical resonator due to fluctuations of the intrinsic mechanical bath. 

To accurately utilize the expressions for the noise power spectral density, it is crucial to take into account the distinction between broadening due to frequency jitter and radiative coupling to an intrinsic bath for the mechanical resonator. To this end, we note that the area underneath $S_m(\delta)$ is proportional to the total Purcell enhanced emission from mechanics, which is given as $\Gamma_\text{em}\tilde{n}_\text{m}$. Hence, we can see that \cref{eq:mech_emission} represents the emission for a mechanical resonator that has a total linewidth $\gamma$ with arbitrary frequency jitter. To take the distinction between jitter and decay into account, once we have extracted  $\tilde{n}_\text{m}$, we should use only the mechanical intrinsic decay rate $\Gamma_i$ to calculate the bath thermal occupancy as 
\begin{equation}
    n_\text{b,m} = \tilde{n}_\text{m} \frac{\Gamma_\text{em} + \Gamma_i}{\Gamma_i}.
    \label{eq:mech_bath_occupancy}
\end{equation}

We note that due to the small non-zero microwave bath occupancy, we have some emission from the mechanical resonator due to microwave thermal fluctuations entering the mechanics by electromechanical back-action. This emission interferes with that of the microwave resonator and leads to noise squashing as can be seen in the term in \cref{eq:therm_full} proportional to $n_\text{b,r}$ \cite{safavi-naeini2013}. This noise squashing is also taken into account in our analysis in order to correctly calculate our mechanical bath occupancy.

\subsection{Driven Response}

We generally do not see significant emission from the mechanical resonator during thermometry, due to the mechanics being deep in its motional ground state. Due to the noise introduced by our HEMT, the spectrum analyzer traces are noisy to an extent that precludes numerically fitting the mechanical emission. Thus, in order to accurately analyze this noisy data and extract the mechanical thermal occupancy, it is crucial to find the mechanical frequency, linewidth and decay rate at a given voltage. 

We achieve this by investigating the driven response of our resonators.  Following the application of a coherent tone in resonance with our mechanics, the drive tone is elastically scattered, leading to a delta-like emission from the mechanics and the generation of a coherent phonon population defined as $n_\text{coh} = |\langle \hat{a} \rangle|^2
$. This coherent population dynamics is the origin of the EIT response which can be detected by a VNA. However, the frequency jitter of our system also leads to inelastic scattering of the drive tone. For frequency noise which has a correlation time smaller than the decay rate, we get an emission with the cavity lineshape as the absorbed incoherent phonons lose memory of the drive frequency due to frequency jitter \cite{zhang2014}. This inelastic scattering is due to the generation of a number of incoherent phonons in the cavity, whose population is $n_\text{inc} = \langle \hat{a}^\dagger\hat{a} \rangle- |\langle \hat{a} \rangle|^2$. The incoherent emission enables us to extract the cavity linewidth and frequency via fitting the Lorentzian response. This routine for parameter extraction is repeated prior to mechanics  thermometry for each voltage to facilitate accurate calculations with \cref{eq:mech_emission} and center the spectrum analyzer detection window. 

Apart from extracting the lineshape of the mechanics, we can also use the driven response to make non time-resolved measurement of the intrinsic decay rate $\Gamma_i$. The total coherent and incoherent phonon populations are related to the total decay rate $\Gamma_\text{d} = \Gamma_i + \Gamma_\text{em}$ and the total linewidth $\gamma$ as follows \cite{koshino2011} \begin{equation}
    \frac{n_\text{inc}}{n_\text{coh}} = \frac{\gamma}{\Gamma_\text{d}}-1.
\end{equation}
These phonon populations can be further related to the areas detected via the spectrum analyzer, where $S_\delta$ is the area underneath the coherent emission, $S_\text{bb}$ ($S_\text{nb}$) is the area underneath the incoherent emission due to broadband (narrow-band) frequency noise.  Therefore, the relation between the decay rates can be expressed as
\begin{equation}
    \frac{\gamma}{\Gamma_\text{d}} = 1 + \frac{S_\text{bb}}{S_\delta}\left(1-\frac{S_\text{nb}}{S_\delta}\right),
\end{equation}
where the broadband frequency noise can be arbitrarily strong and the narrow-band noise is weak \cite{zhang2014}. Subtracting the electromechanical readoıut rate from the total decay rate, we can calculate $\Gamma_i$. The $\Gamma_i$ obtained in this manner is used to extract the mechanical bath occupancy via \cref{eq:mech_bath_occupancy}.

\subsection{Line Calibration}

We calibrate the total gain of the output line using thermometry of a 50 $\Omega$ cryogenic terminator that is thermalized to the mixing (MXC) stage of the cryostat. The output line consists of a HEMT amplifier (LNF-LNC48C) thermalized to the 4 K stage and a room temperature amplifier. The MXC stage temperature is raised by reducing cooling power by turning off the turbo to reduce $^3$He/$^4$He mixture flow, and applying heat using the MXC stage heater. With no external input power, we measure the output power from the amplifier chain with a spectrum analyzer at different mixing stage temperatures $T_\text{MXC}$. The measured output power has contributions from the thermal noise of the resistor thermalized to the MXC stage, and the HEMT noise characterized by a fixed noise temperature $T_\text{HEMT}$. The total power measured in an IF bandwidth $\Delta\nu_\text{IF}$ on the spectrum analyzer is equal to the sum of the Johnson-Nyquist noise from the two sources and is given by, 
\begin{align}
P_{\text{OUT}} = \Delta\nu_{\text{IF}} k_\text{B}  G_\text{A} \left(T_{\text{MXC}} + T_{\text{HEMT}}\right)
\label{eq:thermalnoise}
\end{align}
Here, $G_A$ is the absolute (net) gain factor of the output line, and is a combination of the total gain due to the amplifier chain and losses due to coaxial cables. At $T_\text{MXC} = 10$ mK, the measured output power is dominated by HEMT noise. The output gain $G_A$ can be calculated by subtracting the contribution of the HEMT noise from the total output power measured at various $T_\text{MXC}$. We perform this measurement at multiple different MXC temperatures between 730 mK and 1.05 K. The mean and standard deviation of these measurements are used to obtain $G_A$. We use a factor $\eta = (h\nu/kT)/\left(\exp{\left(h\nu/kT\right)} - 1\right)$ to account for corrections to \cref{eq:thermalnoise} due to the Bose-Einstein distribution in the regime $h\nu \sim k_BT$ \cite{nyquist1928}. Using this calibration method, we obtain a net gain of 65.6 $\pm$ 0.4 dB for the output line. 

\subsection{Pull-in Instability}
\begin{figure}[h]
    \centering
    \includegraphics[width = 0.7\columnwidth]{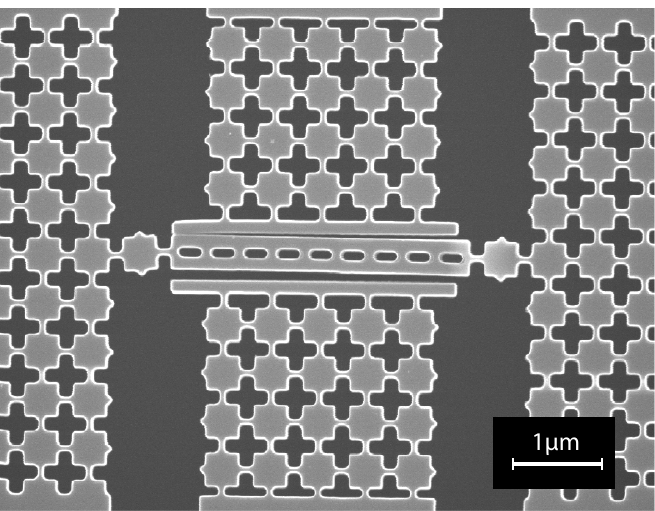}
    \caption{Post measurement SEM image of a transducer device which had permanently broken down. The pull-in phenomenon can be observed at the top-left corner of the electromechanical capacitor where the vacuum gap has become shut.}
    \label{fig:pullin}
\end{figure}

Post measurement imaging of the devices which were subject to breakdown indicates that the breakdown behavior is caused by pull-in of the inner electrode to an outer electrode, as seen in \cref{fig:pullin}. Once the capacitor gap becomes shut, a short resistive leakage path appears, leading to substantial current flow as observed in our leakage current measurements. This `pull-in' phenomenon is commonly observed in electrostatic actuators. Increasing the bias voltage, the strong electrostatic forces can no longer be offset by the mechanical spring force following a certain gap shrinkage, leading to unstable mechanical dynamics \cite{johnstone2004}. Furthermore, stiction can render this phenomenon irreversible as we have observed, where removal of the external voltage does not lead to recovery of the device. 

The observed pull-in behavior in our devices is not fully explained with common behavior of larger electrostatic actuators. For these MEMS devices, the capacitor gap start to shrink gradually and the onset of instability occurs once the gap has shrunk by about a third of its initial value for parallel plate geometries \cite{janschek2011mechatronic}. However, we observe excellent linearity of $g$ vs $V_\text{DC}$, which indicates the absence of any significant continuous shrinkage of the gap. At the moment, the precise mechanism of the observed pull-in instability is not clear to us and will be the subject of further studies.

\section{Mechanical Coherence}
\label{APP.5}

\begin{figure*}[]
    \centering
    \includegraphics[width = 14cm]{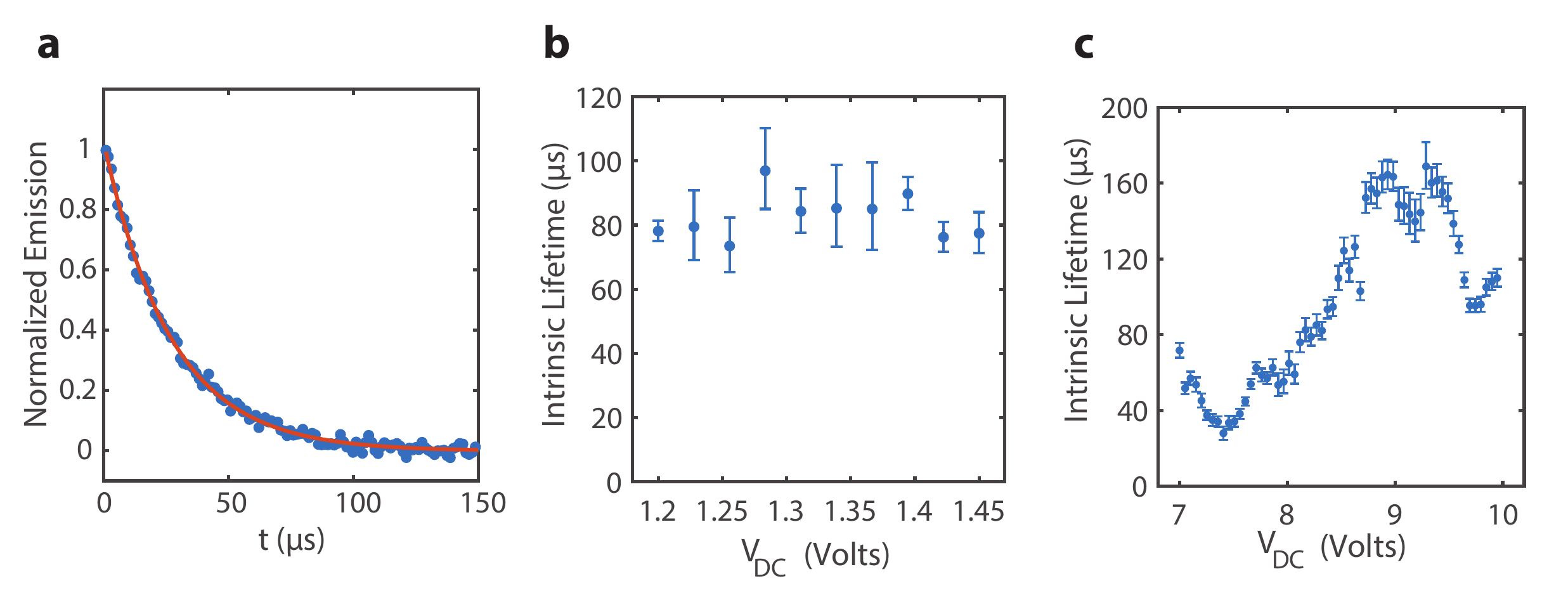}
    \caption{(a) Typical ringdown with an intrinsic lifetime of $\sim 30~\mu$s. (b) Intrinsic lifetime vs voltage for device A with around 1000 phonons inside the cavity. (c) Intrinsic lifetime vs voltage for device B with around 3000 phonons inside the cavity.}
    \label{fig:mech_sup_coh1}
\end{figure*}
\subsection{Ringdown Measurements}

In performing the ringdown measurements, we use a constant DC drive at a selected voltage level and control the external readout rate via setting the detuning between the microwave resonator and the mechanics. We populate the mechanical cavity via sending microwave pulses resonant with the mechanical mode. The pulses are synthesized with an arbitrary waveform generator (AWG), and their length is chosen to be sufficiently long to ensure the mechanical population can reach the steady state. Following the drive pulse, we detect the emitted power in a given interval by processing the downconverted signal from a digitizer. A power measurement (as opposed to field-quadrature) is obtained by summing the square of the demodulated I and Q quadrature values in a detection window in an FPGA. This detection window length is set to be much shorter than the reciprocal of mechanical linewidth to ensure we detect all the phonons emitted from the resonator. The multiple consecutive detection windows during a single measurement gives us a ringdown curve. We then proceed to average many instances of the experiment to improve our signal-to-noise (SNR) ratio and obtain our final data. The AWG, digitizer, and FPGA functionalities are realized using a Quantum Machines OPX+ module. We calibrate the digitizer output using the calibrated gain of our output lines and refer the detected voltage levels to the number of phonons in the cavity.

In finding the optimal lifetime in a voltage range, we carry out multiple ringdowns while sweeping the voltage. These ringdowns are performed at a large number of phonons in order to improve the SNR and make the measurements more tractable. Apart from showing signatures of spectral collisions with TLS that is manifested as deteriorating lifetimes, these measurements enable us to extract statistics about our lifetimes. We can see that for the two devices we have measured, we can reliably obtain lifetimes around 30 $\mu$s by slightly optimizing our voltage level. Such a typical ringdown is visualized in \cref{fig:mech_sup_coh1}a. We can further see that the points where we have improved coherence properties do not exhibit extreme sensitivity to the applied voltage. For example, for the voltage value where we have obtained our best lifetime on device A, we can obtain similar lifetimes in a 250 mV range as shown in \cref{fig:mech_sup_coh1}b. 
We can see similar broad regions where the lifetime is enhanced for device B too, as depicted in \cref{fig:mech_sup_coh1}c. 

\subsection{Linewidth Characterization}

\begin{figure*}[!t]
    \centering
    \includegraphics[width = 14cm]{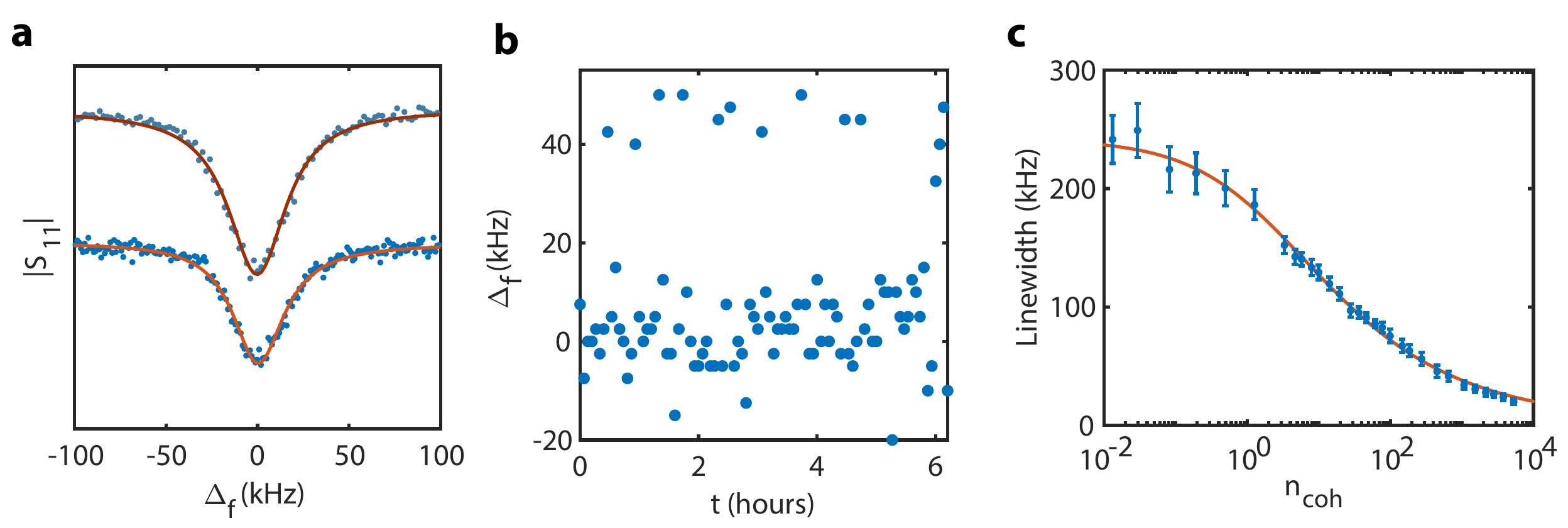}
    \caption{(a) Reflection measurements for the mechanics of device A at 1.2 V. The upper trace is at more than 500 phonons whereas the lower trace is at around 4 phonons. The variation in the depths of the resonances is due to differences in electromechanical readout strength. The upper trace is shifted for greater visibility. (b) Mechanical central frequency measurements showing switching behavior. Each data-point takes 5 s to acquire and the measurement is repeated every 3 mins. (c) Dependence of linewidth on the number of coherent phonons for device B at 3 V. The solid line is a fit to the TLS model. }
    \label{fig:mech_sup_coh2}
\end{figure*}

The reflection spectrum measurements enable us to investigate the mechanical linewidth at different power levels in order to investigate their  power dependence. For device A, these results corroborate our previous observations concerning the absence of saturable TLS dependent losses at our long lifetime point. As shown in \cref{fig:mech_sup_coh2}a, the linewidth does not show any saturation behavior when the phonon number varies between 4-500.

During our ringdown measurements, we keep track of our mechanical resonance frequency in long timescales and observe telegraphic frequency jumps as shown in \cref{fig:mech_sup_coh2}b. A potential model for this behavior could be that of a mechanical resonator directly coupled to a low frequency TLS that has non-negligible thermal population, which is sometimes referred to as a thermal fluctuator (TF) \cite{klimov2018}. The TF jumping to its excited state leads to a dispersive shift for our device, giving rise to to central mechanical frequencies which are bunched at a higher frequency. Using this data, once can infer the frequency of the TF, its switching rate and the coupling strength of it to the mechanics, which is consistent with our first principle calculations and previous literature \cite{schlor2019,klimov2018,meissner2018}. 

In contrast to the previously discussed non-saturable behavior, when we go to a operation point  in device B with signatures of TLS spectral collision, we can recover the familiar saturable behavior of the linewidth for as seen in \cref{fig:mech_sup_coh2}c. We can fit this data to a TLS continuum model, where the total linewidth $\gamma$ can be expressed as \cite{sage2011}
\begin{equation}
    \gamma = \frac{F\,\gamma_\text{TLS}}{\tanh\left(\frac{\hbar\omega}{2k_BT}\right)} \sqrt{1+\left(\frac{n}{n_c}\right)^\beta} + \gamma_0,
\end{equation}
where $n$ is the number of phonons inside the cavity, $F$ is the TLS participation ratio, $\gamma_\text{TLS}$ is the TLS decay rate, $n_c$ is the critical phonon number, $\beta$ is a fit parameter and $\gamma_0$ is the decay rate from power independent broadening mechanisms. Due to ambiguity in the phonon number inside the cavity for VNA measurements, we use the coherent phonon number $n_\text{coh}$ in our fit (see \cref{APP.4} for this distinction). 
We cannot see the complete saturation of the linewidth at large phonon numbers due to our mechanical nonlinearity which leads to narrowing down of the linewidth and deviation from Lorentzian lineshape. The fit gives us $\gamma_0/2\pi \approx$ 30 kHz. 

\subsection{TLS Model}
\label{tls_model_sec}
In analyzing the interactions of TLS with acoustic and electrical fields, we use the standard tunnelling model \cite{muller2019}. The TLS is modeled as two potential wells having an asymmetry energy $\epsilon$ and a tunnelling energy $\Delta$, which leads to a TLS energy of $E = \sqrt{\Delta^2 + \epsilon^2}$. The interaction of TLS with external fields is via modification of its asymmetry energy  
\begin{equation}
    \epsilon = 2\gamma \cdot \mathbf{S} + 2\mathbf{p} \cdot \mathbf{E} + \epsilon_0,
\end{equation}
where $\gamma$ is the mechanical deformation potential around 1.5eV in magnitude, $\mathbf{S}$ is the external strain field, $\mathbf{p}$ is the TLS dipole moment of roughly 1 Debye, $\mathbf{E}$ is the external electric field and $\epsilon_0$ is the residual asymmetry from the environment. 

The dependence of the asymmetry energy on the electric field enables us to Stark shift the frequency of TLS via the voltage applied on our electromechanical capacitor. The tuning rate can be calculated as
\begin{equation}
    \delta E = 2\frac{\epsilon}{E}  \mathbf{p} \cdot \mathbf{E}
\end{equation}
We use $\epsilon/E $ of 0.5 in our analysis. The narrow vacuum gap capacitors gives rise to  large electric fields approaching $5\times {10}^6$ V/m, leading to a steep tuning rate of $\delta E/h \approx\,$25 GHz/V. 

Apart from Stark shifts, the electrical dipole of the TLS also leads to $\mathbf{p} \cdot \mathbf{E}$ coupling to the microwave fields. The zero-point fluctuations of voltage in our microwave resonator is approximately 10$\mu$V, with maximum electric field values of roughly 50 V/m on the interfaces, leading to a TLS-microwave coupling of 250 kHz. 

Due to substantial acoustic susceptibility of TLS, strain coupling constitutes an important mechanism for mechanics-TLS interactions. The coupling strength can be written as \begin{equation}
    \hbar \lambda = \gamma \frac{\epsilon}{E} S_\text{zpf}
\end{equation}
where $S_\text{zpf}$ is the zero-point fluctuations of strain associated with our mechanical mode \cite{ramos2013}. This quantity is related to the strain mode volume of our mechanical mode 
\begin{equation}
    S_\text{zpf} = \left(\frac{\hbar\omega_m}{2\mathcal{E}V_m}\right)^{1/2}
\end{equation}
where $\mathcal{E}$ is the Young's modulus and $V_m$ is the strain mode volume. Due to the small physical dimensions of our mechanical resonator, we have an extremely small strain mode volume of $6\times 10^{-3} \,\mu \text{m}^3$ and a corresponding $S_\text{zpf}$ of $4\times 10^{-8}\,$m/m. This causes a strain coupling strength of $\lambda/2\pi = $ 13 MHz, which clearly dominates other coupling mechanisms to TLS for the mechanical resonator.

\bibliography{EMC_paper_ref.bib}

\begin{thebibliography}{61}%
\makeatletter
\providecommand \@ifxundefined [1]{%
 \@ifx{#1\undefined}
}%
\providecommand \@ifnum [1]{%
 \ifnum #1\expandafter \@firstoftwo
 \else \expandafter \@secondoftwo
 \fi
}%
\providecommand \@ifx [1]{%
 \ifx #1\expandafter \@firstoftwo
 \else \expandafter \@secondoftwo
 \fi
}%
\providecommand \natexlab [1]{#1}%
\providecommand \enquote  [1]{``#1''}%
\providecommand \bibnamefont  [1]{#1}%
\providecommand \bibfnamefont [1]{#1}%
\providecommand \citenamefont [1]{#1}%
\providecommand \href@noop [0]{\@secondoftwo}%
\providecommand \href [0]{\begingroup \@sanitize@url \@href}%
\providecommand \@href[1]{\@@startlink{#1}\@@href}%
\providecommand \@@href[1]{\endgroup#1\@@endlink}%
\providecommand \@sanitize@url [0]{\catcode `\\12\catcode `\$12\catcode
  `\&12\catcode `\#12\catcode `\^12\catcode `\_12\catcode `\%12\relax}%
\providecommand \@@startlink[1]{}%
\providecommand \@@endlink[0]{}%
\providecommand \url  [0]{\begingroup\@sanitize@url \@url }%
\providecommand \@url [1]{\endgroup\@href {#1}{\urlprefix }}%
\providecommand \urlprefix  [0]{URL }%
\providecommand \Eprint [0]{\href }%
\providecommand \doibase [0]{https://doi.org/}%
\providecommand \selectlanguage [0]{\@gobble}%
\providecommand \bibinfo  [0]{\@secondoftwo}%
\providecommand \bibfield  [0]{\@secondoftwo}%
\providecommand \translation [1]{[#1]}%
\providecommand \BibitemOpen [0]{}%
\providecommand \bibitemStop [0]{}%
\providecommand \bibitemNoStop [0]{.\EOS\space}%
\providecommand \EOS [0]{\spacefactor3000\relax}%
\providecommand \BibitemShut  [1]{\csname bibitem#1\endcsname}%
\let\auto@bib@innerbib\@empty
\bibitem [{\citenamefont {McGuigan}\ \emph {et~al.}(1978)\citenamefont
  {McGuigan}, \citenamefont {Lam}, \citenamefont {Gram}, \citenamefont
  {Hoffman}, \citenamefont {Douglass},\ and\ \citenamefont
  {Gutche}}]{mcguigan1978}%
  \BibitemOpen
  \bibfield  {author} {\bibinfo {author} {\bibfnamefont {D.~F.}\ \bibnamefont
  {McGuigan}}, \bibinfo {author} {\bibfnamefont {C.~C.}\ \bibnamefont {Lam}},
  \bibinfo {author} {\bibfnamefont {R.~Q.}\ \bibnamefont {Gram}}, \bibinfo
  {author} {\bibfnamefont {A.~W.}\ \bibnamefont {Hoffman}}, \bibinfo {author}
  {\bibfnamefont {D.~H.}\ \bibnamefont {Douglass}},\ and\ \bibinfo {author}
  {\bibfnamefont {H.~W.}\ \bibnamefont {Gutche}},\ }\bibfield  {title}
  {\bibinfo {title} {Measurements of the mechanical {{Q}} of single-crystal
  silicon at low temperatures},\ }\href {https://doi.org/10.1007/BF00116202}
  {\bibfield  {journal} {\bibinfo  {journal} {Journal of Low Temperature
  Physics}\ }\textbf {\bibinfo {volume} {30}},\ \bibinfo {pages} {621}
  (\bibinfo {year} {1978})}\BibitemShut {NoStop}%
\bibitem [{\citenamefont {Renninger}\ \emph {et~al.}(2018)\citenamefont
  {Renninger}, \citenamefont {Kharel}, \citenamefont {Behunin},\ and\
  \citenamefont {Rakich}}]{renninger2018}%
  \BibitemOpen
  \bibfield  {author} {\bibinfo {author} {\bibfnamefont {W.~H.}\ \bibnamefont
  {Renninger}}, \bibinfo {author} {\bibfnamefont {P.}~\bibnamefont {Kharel}},
  \bibinfo {author} {\bibfnamefont {R.~O.}\ \bibnamefont {Behunin}},\ and\
  \bibinfo {author} {\bibfnamefont {P.~T.}\ \bibnamefont {Rakich}},\ }\bibfield
   {title} {\bibinfo {title} {Bulk crystalline optomechanics},\ }\href
  {https://doi.org/10.1038/s41567-018-0090-3} {\bibfield  {journal} {\bibinfo
  {journal} {Nature Physics}\ }\textbf {\bibinfo {volume} {14}},\ \bibinfo
  {pages} {601} (\bibinfo {year} {2018})}\BibitemShut {NoStop}%
\bibitem [{\citenamefont {Beccari}\ \emph {et~al.}(2022)\citenamefont
  {Beccari}, \citenamefont {Visani}, \citenamefont {Fedorov}, \citenamefont
  {Bereyhi}, \citenamefont {Boureau}, \citenamefont {Engelsen},\ and\
  \citenamefont {Kippenberg}}]{beccari2022b}%
  \BibitemOpen
  \bibfield  {author} {\bibinfo {author} {\bibfnamefont {A.}~\bibnamefont
  {Beccari}}, \bibinfo {author} {\bibfnamefont {D.~A.}\ \bibnamefont {Visani}},
  \bibinfo {author} {\bibfnamefont {S.~A.}\ \bibnamefont {Fedorov}}, \bibinfo
  {author} {\bibfnamefont {M.~J.}\ \bibnamefont {Bereyhi}}, \bibinfo {author}
  {\bibfnamefont {V.}~\bibnamefont {Boureau}}, \bibinfo {author} {\bibfnamefont
  {N.~J.}\ \bibnamefont {Engelsen}},\ and\ \bibinfo {author} {\bibfnamefont
  {T.~J.}\ \bibnamefont {Kippenberg}},\ }\bibfield  {title} {\bibinfo {title}
  {Strained crystalline nanomechanical resonators with quality factors above 10
  billion},\ }\href {https://doi.org/10.1038/s41567-021-01498-4} {\bibfield
  {journal} {\bibinfo  {journal} {Nature Physics}\ }\textbf {\bibinfo {volume}
  {18}},\ \bibinfo {pages} {436} (\bibinfo {year} {2022})}\BibitemShut
  {NoStop}%
\bibitem [{\citenamefont {MacCabe}\ \emph {et~al.}(2020)\citenamefont
  {MacCabe}, \citenamefont {Ren}, \citenamefont {Luo}, \citenamefont {Cohen},
  \citenamefont {Zhou}, \citenamefont {Sipahigil}, \citenamefont
  {Mirhosseini},\ and\ \citenamefont {Painter}}]{maccabe2020b}%
  \BibitemOpen
  \bibfield  {author} {\bibinfo {author} {\bibfnamefont {G.~S.}\ \bibnamefont
  {MacCabe}}, \bibinfo {author} {\bibfnamefont {H.}~\bibnamefont {Ren}},
  \bibinfo {author} {\bibfnamefont {J.}~\bibnamefont {Luo}}, \bibinfo {author}
  {\bibfnamefont {J.~D.}\ \bibnamefont {Cohen}}, \bibinfo {author}
  {\bibfnamefont {H.}~\bibnamefont {Zhou}}, \bibinfo {author} {\bibfnamefont
  {A.}~\bibnamefont {Sipahigil}}, \bibinfo {author} {\bibfnamefont
  {M.}~\bibnamefont {Mirhosseini}},\ and\ \bibinfo {author} {\bibfnamefont
  {O.}~\bibnamefont {Painter}},\ }\bibfield  {title} {\bibinfo {title}
  {Nano-acoustic resonator with ultralong phonon lifetime},\ }\href
  {https://doi.org/10.1126/science.abc7312} {\bibfield  {journal} {\bibinfo
  {journal} {Science}\ }\textbf {\bibinfo {volume} {370}},\ \bibinfo {pages}
  {840} (\bibinfo {year} {2020})}\BibitemShut {NoStop}%
\bibitem [{\citenamefont {Clerk}\ \emph {et~al.}(2020)\citenamefont {Clerk},
  \citenamefont {Lehnert}, \citenamefont {Bertet}, \citenamefont {Petta},\ and\
  \citenamefont {Nakamura}}]{10.1038/s41567-020-0797-9}%
  \BibitemOpen
  \bibfield  {author} {\bibinfo {author} {\bibfnamefont {A.~A.}\ \bibnamefont
  {Clerk}}, \bibinfo {author} {\bibfnamefont {K.~W.}\ \bibnamefont {Lehnert}},
  \bibinfo {author} {\bibfnamefont {P.}~\bibnamefont {Bertet}}, \bibinfo
  {author} {\bibfnamefont {J.~R.}\ \bibnamefont {Petta}},\ and\ \bibinfo
  {author} {\bibfnamefont {Y.}~\bibnamefont {Nakamura}},\ }\bibfield  {title}
  {\bibinfo {title} {{Hybrid quantum systems with circuit quantum
  electrodynamics}},\ }\href {https://doi.org/10.1038/s41567-020-0797-9}
  {\bibfield  {journal} {\bibinfo  {journal} {Nature Physics}\ }\textbf
  {\bibinfo {volume} {16}},\ \bibinfo {pages} {257} (\bibinfo {year}
  {2020})}\BibitemShut {NoStop}%
\bibitem [{\citenamefont {Safavi-Naeini}\ \emph {et~al.}(2019)\citenamefont
  {Safavi-Naeini}, \citenamefont {Thourhout}, \citenamefont {Baets},\ and\
  \citenamefont {Laer}}]{SafaviNaeini:2019ci}%
  \BibitemOpen
  \bibfield  {author} {\bibinfo {author} {\bibfnamefont {A.~H.}\ \bibnamefont
  {Safavi-Naeini}}, \bibinfo {author} {\bibfnamefont {D.~V.}\ \bibnamefont
  {Thourhout}}, \bibinfo {author} {\bibfnamefont {R.}~\bibnamefont {Baets}},\
  and\ \bibinfo {author} {\bibfnamefont {R.~V.}\ \bibnamefont {Laer}},\
  }\bibfield  {title} {\bibinfo {title} {{Controlling phonons and photons at
  the wavelength scale: integrated photonics meets integrated phononics}},\
  }\href {https://doi.org/10.1364/optica.6.000213} {\bibfield  {journal}
  {\bibinfo  {journal} {Optica}\ }\textbf {\bibinfo {volume} {6}},\ \bibinfo
  {pages} {213} (\bibinfo {year} {2019})},\ \Eprint
  {https://arxiv.org/abs/1810.03217} {1810.03217} \BibitemShut {NoStop}%
\bibitem [{\citenamefont {Han}\ \emph {et~al.}(2021)\citenamefont {Han},
  \citenamefont {Fu}, \citenamefont {Zou}, \citenamefont {Jiang},\ and\
  \citenamefont {Tang}}]{10.1364/optica.425414}%
  \BibitemOpen
  \bibfield  {author} {\bibinfo {author} {\bibfnamefont {X.}~\bibnamefont
  {Han}}, \bibinfo {author} {\bibfnamefont {W.}~\bibnamefont {Fu}}, \bibinfo
  {author} {\bibfnamefont {C.-L.}\ \bibnamefont {Zou}}, \bibinfo {author}
  {\bibfnamefont {L.}~\bibnamefont {Jiang}},\ and\ \bibinfo {author}
  {\bibfnamefont {H.~X.}\ \bibnamefont {Tang}},\ }\bibfield  {title} {\bibinfo
  {title} {{Microwave-optical quantum frequency conversion}},\ }\href
  {https://doi.org/10.1364/optica.425414} {\bibfield  {journal} {\bibinfo
  {journal} {Optica}\ }\textbf {\bibinfo {volume} {8}},\ \bibinfo {pages}
  {1050} (\bibinfo {year} {2021})}\BibitemShut {NoStop}%
\bibitem [{\citenamefont {Wallucks}\ \emph {et~al.}(2020)\citenamefont
  {Wallucks}, \citenamefont {Marinkovi{\'c}}, \citenamefont {Hensen},
  \citenamefont {Stockill},\ and\ \citenamefont
  {Gr{\"o}blacher}}]{wallucks2020}%
  \BibitemOpen
  \bibfield  {author} {\bibinfo {author} {\bibfnamefont {A.}~\bibnamefont
  {Wallucks}}, \bibinfo {author} {\bibfnamefont {I.}~\bibnamefont
  {Marinkovi{\'c}}}, \bibinfo {author} {\bibfnamefont {B.}~\bibnamefont
  {Hensen}}, \bibinfo {author} {\bibfnamefont {R.}~\bibnamefont {Stockill}},\
  and\ \bibinfo {author} {\bibfnamefont {S.}~\bibnamefont {Gr{\"o}blacher}},\
  }\bibfield  {title} {\bibinfo {title} {A quantum memory at telecom
  wavelengths},\ }\href {https://doi.org/10.1038/s41567-020-0891-z} {\bibfield
  {journal} {\bibinfo  {journal} {Nature Physics}\ }\textbf {\bibinfo {volume}
  {16}},\ \bibinfo {pages} {772} (\bibinfo {year} {2020})}\BibitemShut
  {NoStop}%
\bibitem [{\citenamefont {Satzinger}\ \emph {et~al.}(2018)\citenamefont
  {Satzinger}, \citenamefont {Zhong}, \citenamefont {Chang}, \citenamefont
  {Peairs}, \citenamefont {Bienfait}, \citenamefont {Chou}, \citenamefont
  {Cleland}, \citenamefont {Conner}, \citenamefont {Dumur}, \citenamefont
  {Grebel} \emph {et~al.}}]{10.1038/s41586-018-0719-5}%
  \BibitemOpen
  \bibfield  {author} {\bibinfo {author} {\bibfnamefont {K.~J.}\ \bibnamefont
  {Satzinger}}, \bibinfo {author} {\bibfnamefont {Y.}~\bibnamefont {Zhong}},
  \bibinfo {author} {\bibfnamefont {H.-S.}\ \bibnamefont {Chang}}, \bibinfo
  {author} {\bibfnamefont {G.~A.}\ \bibnamefont {Peairs}}, \bibinfo {author}
  {\bibfnamefont {A.}~\bibnamefont {Bienfait}}, \bibinfo {author}
  {\bibfnamefont {M.-H.}\ \bibnamefont {Chou}}, \bibinfo {author}
  {\bibfnamefont {A.}~\bibnamefont {Cleland}}, \bibinfo {author} {\bibfnamefont
  {C.~R.}\ \bibnamefont {Conner}}, \bibinfo {author} {\bibfnamefont
  {{\'E}.}~\bibnamefont {Dumur}}, \bibinfo {author} {\bibfnamefont
  {J.}~\bibnamefont {Grebel}}, \emph {et~al.},\ }\bibfield  {title} {\bibinfo
  {title} {Quantum control of surface acoustic-wave phonons},\ }\href@noop {}
  {\bibfield  {journal} {\bibinfo  {journal} {Nature}\ }\textbf {\bibinfo
  {volume} {563}},\ \bibinfo {pages} {661} (\bibinfo {year}
  {2018})}\BibitemShut {NoStop}%
\bibitem [{\citenamefont {Lüpke}\ \emph {et~al.}(2022)\citenamefont {Lüpke},
  \citenamefont {Yang}, \citenamefont {Bild}, \citenamefont {Michaud},
  \citenamefont {Fadel},\ and\ \citenamefont
  {Chu}}]{10.1038/s41567-022-01591-2}%
  \BibitemOpen
  \bibfield  {author} {\bibinfo {author} {\bibfnamefont {U.~v.}\ \bibnamefont
  {Lüpke}}, \bibinfo {author} {\bibfnamefont {Y.}~\bibnamefont {Yang}},
  \bibinfo {author} {\bibfnamefont {M.}~\bibnamefont {Bild}}, \bibinfo {author}
  {\bibfnamefont {L.}~\bibnamefont {Michaud}}, \bibinfo {author} {\bibfnamefont
  {M.}~\bibnamefont {Fadel}},\ and\ \bibinfo {author} {\bibfnamefont
  {Y.}~\bibnamefont {Chu}},\ }\bibfield  {title} {\bibinfo {title} {{Parity
  measurement in the strong dispersive regime of circuit quantum
  acoustodynamics}},\ }\href {https://doi.org/10.1038/s41567-022-01591-2}
  {\bibfield  {journal} {\bibinfo  {journal} {Nature Physics}\ ,\ \bibinfo
  {pages} {1}} (\bibinfo {year} {2022})},\ \Eprint
  {https://arxiv.org/abs/2110.00263} {2110.00263} \BibitemShut {NoStop}%
\bibitem [{\citenamefont {Wollack}\ \emph {et~al.}(2022)\citenamefont
  {Wollack}, \citenamefont {Cleland}, \citenamefont {Gruenke}, \citenamefont
  {Wang}, \citenamefont {{Arrangoiz-Arriola}},\ and\ \citenamefont
  {{Safavi-Naeini}}}]{wollack2022}%
  \BibitemOpen
  \bibfield  {author} {\bibinfo {author} {\bibfnamefont {E.~A.}\ \bibnamefont
  {Wollack}}, \bibinfo {author} {\bibfnamefont {A.~Y.}\ \bibnamefont
  {Cleland}}, \bibinfo {author} {\bibfnamefont {R.~G.}\ \bibnamefont
  {Gruenke}}, \bibinfo {author} {\bibfnamefont {Z.}~\bibnamefont {Wang}},
  \bibinfo {author} {\bibfnamefont {P.}~\bibnamefont {{Arrangoiz-Arriola}}},\
  and\ \bibinfo {author} {\bibfnamefont {A.~H.}\ \bibnamefont
  {{Safavi-Naeini}}},\ }\bibfield  {title} {\bibinfo {title} {Quantum state
  preparation and tomography of entangled mechanical resonators},\ }\href
  {https://doi.org/10.1038/s41586-022-04500-y} {\bibfield  {journal} {\bibinfo
  {journal} {Nature}\ }\textbf {\bibinfo {volume} {604}},\ \bibinfo {pages}
  {463} (\bibinfo {year} {2022})}\BibitemShut {NoStop}%
\bibitem [{\citenamefont {Wollack}\ \emph {et~al.}(2021)\citenamefont
  {Wollack}, \citenamefont {Cleland}, \citenamefont {{Arrangoiz-Arriola}},
  \citenamefont {McKenna}, \citenamefont {Gruenke}, \citenamefont {Patel},
  \citenamefont {Jiang}, \citenamefont {Sarabalis},\ and\ \citenamefont
  {{Safavi-Naeini}}}]{wollack2021b}%
  \BibitemOpen
  \bibfield  {author} {\bibinfo {author} {\bibfnamefont {E.~A.}\ \bibnamefont
  {Wollack}}, \bibinfo {author} {\bibfnamefont {A.~Y.}\ \bibnamefont
  {Cleland}}, \bibinfo {author} {\bibfnamefont {P.}~\bibnamefont
  {{Arrangoiz-Arriola}}}, \bibinfo {author} {\bibfnamefont {T.~P.}\
  \bibnamefont {McKenna}}, \bibinfo {author} {\bibfnamefont {R.~G.}\
  \bibnamefont {Gruenke}}, \bibinfo {author} {\bibfnamefont {R.~N.}\
  \bibnamefont {Patel}}, \bibinfo {author} {\bibfnamefont {W.}~\bibnamefont
  {Jiang}}, \bibinfo {author} {\bibfnamefont {C.~J.}\ \bibnamefont
  {Sarabalis}},\ and\ \bibinfo {author} {\bibfnamefont {A.~H.}\ \bibnamefont
  {{Safavi-Naeini}}},\ }\bibfield  {title} {\bibinfo {title} {Loss channels
  affecting lithium niobate phononic crystal resonators at cryogenic
  temperature},\ }\href {https://doi.org/10.1063/5.0034909} {\bibfield
  {journal} {\bibinfo  {journal} {Applied Physics Letters}\ }\textbf {\bibinfo
  {volume} {118}},\ \bibinfo {pages} {123501} (\bibinfo {year}
  {2021})}\BibitemShut {NoStop}%
\bibitem [{\citenamefont {Mirhosseini}\ \emph {et~al.}(2020)\citenamefont
  {Mirhosseini}, \citenamefont {Sipahigil}, \citenamefont {Kalaee},\ and\
  \citenamefont {Painter}}]{10.1038/s41586-020-3038-6}%
  \BibitemOpen
  \bibfield  {author} {\bibinfo {author} {\bibfnamefont {M.}~\bibnamefont
  {Mirhosseini}}, \bibinfo {author} {\bibfnamefont {A.}~\bibnamefont
  {Sipahigil}}, \bibinfo {author} {\bibfnamefont {M.}~\bibnamefont {Kalaee}},\
  and\ \bibinfo {author} {\bibfnamefont {O.}~\bibnamefont {Painter}},\
  }\bibfield  {title} {\bibinfo {title} {{Superconducting qubit to optical
  photon transduction}},\ }\href {https://doi.org/10.1038/s41586-020-3038-6}
  {\bibfield  {journal} {\bibinfo  {journal} {Nature}\ }\textbf {\bibinfo
  {volume} {588}},\ \bibinfo {pages} {599} (\bibinfo {year}
  {2020})}\BibitemShut {NoStop}%
\bibitem [{\citenamefont {Chu}\ and\ \citenamefont
  {Gr{\"o}blacher}(2020)}]{chu2020}%
  \BibitemOpen
  \bibfield  {author} {\bibinfo {author} {\bibfnamefont {Y.}~\bibnamefont
  {Chu}}\ and\ \bibinfo {author} {\bibfnamefont {S.}~\bibnamefont
  {Gr{\"o}blacher}},\ }\bibfield  {title} {\bibinfo {title} {A perspective on
  hybrid quantum opto- and electromechanical systems},\ }\href
  {https://doi.org/10.1063/5.0021088} {\bibfield  {journal} {\bibinfo
  {journal} {Applied Physics Letters}\ }\textbf {\bibinfo {volume} {117}},\
  \bibinfo {pages} {150503} (\bibinfo {year} {2020})}\BibitemShut {NoStop}%
\bibitem [{\citenamefont {Teufel}\ \emph
  {et~al.}(2011{\natexlab{a}})\citenamefont {Teufel}, \citenamefont {Donner},
  \citenamefont {Li}, \citenamefont {Harlow}, \citenamefont {Allman},
  \citenamefont {Cicak}, \citenamefont {Sirois}, \citenamefont {Whittaker},
  \citenamefont {Lehnert},\ and\ \citenamefont {Simmonds}}]{teufel2011b}%
  \BibitemOpen
  \bibfield  {author} {\bibinfo {author} {\bibfnamefont {J.~D.}\ \bibnamefont
  {Teufel}}, \bibinfo {author} {\bibfnamefont {T.}~\bibnamefont {Donner}},
  \bibinfo {author} {\bibfnamefont {D.}~\bibnamefont {Li}}, \bibinfo {author}
  {\bibfnamefont {J.~W.}\ \bibnamefont {Harlow}}, \bibinfo {author}
  {\bibfnamefont {M.~S.}\ \bibnamefont {Allman}}, \bibinfo {author}
  {\bibfnamefont {K.}~\bibnamefont {Cicak}}, \bibinfo {author} {\bibfnamefont
  {A.~J.}\ \bibnamefont {Sirois}}, \bibinfo {author} {\bibfnamefont {J.~D.}\
  \bibnamefont {Whittaker}}, \bibinfo {author} {\bibfnamefont {K.~W.}\
  \bibnamefont {Lehnert}},\ and\ \bibinfo {author} {\bibfnamefont {R.~W.}\
  \bibnamefont {Simmonds}},\ }\bibfield  {title} {\bibinfo {title} {Sideband
  cooling of micromechanical motion to the quantum ground state},\ }\href
  {https://doi.org/10.1038/nature10261} {\bibfield  {journal} {\bibinfo
  {journal} {Nature}\ }\textbf {\bibinfo {volume} {475}},\ \bibinfo {pages}
  {359} (\bibinfo {year} {2011}{\natexlab{a}})}\BibitemShut {NoStop}%
\bibitem [{\citenamefont {Peterson}\ \emph {et~al.}(2019)\citenamefont
  {Peterson}, \citenamefont {Kotler}, \citenamefont {Lecocq}, \citenamefont
  {Cicak}, \citenamefont {Jin}, \citenamefont {Simmonds}, \citenamefont
  {Aumentado},\ and\ \citenamefont {Teufel}}]{peterson2019a}%
  \BibitemOpen
  \bibfield  {author} {\bibinfo {author} {\bibfnamefont {G.~A.}\ \bibnamefont
  {Peterson}}, \bibinfo {author} {\bibfnamefont {S.}~\bibnamefont {Kotler}},
  \bibinfo {author} {\bibfnamefont {F.}~\bibnamefont {Lecocq}}, \bibinfo
  {author} {\bibfnamefont {K.}~\bibnamefont {Cicak}}, \bibinfo {author}
  {\bibfnamefont {X.~Y.}\ \bibnamefont {Jin}}, \bibinfo {author} {\bibfnamefont
  {R.~W.}\ \bibnamefont {Simmonds}}, \bibinfo {author} {\bibfnamefont
  {J.}~\bibnamefont {Aumentado}},\ and\ \bibinfo {author} {\bibfnamefont
  {J.~D.}\ \bibnamefont {Teufel}},\ }\bibfield  {title} {\bibinfo {title}
  {Ultrastrong {{Parametric Coupling}} between a {{Superconducting Cavity}} and
  a {{Mechanical Resonator}}},\ }\href
  {https://doi.org/10.1103/PhysRevLett.123.247701} {\bibfield  {journal}
  {\bibinfo  {journal} {Physical Review Letters}\ }\textbf {\bibinfo {volume}
  {123}},\ \bibinfo {pages} {247701} (\bibinfo {year} {2019})}\BibitemShut
  {NoStop}%
\bibitem [{\citenamefont {Kalaee}\ \emph {et~al.}(2019)\citenamefont {Kalaee},
  \citenamefont {Mirhosseini}, \citenamefont {Dieterle}, \citenamefont
  {Peruzzo}, \citenamefont {Fink},\ and\ \citenamefont {Painter}}]{kalaee2019}%
  \BibitemOpen
  \bibfield  {author} {\bibinfo {author} {\bibfnamefont {M.}~\bibnamefont
  {Kalaee}}, \bibinfo {author} {\bibfnamefont {M.}~\bibnamefont {Mirhosseini}},
  \bibinfo {author} {\bibfnamefont {P.~B.}\ \bibnamefont {Dieterle}}, \bibinfo
  {author} {\bibfnamefont {M.}~\bibnamefont {Peruzzo}}, \bibinfo {author}
  {\bibfnamefont {J.~M.}\ \bibnamefont {Fink}},\ and\ \bibinfo {author}
  {\bibfnamefont {O.}~\bibnamefont {Painter}},\ }\bibfield  {title} {\bibinfo
  {title} {Quantum electromechanics of a hypersonic crystal},\ }\href
  {https://doi.org/10.1038/s41565-019-0377-2} {\bibfield  {journal} {\bibinfo
  {journal} {Nature Nanotechnology}\ }\textbf {\bibinfo {volume} {14}},\
  \bibinfo {pages} {334} (\bibinfo {year} {2019})}\BibitemShut {NoStop}%
\bibitem [{\citenamefont {Rouxinol}\ \emph {et~al.}(2016)\citenamefont
  {Rouxinol}, \citenamefont {Hao}, \citenamefont {Brito}, \citenamefont
  {Caldeira}, \citenamefont {Irish},\ and\ \citenamefont
  {LaHaye}}]{rouxinol2016a}%
  \BibitemOpen
  \bibfield  {author} {\bibinfo {author} {\bibfnamefont {F.}~\bibnamefont
  {Rouxinol}}, \bibinfo {author} {\bibfnamefont {Y.}~\bibnamefont {Hao}},
  \bibinfo {author} {\bibfnamefont {F.}~\bibnamefont {Brito}}, \bibinfo
  {author} {\bibfnamefont {A.~O.}\ \bibnamefont {Caldeira}}, \bibinfo {author}
  {\bibfnamefont {E.~K.}\ \bibnamefont {Irish}},\ and\ \bibinfo {author}
  {\bibfnamefont {M.~D.}\ \bibnamefont {LaHaye}},\ }\bibfield  {title}
  {\bibinfo {title} {Measurements of nanoresonator-qubit interactions in a
  hybrid quantum electromechanical system},\ }\href
  {https://doi.org/10.1088/0957-4484/27/36/364003} {\bibfield  {journal}
  {\bibinfo  {journal} {Nanotechnology}\ }\textbf {\bibinfo {volume} {27}},\
  \bibinfo {pages} {364003} (\bibinfo {year} {2016})}\BibitemShut {NoStop}%
\bibitem [{\citenamefont {Van~Laer}\ \emph {et~al.}(2018)\citenamefont
  {Van~Laer}, \citenamefont {Patel}, \citenamefont {McKenna}, \citenamefont
  {Witmer},\ and\ \citenamefont {{Safavi-Naeini}}}]{vanlaer2018}%
  \BibitemOpen
  \bibfield  {author} {\bibinfo {author} {\bibfnamefont {R.}~\bibnamefont
  {Van~Laer}}, \bibinfo {author} {\bibfnamefont {R.~N.}\ \bibnamefont {Patel}},
  \bibinfo {author} {\bibfnamefont {T.~P.}\ \bibnamefont {McKenna}}, \bibinfo
  {author} {\bibfnamefont {J.~D.}\ \bibnamefont {Witmer}},\ and\ \bibinfo
  {author} {\bibfnamefont {A.~H.}\ \bibnamefont {{Safavi-Naeini}}},\ }\bibfield
   {title} {\bibinfo {title} {Electrical driving of {{X-band}} mechanical waves
  in a silicon photonic circuit},\ }\href {https://doi.org/10.1063/1.5042428}
  {\bibfield  {journal} {\bibinfo  {journal} {APL Photonics}\ }\textbf
  {\bibinfo {volume} {3}},\ \bibinfo {pages} {086102} (\bibinfo {year}
  {2018})}\BibitemShut {NoStop}%
\bibitem [{\citenamefont {Teufel}\ \emph
  {et~al.}(2011{\natexlab{b}})\citenamefont {Teufel}, \citenamefont {Li},
  \citenamefont {Allman}, \citenamefont {Cicak}, \citenamefont {Sirois},
  \citenamefont {Whittaker},\ and\ \citenamefont {Simmonds}}]{teufel2011a}%
  \BibitemOpen
  \bibfield  {author} {\bibinfo {author} {\bibfnamefont {J.~D.}\ \bibnamefont
  {Teufel}}, \bibinfo {author} {\bibfnamefont {D.}~\bibnamefont {Li}}, \bibinfo
  {author} {\bibfnamefont {M.~S.}\ \bibnamefont {Allman}}, \bibinfo {author}
  {\bibfnamefont {K.}~\bibnamefont {Cicak}}, \bibinfo {author} {\bibfnamefont
  {A.~J.}\ \bibnamefont {Sirois}}, \bibinfo {author} {\bibfnamefont {J.~D.}\
  \bibnamefont {Whittaker}},\ and\ \bibinfo {author} {\bibfnamefont {R.~W.}\
  \bibnamefont {Simmonds}},\ }\bibfield  {title} {\bibinfo {title} {Circuit
  cavity electromechanics in the strong-coupling regime},\ }\href
  {https://doi.org/10.1038/nature09898} {\bibfield  {journal} {\bibinfo
  {journal} {Nature}\ }\textbf {\bibinfo {volume} {471}},\ \bibinfo {pages}
  {204} (\bibinfo {year} {2011}{\natexlab{b}})}\BibitemShut {NoStop}%
\bibitem [{\citenamefont {Mason}\ and\ \citenamefont
  {McSkimin}(1947)}]{mason1947}%
  \BibitemOpen
  \bibfield  {author} {\bibinfo {author} {\bibfnamefont {W.~P.}\ \bibnamefont
  {Mason}}\ and\ \bibinfo {author} {\bibfnamefont {H.~J.}\ \bibnamefont
  {McSkimin}},\ }\bibfield  {title} {\bibinfo {title} {Attenuation and
  {{Scattering}} of {{High Frequency Sound Waves}} in {{Metals}} and
  {{Glasses}}},\ }\href {https://doi.org/10.1121/1.1916504} {\bibfield
  {journal} {\bibinfo  {journal} {The Journal of the Acoustical Society of
  America}\ }\textbf {\bibinfo {volume} {19}},\ \bibinfo {pages} {464}
  (\bibinfo {year} {1947})}\BibitemShut {NoStop}%
\bibitem [{\citenamefont {Zeng}\ \emph {et~al.}(2010)\citenamefont {Zeng},
  \citenamefont {Agnew}, \citenamefont {Raeisinia},\ and\ \citenamefont
  {Myneni}}]{zeng2010}%
  \BibitemOpen
  \bibfield  {author} {\bibinfo {author} {\bibfnamefont {F.}~\bibnamefont
  {Zeng}}, \bibinfo {author} {\bibfnamefont {S.~R.}\ \bibnamefont {Agnew}},
  \bibinfo {author} {\bibfnamefont {B.}~\bibnamefont {Raeisinia}},\ and\
  \bibinfo {author} {\bibfnamefont {G.~R.}\ \bibnamefont {Myneni}},\ }\bibfield
   {title} {\bibinfo {title} {Ultrasonic {{Attenuation Due}} to {{Grain
  Boundary Scattering}} in {{Pure Niobium}}},\ }\href
  {https://doi.org/10.1007/s10921-010-0068-2} {\bibfield  {journal} {\bibinfo
  {journal} {Journal of Nondestructive Evaluation}\ }\textbf {\bibinfo {volume}
  {29}},\ \bibinfo {pages} {93} (\bibinfo {year} {2010})}\BibitemShut {NoStop}%
\bibitem [{\citenamefont {Shearrow}\ \emph {et~al.}(2018)\citenamefont
  {Shearrow}, \citenamefont {Koolstra}, \citenamefont {Whiteley}, \citenamefont
  {Earnest}, \citenamefont {Barry}, \citenamefont {Heremans}, \citenamefont
  {Awschalom}, \citenamefont {Shirokoff},\ and\ \citenamefont
  {Schuster}}]{shearrow2018a}%
  \BibitemOpen
  \bibfield  {author} {\bibinfo {author} {\bibfnamefont {A.}~\bibnamefont
  {Shearrow}}, \bibinfo {author} {\bibfnamefont {G.}~\bibnamefont {Koolstra}},
  \bibinfo {author} {\bibfnamefont {S.~J.}\ \bibnamefont {Whiteley}}, \bibinfo
  {author} {\bibfnamefont {N.}~\bibnamefont {Earnest}}, \bibinfo {author}
  {\bibfnamefont {P.~S.}\ \bibnamefont {Barry}}, \bibinfo {author}
  {\bibfnamefont {F.~J.}\ \bibnamefont {Heremans}}, \bibinfo {author}
  {\bibfnamefont {D.~D.}\ \bibnamefont {Awschalom}}, \bibinfo {author}
  {\bibfnamefont {E.}~\bibnamefont {Shirokoff}},\ and\ \bibinfo {author}
  {\bibfnamefont {D.~I.}\ \bibnamefont {Schuster}},\ }\bibfield  {title}
  {\bibinfo {title} {Atomic layer deposition of titanium nitride for quantum
  circuits},\ }\href {https://doi.org/10.1063/1.5053461} {\bibfield  {journal}
  {\bibinfo  {journal} {Applied Physics Letters}\ }\textbf {\bibinfo {volume}
  {113}},\ \bibinfo {pages} {212601} (\bibinfo {year} {2018})},\ \Eprint
  {https://arxiv.org/abs/1808.06009} {arXiv:1808.06009} \BibitemShut {NoStop}%
\bibitem [{\citenamefont {Hazard}\ \emph {et~al.}(2019)\citenamefont {Hazard},
  \citenamefont {Gyenis}, \citenamefont {Paolo}, \citenamefont {Asfaw},
  \citenamefont {Lyon}, \citenamefont {Blais},\ and\ \citenamefont
  {Houck}}]{10.1103/PhysRevLett.122.010504}%
  \BibitemOpen
  \bibfield  {author} {\bibinfo {author} {\bibfnamefont {T.~M.}\ \bibnamefont
  {Hazard}}, \bibinfo {author} {\bibfnamefont {A.}~\bibnamefont {Gyenis}},
  \bibinfo {author} {\bibfnamefont {A.~D.}\ \bibnamefont {Paolo}}, \bibinfo
  {author} {\bibfnamefont {A.~T.}\ \bibnamefont {Asfaw}}, \bibinfo {author}
  {\bibfnamefont {S.~A.}\ \bibnamefont {Lyon}}, \bibinfo {author}
  {\bibfnamefont {A.}~\bibnamefont {Blais}},\ and\ \bibinfo {author}
  {\bibfnamefont {A.~A.}\ \bibnamefont {Houck}},\ }\bibfield  {title} {\bibinfo
  {title} {{Nanowire Superinductance Fluxonium Qubit}},\ }\href
  {https://doi.org/10.1103/PhysRevLett.122.010504} {\bibfield  {journal}
  {\bibinfo  {journal} {Physical Review Letters}\ }\textbf {\bibinfo {volume}
  {122}},\ \bibinfo {pages} {010504} (\bibinfo {year} {2019})},\ \Eprint
  {https://arxiv.org/abs/1805.00938} {1805.00938} \BibitemShut {NoStop}%
\bibitem [{\citenamefont {Pechenezhskiy}\ \emph {et~al.}(2020)\citenamefont
  {Pechenezhskiy}, \citenamefont {Mencia}, \citenamefont {Nguyen},
  \citenamefont {Lin},\ and\ \citenamefont
  {Manucharyan}}]{10.1038/s41586-020-2687-9}%
  \BibitemOpen
  \bibfield  {author} {\bibinfo {author} {\bibfnamefont {I.~V.}\ \bibnamefont
  {Pechenezhskiy}}, \bibinfo {author} {\bibfnamefont {R.~A.}\ \bibnamefont
  {Mencia}}, \bibinfo {author} {\bibfnamefont {L.~B.}\ \bibnamefont {Nguyen}},
  \bibinfo {author} {\bibfnamefont {Y.-H.}\ \bibnamefont {Lin}},\ and\ \bibinfo
  {author} {\bibfnamefont {V.~E.}\ \bibnamefont {Manucharyan}},\ }\bibfield
  {title} {\bibinfo {title} {{The superconducting quasicharge qubit}},\ }\href
  {https://doi.org/10.1038/s41586-020-2687-9} {\bibfield  {journal} {\bibinfo
  {journal} {Nature}\ }\textbf {\bibinfo {volume} {585}},\ \bibinfo {pages}
  {368} (\bibinfo {year} {2020})},\ \Eprint {https://arxiv.org/abs/1907.02937}
  {1907.02937} \BibitemShut {NoStop}%
\bibitem [{\citenamefont {Zmuidzinas}(2012)}]{zmuidzinas2012}%
  \BibitemOpen
  \bibfield  {author} {\bibinfo {author} {\bibfnamefont {J.}~\bibnamefont
  {Zmuidzinas}},\ }\bibfield  {title} {\bibinfo {title} {Superconducting
  {{Microresonators}}: {{Physics}} and {{Applications}}},\ }\href
  {https://doi.org/10.1146/annurev-conmatphys-020911-125022} {\bibfield
  {journal} {\bibinfo  {journal} {Annual Review of Condensed Matter Physics}\
  }\textbf {\bibinfo {volume} {3}},\ \bibinfo {pages} {169} (\bibinfo {year}
  {2012})}\BibitemShut {NoStop}%
\bibitem [{\citenamefont {Xu}\ \emph {et~al.}(2019)\citenamefont {Xu},
  \citenamefont {Han}, \citenamefont {Fu}, \citenamefont {Zou},\ and\
  \citenamefont {Tang}}]{xu2019a}%
  \BibitemOpen
  \bibfield  {author} {\bibinfo {author} {\bibfnamefont {M.}~\bibnamefont
  {Xu}}, \bibinfo {author} {\bibfnamefont {X.}~\bibnamefont {Han}}, \bibinfo
  {author} {\bibfnamefont {W.}~\bibnamefont {Fu}}, \bibinfo {author}
  {\bibfnamefont {C.-L.}\ \bibnamefont {Zou}},\ and\ \bibinfo {author}
  {\bibfnamefont {H.~X.}\ \bibnamefont {Tang}},\ }\bibfield  {title} {\bibinfo
  {title} {Frequency-tunable high- {{{\emph{Q}}}} superconducting resonators
  via wireless control of nonlinear kinetic inductance},\ }\href
  {https://doi.org/10.1063/1.5098466} {\bibfield  {journal} {\bibinfo
  {journal} {Applied Physics Letters}\ }\textbf {\bibinfo {volume} {114}},\
  \bibinfo {pages} {192601} (\bibinfo {year} {2019})}\BibitemShut {NoStop}%
\bibitem [{\citenamefont {Fleetwood}\ \emph {et~al.}(1993)\citenamefont
  {Fleetwood}, \citenamefont {Winokur}, \citenamefont {Reber}, \citenamefont
  {Meisenheimer}, \citenamefont {Schwank}, \citenamefont {Shaneyfelt},\ and\
  \citenamefont {Riewe}}]{fleetwood1993}%
  \BibitemOpen
  \bibfield  {author} {\bibinfo {author} {\bibfnamefont {D.~M.}\ \bibnamefont
  {Fleetwood}}, \bibinfo {author} {\bibfnamefont {P.~S.}\ \bibnamefont
  {Winokur}}, \bibinfo {author} {\bibfnamefont {R.~A.}\ \bibnamefont {Reber}},
  \bibinfo {author} {\bibfnamefont {T.~L.}\ \bibnamefont {Meisenheimer}},
  \bibinfo {author} {\bibfnamefont {J.~R.}\ \bibnamefont {Schwank}}, \bibinfo
  {author} {\bibfnamefont {M.~R.}\ \bibnamefont {Shaneyfelt}},\ and\ \bibinfo
  {author} {\bibfnamefont {L.~C.}\ \bibnamefont {Riewe}},\ }\bibfield  {title}
  {\bibinfo {title} {Effects of oxide traps, interface traps, and ``border
  traps'' on metal-oxide-semiconductor devices},\ }\href
  {https://doi.org/10.1063/1.353777} {\bibfield  {journal} {\bibinfo  {journal}
  {Journal of Applied Physics}\ }\textbf {\bibinfo {volume} {73}},\ \bibinfo
  {pages} {5058} (\bibinfo {year} {1993})}\BibitemShut {NoStop}%
\bibitem [{\citenamefont {{Guti{\'e}rrez-D.}}\ \emph
  {et~al.}(2001)\citenamefont {{Guti{\'e}rrez-D.}}, \citenamefont {Deen},\ and\
  \citenamefont {Claeys}}]{gutierrez-d.2001}%
  \BibitemOpen
  \bibfield  {author} {\bibinfo {author} {\bibfnamefont {E.}~\bibnamefont
  {{Guti{\'e}rrez-D.}}}, \bibinfo {author} {\bibfnamefont {J.}~\bibnamefont
  {Deen}},\ and\ \bibinfo {author} {\bibfnamefont {C.}~\bibnamefont {Claeys}},\
  }in\ \href {https://doi.org/10.1016/B978-012310675-9/50000-1} {\emph
  {\bibinfo {booktitle} {Low {{Temperature Electronics}}}}},\ \bibinfo {editor}
  {edited by\ \bibinfo {editor} {\bibfnamefont {E.~A.}\ \bibnamefont
  {{Guti{\'e}rrez-D.}}}, \bibinfo {editor} {\bibfnamefont {M.~J.}\ \bibnamefont
  {Deen}},\ and\ \bibinfo {editor} {\bibfnamefont {C.}~\bibnamefont {Claeys}}}\
  (\bibinfo  {publisher} {{Academic Press}},\ \bibinfo {address} {{San
  Diego}},\ \bibinfo {year} {2001})\BibitemShut {NoStop}%
\bibitem [{\citenamefont {Zhang}\ \emph {et~al.}(2014)\citenamefont {Zhang},
  \citenamefont {Moser}, \citenamefont {G{\"u}ttinger}, \citenamefont
  {Bachtold},\ and\ \citenamefont {Dykman}}]{zhang2014}%
  \BibitemOpen
  \bibfield  {author} {\bibinfo {author} {\bibfnamefont {Y.}~\bibnamefont
  {Zhang}}, \bibinfo {author} {\bibfnamefont {J.}~\bibnamefont {Moser}},
  \bibinfo {author} {\bibfnamefont {J.}~\bibnamefont {G{\"u}ttinger}}, \bibinfo
  {author} {\bibfnamefont {A.}~\bibnamefont {Bachtold}},\ and\ \bibinfo
  {author} {\bibfnamefont {M.~I.}\ \bibnamefont {Dykman}},\ }\bibfield  {title}
  {\bibinfo {title} {Interplay of {{Driving}} and {{Frequency Noise}} in the
  {{Spectra}} of {{Vibrational Systems}}},\ }\href
  {https://doi.org/10.1103/PhysRevLett.113.255502} {\bibfield  {journal}
  {\bibinfo  {journal} {Physical Review Letters}\ }\textbf {\bibinfo {volume}
  {113}},\ \bibinfo {pages} {255502} (\bibinfo {year} {2014})}\BibitemShut
  {NoStop}%
\bibitem [{\citenamefont {Gr{\"u}nhaupt}\ \emph {et~al.}(2018)\citenamefont
  {Gr{\"u}nhaupt}, \citenamefont {Maleeva}, \citenamefont {Skacel},
  \citenamefont {Calvo}, \citenamefont {{Levy-Bertrand}}, \citenamefont
  {Ustinov}, \citenamefont {Rotzinger}, \citenamefont {Monfardini},
  \citenamefont {Catelani},\ and\ \citenamefont {Pop}}]{grunhaupt2018}%
  \BibitemOpen
  \bibfield  {author} {\bibinfo {author} {\bibfnamefont {L.}~\bibnamefont
  {Gr{\"u}nhaupt}}, \bibinfo {author} {\bibfnamefont {N.}~\bibnamefont
  {Maleeva}}, \bibinfo {author} {\bibfnamefont {S.~T.}\ \bibnamefont {Skacel}},
  \bibinfo {author} {\bibfnamefont {M.}~\bibnamefont {Calvo}}, \bibinfo
  {author} {\bibfnamefont {F.}~\bibnamefont {{Levy-Bertrand}}}, \bibinfo
  {author} {\bibfnamefont {A.~V.}\ \bibnamefont {Ustinov}}, \bibinfo {author}
  {\bibfnamefont {H.}~\bibnamefont {Rotzinger}}, \bibinfo {author}
  {\bibfnamefont {A.}~\bibnamefont {Monfardini}}, \bibinfo {author}
  {\bibfnamefont {G.}~\bibnamefont {Catelani}},\ and\ \bibinfo {author}
  {\bibfnamefont {I.~M.}\ \bibnamefont {Pop}},\ }\bibfield  {title} {\bibinfo
  {title} {Loss {{Mechanisms}} and {{Quasiparticle Dynamics}} in
  {{Superconducting Microwave Resonators Made}} of {{Thin-Film Granular
  Aluminum}}},\ }\href {https://doi.org/10.1103/PhysRevLett.121.117001}
  {\bibfield  {journal} {\bibinfo  {journal} {Physical Review Letters}\
  }\textbf {\bibinfo {volume} {121}},\ \bibinfo {pages} {117001} (\bibinfo
  {year} {2018})}\BibitemShut {NoStop}%
\bibitem [{\citenamefont {Serniak}\ \emph {et~al.}(2018)\citenamefont
  {Serniak}, \citenamefont {Hays}, \citenamefont {{de Lange}}, \citenamefont
  {Diamond}, \citenamefont {Shankar}, \citenamefont {Burkhart}, \citenamefont
  {Frunzio}, \citenamefont {Houzet},\ and\ \citenamefont
  {Devoret}}]{serniak2018}%
  \BibitemOpen
  \bibfield  {author} {\bibinfo {author} {\bibfnamefont {K.}~\bibnamefont
  {Serniak}}, \bibinfo {author} {\bibfnamefont {M.}~\bibnamefont {Hays}},
  \bibinfo {author} {\bibfnamefont {G.}~\bibnamefont {{de Lange}}}, \bibinfo
  {author} {\bibfnamefont {S.}~\bibnamefont {Diamond}}, \bibinfo {author}
  {\bibfnamefont {S.}~\bibnamefont {Shankar}}, \bibinfo {author} {\bibfnamefont
  {L.~D.}\ \bibnamefont {Burkhart}}, \bibinfo {author} {\bibfnamefont
  {L.}~\bibnamefont {Frunzio}}, \bibinfo {author} {\bibfnamefont
  {M.}~\bibnamefont {Houzet}},\ and\ \bibinfo {author} {\bibfnamefont {M.~H.}\
  \bibnamefont {Devoret}},\ }\bibfield  {title} {\bibinfo {title} {Hot
  {{Nonequilibrium Quasiparticles}} in {{Transmon Qubits}}},\ }\href
  {https://doi.org/10.1103/PhysRevLett.121.157701} {\bibfield  {journal}
  {\bibinfo  {journal} {Physical Review Letters}\ }\textbf {\bibinfo {volume}
  {121}},\ \bibinfo {pages} {157701} (\bibinfo {year} {2018})}\BibitemShut
  {NoStop}%
\bibitem [{\citenamefont {Phillips}(1987)}]{Phillips:1987ge}%
  \BibitemOpen
  \bibfield  {author} {\bibinfo {author} {\bibfnamefont {W.~A.}\ \bibnamefont
  {Phillips}},\ }\bibfield  {title} {\bibinfo {title} {{Two-level states in
  glasses}},\ }\href {https://doi.org/10.1088/0034-4885/50/12/003} {\bibfield
  {journal} {\bibinfo  {journal} {Reports on Progress in Physics}\ }\textbf
  {\bibinfo {volume} {50}},\ \bibinfo {pages} {1657 1708} (\bibinfo {year}
  {1987})}\BibitemShut {NoStop}%
\bibitem [{\citenamefont {M{\"u}ller}\ \emph {et~al.}(2019)\citenamefont
  {M{\"u}ller}, \citenamefont {Cole},\ and\ \citenamefont
  {Lisenfeld}}]{muller2019}%
  \BibitemOpen
  \bibfield  {author} {\bibinfo {author} {\bibfnamefont {C.}~\bibnamefont
  {M{\"u}ller}}, \bibinfo {author} {\bibfnamefont {J.~H.}\ \bibnamefont
  {Cole}},\ and\ \bibinfo {author} {\bibfnamefont {J.}~\bibnamefont
  {Lisenfeld}},\ }\bibfield  {title} {\bibinfo {title} {Towards understanding
  two-level-systems in amorphous solids: Insights from quantum circuits},\
  }\href {https://doi.org/10.1088/1361-6633/ab3a7e} {\bibfield  {journal}
  {\bibinfo  {journal} {Reports on Progress in Physics}\ }\textbf {\bibinfo
  {volume} {82}},\ \bibinfo {pages} {124501} (\bibinfo {year}
  {2019})}\BibitemShut {NoStop}%
\bibitem [{\citenamefont {Ramos}\ \emph {et~al.}(2013)\citenamefont {Ramos},
  \citenamefont {Sudhir}, \citenamefont {Stannigel}, \citenamefont {Zoller},\
  and\ \citenamefont {Kippenberg}}]{ramos2013}%
  \BibitemOpen
  \bibfield  {author} {\bibinfo {author} {\bibfnamefont {T.}~\bibnamefont
  {Ramos}}, \bibinfo {author} {\bibfnamefont {V.}~\bibnamefont {Sudhir}},
  \bibinfo {author} {\bibfnamefont {K.}~\bibnamefont {Stannigel}}, \bibinfo
  {author} {\bibfnamefont {P.}~\bibnamefont {Zoller}},\ and\ \bibinfo {author}
  {\bibfnamefont {T.~J.}\ \bibnamefont {Kippenberg}},\ }\bibfield  {title}
  {\bibinfo {title} {Nonlinear {{Quantum Optomechanics}} via {{Individual
  Intrinsic Two-Level Defects}}},\ }\href
  {https://doi.org/10.1103/PhysRevLett.110.193602} {\bibfield  {journal}
  {\bibinfo  {journal} {Physical Review Letters}\ }\textbf {\bibinfo {volume}
  {110}},\ \bibinfo {pages} {193602} (\bibinfo {year} {2013})}\BibitemShut
  {NoStop}%
\bibitem [{\citenamefont {Manenti}\ \emph {et~al.}(2016)\citenamefont
  {Manenti}, \citenamefont {Peterer}, \citenamefont {Nersisyan}, \citenamefont
  {Magnusson}, \citenamefont {Patterson},\ and\ \citenamefont
  {Leek}}]{10.1103/physrevb.93.041411}%
  \BibitemOpen
  \bibfield  {author} {\bibinfo {author} {\bibfnamefont {R.}~\bibnamefont
  {Manenti}}, \bibinfo {author} {\bibfnamefont {M.~J.}\ \bibnamefont
  {Peterer}}, \bibinfo {author} {\bibfnamefont {A.}~\bibnamefont {Nersisyan}},
  \bibinfo {author} {\bibfnamefont {E.~B.}\ \bibnamefont {Magnusson}}, \bibinfo
  {author} {\bibfnamefont {A.}~\bibnamefont {Patterson}},\ and\ \bibinfo
  {author} {\bibfnamefont {P.~J.}\ \bibnamefont {Leek}},\ }\bibfield  {title}
  {\bibinfo {title} {{Surface acoustic wave resonators in the quantum
  regime}},\ }\href {https://doi.org/10.1103/physrevb.93.041411} {\bibfield
  {journal} {\bibinfo  {journal} {Physical Review B}\ }\textbf {\bibinfo
  {volume} {93}},\ \bibinfo {pages} {041411} (\bibinfo {year} {2016})},\
  \Eprint {https://arxiv.org/abs/1510.04965} {1510.04965} \BibitemShut
  {NoStop}%
\bibitem [{\citenamefont {Andersson}\ \emph {et~al.}(2021)\citenamefont
  {Andersson}, \citenamefont {Bilobran}, \citenamefont {Scigliuzzo},
  \citenamefont {Lima}, \citenamefont {Cole},\ and\ \citenamefont
  {Delsing}}]{10.1038/s41534-020-00348-0}%
  \BibitemOpen
  \bibfield  {author} {\bibinfo {author} {\bibfnamefont {G.}~\bibnamefont
  {Andersson}}, \bibinfo {author} {\bibfnamefont {A.~L.~O.}\ \bibnamefont
  {Bilobran}}, \bibinfo {author} {\bibfnamefont {M.}~\bibnamefont
  {Scigliuzzo}}, \bibinfo {author} {\bibfnamefont {M.~M.~d.}\ \bibnamefont
  {Lima}}, \bibinfo {author} {\bibfnamefont {J.~H.}\ \bibnamefont {Cole}},\
  and\ \bibinfo {author} {\bibfnamefont {P.}~\bibnamefont {Delsing}},\
  }\bibfield  {title} {\bibinfo {title} {{Acoustic spectral hole-burning in a
  two-level system ensemble}},\ }\href
  {https://doi.org/10.1038/s41534-020-00348-0} {\bibfield  {journal} {\bibinfo
  {journal} {npj Quantum Information}\ }\textbf {\bibinfo {volume} {7}},\
  \bibinfo {pages} {15} (\bibinfo {year} {2021})}\BibitemShut {NoStop}%
\bibitem [{\citenamefont {Lisenfeld}\ \emph {et~al.}(2019)\citenamefont
  {Lisenfeld}, \citenamefont {Bilmes}, \citenamefont {Megrant}, \citenamefont
  {Barends}, \citenamefont {Kelly}, \citenamefont {Klimov}, \citenamefont
  {Weiss}, \citenamefont {Martinis},\ and\ \citenamefont
  {Ustinov}}]{lisenfeld2019}%
  \BibitemOpen
  \bibfield  {author} {\bibinfo {author} {\bibfnamefont {J.}~\bibnamefont
  {Lisenfeld}}, \bibinfo {author} {\bibfnamefont {A.}~\bibnamefont {Bilmes}},
  \bibinfo {author} {\bibfnamefont {A.}~\bibnamefont {Megrant}}, \bibinfo
  {author} {\bibfnamefont {R.}~\bibnamefont {Barends}}, \bibinfo {author}
  {\bibfnamefont {J.}~\bibnamefont {Kelly}}, \bibinfo {author} {\bibfnamefont
  {P.}~\bibnamefont {Klimov}}, \bibinfo {author} {\bibfnamefont
  {G.}~\bibnamefont {Weiss}}, \bibinfo {author} {\bibfnamefont {J.~M.}\
  \bibnamefont {Martinis}},\ and\ \bibinfo {author} {\bibfnamefont {A.~V.}\
  \bibnamefont {Ustinov}},\ }\bibfield  {title} {\bibinfo {title} {Electric
  field spectroscopy of material defects in transmon qubits},\ }\href
  {https://doi.org/10.1038/s41534-019-0224-1} {\bibfield  {journal} {\bibinfo
  {journal} {npj Quantum Information}\ }\textbf {\bibinfo {volume} {5}},\
  \bibinfo {pages} {105} (\bibinfo {year} {2019})}\BibitemShut {NoStop}%
\bibitem [{\citenamefont {Bachtold}\ \emph {et~al.}(2022)\citenamefont
  {Bachtold}, \citenamefont {Moser},\ and\ \citenamefont
  {Dykman}}]{bachtold2022}%
  \BibitemOpen
  \bibfield  {author} {\bibinfo {author} {\bibfnamefont {A.}~\bibnamefont
  {Bachtold}}, \bibinfo {author} {\bibfnamefont {J.}~\bibnamefont {Moser}},\
  and\ \bibinfo {author} {\bibfnamefont {M.~I.}\ \bibnamefont {Dykman}},\
  }\href@noop {} {\bibinfo {title} {Mesoscopic physics of nanomechanical
  systems}} (\bibinfo {year} {2022}),\ \Eprint
  {https://arxiv.org/abs/2202.01819} {arXiv:2202.01819 [cond-mat,
  physics:quant-ph]} \BibitemShut {NoStop}%
\bibitem [{\citenamefont {Machielse}\ \emph {et~al.}(2019)\citenamefont
  {Machielse}, \citenamefont {Bogdanovic}, \citenamefont {Meesala},
  \citenamefont {Gauthier}, \citenamefont {Burek}, \citenamefont {Joe},
  \citenamefont {Chalupnik}, \citenamefont {Sohn}, \citenamefont {Holzgrafe},
  \citenamefont {Evans}, \citenamefont {Chia}, \citenamefont {Atikian},
  \citenamefont {Bhaskar}, \citenamefont {Sukachev}, \citenamefont {Shao},
  \citenamefont {Maity}, \citenamefont {Lukin},\ and\ \citenamefont
  {Lončar}}]{10.1103/physrevx.9.031022}%
  \BibitemOpen
  \bibfield  {author} {\bibinfo {author} {\bibfnamefont {B.}~\bibnamefont
  {Machielse}}, \bibinfo {author} {\bibfnamefont {S.}~\bibnamefont
  {Bogdanovic}}, \bibinfo {author} {\bibfnamefont {S.}~\bibnamefont {Meesala}},
  \bibinfo {author} {\bibfnamefont {S.}~\bibnamefont {Gauthier}}, \bibinfo
  {author} {\bibfnamefont {M.~J.}\ \bibnamefont {Burek}}, \bibinfo {author}
  {\bibfnamefont {G.}~\bibnamefont {Joe}}, \bibinfo {author} {\bibfnamefont
  {M.}~\bibnamefont {Chalupnik}}, \bibinfo {author} {\bibfnamefont {Y.~I.}\
  \bibnamefont {Sohn}}, \bibinfo {author} {\bibfnamefont {J.}~\bibnamefont
  {Holzgrafe}}, \bibinfo {author} {\bibfnamefont {R.~E.}\ \bibnamefont
  {Evans}}, \bibinfo {author} {\bibfnamefont {C.}~\bibnamefont {Chia}},
  \bibinfo {author} {\bibfnamefont {H.}~\bibnamefont {Atikian}}, \bibinfo
  {author} {\bibfnamefont {M.~K.}\ \bibnamefont {Bhaskar}}, \bibinfo {author}
  {\bibfnamefont {D.~D.}\ \bibnamefont {Sukachev}}, \bibinfo {author}
  {\bibfnamefont {L.}~\bibnamefont {Shao}}, \bibinfo {author} {\bibfnamefont
  {S.}~\bibnamefont {Maity}}, \bibinfo {author} {\bibfnamefont {M.~D.}\
  \bibnamefont {Lukin}},\ and\ \bibinfo {author} {\bibfnamefont
  {M.}~\bibnamefont {Lončar}},\ }\bibfield  {title} {\bibinfo {title}
  {{Quantum Interference of Electromechanically Stabilized Emitters in
  Nanophotonic Devices}},\ }\href {https://doi.org/10.1103/physrevx.9.031022}
  {\bibfield  {journal} {\bibinfo  {journal} {Physical Review X}\ }\textbf
  {\bibinfo {volume} {9}},\ \bibinfo {pages} {031022} (\bibinfo {year}
  {2019})},\ \Eprint {https://arxiv.org/abs/1901.09103} {1901.09103}
  \BibitemShut {NoStop}%
\bibitem [{\citenamefont {Peruzzo}\ \emph {et~al.}(2020)\citenamefont
  {Peruzzo}, \citenamefont {Trioni}, \citenamefont {Hassani}, \citenamefont
  {Zemlicka},\ and\ \citenamefont {Fink}}]{10.1103/physrevapplied.14.044055}%
  \BibitemOpen
  \bibfield  {author} {\bibinfo {author} {\bibfnamefont {M.}~\bibnamefont
  {Peruzzo}}, \bibinfo {author} {\bibfnamefont {A.}~\bibnamefont {Trioni}},
  \bibinfo {author} {\bibfnamefont {F.}~\bibnamefont {Hassani}}, \bibinfo
  {author} {\bibfnamefont {M.}~\bibnamefont {Zemlicka}},\ and\ \bibinfo
  {author} {\bibfnamefont {J.~M.}\ \bibnamefont {Fink}},\ }\bibfield  {title}
  {\bibinfo {title} {{Surpassing the Resistance Quantum with a Geometric
  Superinductor}},\ }\href {https://doi.org/10.1103/physrevapplied.14.044055}
  {\bibfield  {journal} {\bibinfo  {journal} {Physical Review Applied}\
  }\textbf {\bibinfo {volume} {14}},\ \bibinfo {pages} {044055} (\bibinfo
  {year} {2020})},\ \Eprint {https://arxiv.org/abs/2007.01644} {2007.01644}
  \BibitemShut {NoStop}%
\bibitem [{\citenamefont {Chamberland}\ \emph {et~al.}(2022)\citenamefont
  {Chamberland}, \citenamefont {Noh}, \citenamefont {Arrangoiz-Arriola},
  \citenamefont {Campbell}, \citenamefont {Hann}, \citenamefont {Iverson},
  \citenamefont {Putterman}, \citenamefont {Bohdanowicz}, \citenamefont
  {Flammia}, \citenamefont {Keller}, \citenamefont {Refael}, \citenamefont
  {Preskill}, \citenamefont {Jiang}, \citenamefont {Safavi-Naeini},
  \citenamefont {Painter},\ and\ \citenamefont
  {Brandão}}]{10.1103/prxquantum.3.010329}%
  \BibitemOpen
  \bibfield  {author} {\bibinfo {author} {\bibfnamefont {C.}~\bibnamefont
  {Chamberland}}, \bibinfo {author} {\bibfnamefont {K.}~\bibnamefont {Noh}},
  \bibinfo {author} {\bibfnamefont {P.}~\bibnamefont {Arrangoiz-Arriola}},
  \bibinfo {author} {\bibfnamefont {E.~T.}\ \bibnamefont {Campbell}}, \bibinfo
  {author} {\bibfnamefont {C.~T.}\ \bibnamefont {Hann}}, \bibinfo {author}
  {\bibfnamefont {J.}~\bibnamefont {Iverson}}, \bibinfo {author} {\bibfnamefont
  {H.}~\bibnamefont {Putterman}}, \bibinfo {author} {\bibfnamefont {T.~C.}\
  \bibnamefont {Bohdanowicz}}, \bibinfo {author} {\bibfnamefont {S.~T.}\
  \bibnamefont {Flammia}}, \bibinfo {author} {\bibfnamefont {A.}~\bibnamefont
  {Keller}}, \bibinfo {author} {\bibfnamefont {G.}~\bibnamefont {Refael}},
  \bibinfo {author} {\bibfnamefont {J.}~\bibnamefont {Preskill}}, \bibinfo
  {author} {\bibfnamefont {L.}~\bibnamefont {Jiang}}, \bibinfo {author}
  {\bibfnamefont {A.~H.}\ \bibnamefont {Safavi-Naeini}}, \bibinfo {author}
  {\bibfnamefont {O.}~\bibnamefont {Painter}},\ and\ \bibinfo {author}
  {\bibfnamefont {F.~G.}\ \bibnamefont {Brandão}},\ }\bibfield  {title}
  {\bibinfo {title} {{Building a Fault-Tolerant Quantum Computer Using
  Concatenated Cat Codes}},\ }\href
  {https://doi.org/10.1103/prxquantum.3.010329} {\bibfield  {journal} {\bibinfo
   {journal} {PRX Quantum}\ }\textbf {\bibinfo {volume} {3}},\ \bibinfo {pages}
  {010329} (\bibinfo {year} {2022})},\ \Eprint
  {https://arxiv.org/abs/2012.04108} {2012.04108} \BibitemShut {NoStop}%
\bibitem [{\citenamefont {Hann}\ \emph {et~al.}(2019)\citenamefont {Hann},
  \citenamefont {Zou}, \citenamefont {Zhang}, \citenamefont {Chu},
  \citenamefont {Schoelkopf}, \citenamefont {Girvin},\ and\ \citenamefont
  {Jiang}}]{10.1103/physrevlett.123.250501}%
  \BibitemOpen
  \bibfield  {author} {\bibinfo {author} {\bibfnamefont {C.~T.}\ \bibnamefont
  {Hann}}, \bibinfo {author} {\bibfnamefont {C.-L.}\ \bibnamefont {Zou}},
  \bibinfo {author} {\bibfnamefont {Y.}~\bibnamefont {Zhang}}, \bibinfo
  {author} {\bibfnamefont {Y.}~\bibnamefont {Chu}}, \bibinfo {author}
  {\bibfnamefont {R.~J.}\ \bibnamefont {Schoelkopf}}, \bibinfo {author}
  {\bibfnamefont {S.~M.}\ \bibnamefont {Girvin}},\ and\ \bibinfo {author}
  {\bibfnamefont {L.}~\bibnamefont {Jiang}},\ }\bibfield  {title} {\bibinfo
  {title} {{Hardware-Efficient Quantum Random Access Memory with Hybrid Quantum
  Acoustic Systems}},\ }\href {https://doi.org/10.1103/physrevlett.123.250501}
  {\bibfield  {journal} {\bibinfo  {journal} {Physical Review Letters}\
  }\textbf {\bibinfo {volume} {123}},\ \bibinfo {pages} {250501} (\bibinfo
  {year} {2019})},\ \Eprint {https://arxiv.org/abs/1906.11340} {1906.11340}
  \BibitemShut {NoStop}%
\bibitem [{\citenamefont {Pechal}\ \emph {et~al.}(2018)\citenamefont {Pechal},
  \citenamefont {Arrangoiz-Arriola},\ and\ \citenamefont
  {Safavi-Naeini}}]{Pechal:tm}%
  \BibitemOpen
  \bibfield  {author} {\bibinfo {author} {\bibfnamefont {M.}~\bibnamefont
  {Pechal}}, \bibinfo {author} {\bibfnamefont {P.}~\bibnamefont
  {Arrangoiz-Arriola}},\ and\ \bibinfo {author} {\bibfnamefont {A.~H.}\
  \bibnamefont {Safavi-Naeini}},\ }\bibfield  {title} {\bibinfo {title}
  {{Superconducting circuit quantum computing with nanomechanical resonators as
  storage}},\ }\href@noop {} {\bibfield  {journal} {\bibinfo  {journal}
  {iopscience.iop.org}\ } (\bibinfo {year} {2018})},\ \Eprint
  {https://arxiv.org/abs/1808.08216} {1808.08216} \BibitemShut {NoStop}%
\bibitem [{\citenamefont {Leduc}\ \emph {et~al.}(2010)\citenamefont {Leduc},
  \citenamefont {Bumble}, \citenamefont {Day}, \citenamefont {Eom},
  \citenamefont {Gao}, \citenamefont {Golwala}, \citenamefont {Mazin},
  \citenamefont {McHugh}, \citenamefont {Merrill}, \citenamefont {Moore},
  \citenamefont {Noroozian}, \citenamefont {Turner},\ and\ \citenamefont
  {Zmuidzinas}}]{leduc2010}%
  \BibitemOpen
  \bibfield  {author} {\bibinfo {author} {\bibfnamefont {H.~G.}\ \bibnamefont
  {Leduc}}, \bibinfo {author} {\bibfnamefont {B.}~\bibnamefont {Bumble}},
  \bibinfo {author} {\bibfnamefont {P.~K.}\ \bibnamefont {Day}}, \bibinfo
  {author} {\bibfnamefont {B.~H.}\ \bibnamefont {Eom}}, \bibinfo {author}
  {\bibfnamefont {J.}~\bibnamefont {Gao}}, \bibinfo {author} {\bibfnamefont
  {S.}~\bibnamefont {Golwala}}, \bibinfo {author} {\bibfnamefont {B.~A.}\
  \bibnamefont {Mazin}}, \bibinfo {author} {\bibfnamefont {S.}~\bibnamefont
  {McHugh}}, \bibinfo {author} {\bibfnamefont {A.}~\bibnamefont {Merrill}},
  \bibinfo {author} {\bibfnamefont {D.~C.}\ \bibnamefont {Moore}}, \bibinfo
  {author} {\bibfnamefont {O.}~\bibnamefont {Noroozian}}, \bibinfo {author}
  {\bibfnamefont {A.~D.}\ \bibnamefont {Turner}},\ and\ \bibinfo {author}
  {\bibfnamefont {J.}~\bibnamefont {Zmuidzinas}},\ }\bibfield  {title}
  {\bibinfo {title} {Titanium nitride films for ultrasensitive microresonator
  detectors},\ }\href {https://doi.org/10.1063/1.3480420} {\bibfield  {journal}
  {\bibinfo  {journal} {Applied Physics Letters}\ }\textbf {\bibinfo {volume}
  {97}},\ \bibinfo {pages} {102509} (\bibinfo {year} {2010})}\BibitemShut
  {NoStop}%
\bibitem [{\citenamefont {Pitanti}\ \emph {et~al.}(2015)\citenamefont
  {Pitanti}, \citenamefont {Fink}, \citenamefont {{Safavi-Naeini}},
  \citenamefont {Hill}, \citenamefont {Lei}, \citenamefont {Tredicucci},\ and\
  \citenamefont {Painter}}]{pitanti2015}%
  \BibitemOpen
  \bibfield  {author} {\bibinfo {author} {\bibfnamefont {A.}~\bibnamefont
  {Pitanti}}, \bibinfo {author} {\bibfnamefont {J.~M.}\ \bibnamefont {Fink}},
  \bibinfo {author} {\bibfnamefont {A.~H.}\ \bibnamefont {{Safavi-Naeini}}},
  \bibinfo {author} {\bibfnamefont {J.~T.}\ \bibnamefont {Hill}}, \bibinfo
  {author} {\bibfnamefont {C.~U.}\ \bibnamefont {Lei}}, \bibinfo {author}
  {\bibfnamefont {A.}~\bibnamefont {Tredicucci}},\ and\ \bibinfo {author}
  {\bibfnamefont {O.}~\bibnamefont {Painter}},\ }\bibfield  {title} {\bibinfo
  {title} {Strong opto-electro-mechanical coupling in a silicon photonic
  crystal cavity},\ }\href {https://doi.org/10.1364/OE.23.003196} {\bibfield
  {journal} {\bibinfo  {journal} {Optics Express}\ }\textbf {\bibinfo {volume}
  {23}},\ \bibinfo {pages} {3196} (\bibinfo {year} {2015})}\BibitemShut
  {NoStop}%
\bibitem [{\citenamefont {Beckers}\ \emph {et~al.}(2019)\citenamefont
  {Beckers}, \citenamefont {Jazaeri},\ and\ \citenamefont {Enz}}]{beckers2019}%
  \BibitemOpen
  \bibfield  {author} {\bibinfo {author} {\bibfnamefont {A.}~\bibnamefont
  {Beckers}}, \bibinfo {author} {\bibfnamefont {F.}~\bibnamefont {Jazaeri}},\
  and\ \bibinfo {author} {\bibfnamefont {C.}~\bibnamefont {Enz}},\ }\bibfield
  {title} {\bibinfo {title} {Cryogenic {{MOSFET Threshold Voltage Model}}},\
  }in\ \href {https://doi.org/10.1109/ESSDERC.2019.8901806} {\emph {\bibinfo
  {booktitle} {{{ESSDERC}} 2019 - 49th {{European Solid-State Device Research
  Conference}} ({{ESSDERC}})}}}\ (\bibinfo {year} {2019})\ pp.\ \bibinfo
  {pages} {94--97}\BibitemShut {NoStop}%
\bibitem [{\citenamefont {Chan}\ \emph {et~al.}(2012)\citenamefont {Chan},
  \citenamefont {{Safavi-Naeini}}, \citenamefont {Hill}, \citenamefont
  {Meenehan},\ and\ \citenamefont {Painter}}]{chan2012}%
  \BibitemOpen
  \bibfield  {author} {\bibinfo {author} {\bibfnamefont {J.}~\bibnamefont
  {Chan}}, \bibinfo {author} {\bibfnamefont {A.~H.}\ \bibnamefont
  {{Safavi-Naeini}}}, \bibinfo {author} {\bibfnamefont {J.~T.}\ \bibnamefont
  {Hill}}, \bibinfo {author} {\bibfnamefont {S.}~\bibnamefont {Meenehan}},\
  and\ \bibinfo {author} {\bibfnamefont {O.}~\bibnamefont {Painter}},\
  }\bibfield  {title} {\bibinfo {title} {Optimized optomechanical crystal
  cavity with acoustic radiation shield},\ }\href
  {https://doi.org/10.1063/1.4747726} {\bibfield  {journal} {\bibinfo
  {journal} {Applied Physics Letters}\ }\textbf {\bibinfo {volume} {101}},\
  \bibinfo {pages} {081115} (\bibinfo {year} {2012})}\BibitemShut {NoStop}%
\bibitem [{\citenamefont {Le~Thomas}\ \emph {et~al.}(2011)\citenamefont
  {Le~Thomas}, \citenamefont {Diao}, \citenamefont {Zhang},\ and\ \citenamefont
  {Houdr{\'e}}}]{lethomas2011}%
  \BibitemOpen
  \bibfield  {author} {\bibinfo {author} {\bibfnamefont {N.}~\bibnamefont
  {Le~Thomas}}, \bibinfo {author} {\bibfnamefont {Z.}~\bibnamefont {Diao}},
  \bibinfo {author} {\bibfnamefont {H.}~\bibnamefont {Zhang}},\ and\ \bibinfo
  {author} {\bibfnamefont {R.}~\bibnamefont {Houdr{\'e}}},\ }\bibfield  {title}
  {\bibinfo {title} {Statistical analysis of subnanometer residual disorder in
  photonic crystal waveguides: {{Correlation}} between slow light properties
  and structural properties},\ }\href {https://doi.org/10.1116/1.3622289}
  {\bibfield  {journal} {\bibinfo  {journal} {Journal of Vacuum Science \&
  Technology B}\ }\textbf {\bibinfo {volume} {29}},\ \bibinfo {pages} {051601}
  (\bibinfo {year} {2011})}\BibitemShut {NoStop}%
\bibitem [{\citenamefont {Minkov}\ \emph {et~al.}(2013)\citenamefont {Minkov},
  \citenamefont {Dharanipathy}, \citenamefont {Houdr{\'e}},\ and\ \citenamefont
  {Savona}}]{minkov2013}%
  \BibitemOpen
  \bibfield  {author} {\bibinfo {author} {\bibfnamefont {M.}~\bibnamefont
  {Minkov}}, \bibinfo {author} {\bibfnamefont {U.~P.}\ \bibnamefont
  {Dharanipathy}}, \bibinfo {author} {\bibfnamefont {R.}~\bibnamefont
  {Houdr{\'e}}},\ and\ \bibinfo {author} {\bibfnamefont {V.}~\bibnamefont
  {Savona}},\ }\bibfield  {title} {\bibinfo {title} {Statistics of the
  disorder-induced losses of high-{{Q}} photonic crystal cavities},\ }\href
  {https://doi.org/10.1364/OE.21.028233} {\bibfield  {journal} {\bibinfo
  {journal} {Optics Express}\ }\textbf {\bibinfo {volume} {21}},\ \bibinfo
  {pages} {28233} (\bibinfo {year} {2013})}\BibitemShut {NoStop}%
\bibitem [{\citenamefont {MacCabe}(2019)}]{maccabe2019}%
  \BibitemOpen
  \bibfield  {author} {\bibinfo {author} {\bibfnamefont {G.~S.}\ \bibnamefont
  {MacCabe}},\ }\emph {\bibinfo {title} {Phonon {{Dynamics}} and {{Damping}} in
  {{Three-Dimensional Acoustic Bandgap Cavity-Optomechanical Resonators}}}},\
  \href {https://doi.org/10.7907/7R9W-EV53} {Ph.D. thesis},\ \bibinfo  {school}
  {California Institute of Technology} (\bibinfo {year} {2019})\BibitemShut
  {NoStop}%
\bibitem [{\citenamefont {Fink}\ \emph {et~al.}(2016)\citenamefont {Fink},
  \citenamefont {Kalaee}, \citenamefont {Pitanti}, \citenamefont {Norte},
  \citenamefont {Heinzle}, \citenamefont {Davan{\c c}o}, \citenamefont
  {Srinivasan},\ and\ \citenamefont {Painter}}]{fink2016}%
  \BibitemOpen
  \bibfield  {author} {\bibinfo {author} {\bibfnamefont {J.~M.}\ \bibnamefont
  {Fink}}, \bibinfo {author} {\bibfnamefont {M.}~\bibnamefont {Kalaee}},
  \bibinfo {author} {\bibfnamefont {A.}~\bibnamefont {Pitanti}}, \bibinfo
  {author} {\bibfnamefont {R.}~\bibnamefont {Norte}}, \bibinfo {author}
  {\bibfnamefont {L.}~\bibnamefont {Heinzle}}, \bibinfo {author} {\bibfnamefont
  {M.}~\bibnamefont {Davan{\c c}o}}, \bibinfo {author} {\bibfnamefont
  {K.}~\bibnamefont {Srinivasan}},\ and\ \bibinfo {author} {\bibfnamefont
  {O.}~\bibnamefont {Painter}},\ }\bibfield  {title} {\bibinfo {title} {Quantum
  electromechanics on silicon nitride nanomembranes},\ }\href
  {https://doi.org/10.1038/ncomms12396} {\bibfield  {journal} {\bibinfo
  {journal} {Nature Communications}\ }\textbf {\bibinfo {volume} {7}},\
  \bibinfo {pages} {12396} (\bibinfo {year} {2016})}\BibitemShut {NoStop}%
\bibitem [{\citenamefont {{Safavi-Naeini}}\ \emph {et~al.}(2013)\citenamefont
  {{Safavi-Naeini}}, \citenamefont {Chan}, \citenamefont {Hill}, \citenamefont
  {Gr{\"o}blacher}, \citenamefont {Miao}, \citenamefont {Chen}, \citenamefont
  {Aspelmeyer},\ and\ \citenamefont {Painter}}]{safavi-naeini2013}%
  \BibitemOpen
  \bibfield  {author} {\bibinfo {author} {\bibfnamefont {A.~H.}\ \bibnamefont
  {{Safavi-Naeini}}}, \bibinfo {author} {\bibfnamefont {J.}~\bibnamefont
  {Chan}}, \bibinfo {author} {\bibfnamefont {J.~T.}\ \bibnamefont {Hill}},
  \bibinfo {author} {\bibfnamefont {S.}~\bibnamefont {Gr{\"o}blacher}},
  \bibinfo {author} {\bibfnamefont {H.}~\bibnamefont {Miao}}, \bibinfo {author}
  {\bibfnamefont {Y.}~\bibnamefont {Chen}}, \bibinfo {author} {\bibfnamefont
  {M.}~\bibnamefont {Aspelmeyer}},\ and\ \bibinfo {author} {\bibfnamefont
  {O.}~\bibnamefont {Painter}},\ }\bibfield  {title} {\bibinfo {title} {Laser
  noise in cavity-optomechanical cooling and thermometry},\ }\href
  {https://doi.org/10.1088/1367-2630/15/3/035007} {\bibfield  {journal}
  {\bibinfo  {journal} {New Journal of Physics}\ }\textbf {\bibinfo {volume}
  {15}},\ \bibinfo {pages} {035007} (\bibinfo {year} {2013})}\BibitemShut
  {NoStop}%
\bibitem [{\citenamefont {Koshino}(2011)}]{koshino2011}%
  \BibitemOpen
  \bibfield  {author} {\bibinfo {author} {\bibfnamefont {K.}~\bibnamefont
  {Koshino}},\ }\bibfield  {title} {\bibinfo {title} {Theory of resonance
  fluorescence from a solid-state cavity {{QED}} system: {{Effects}} of pure
  dephasing},\ }\href {https://doi.org/10.1103/PhysRevA.84.033824} {\bibfield
  {journal} {\bibinfo  {journal} {Physical Review A}\ }\textbf {\bibinfo
  {volume} {84}},\ \bibinfo {pages} {033824} (\bibinfo {year}
  {2011})}\BibitemShut {NoStop}%
\bibitem [{\citenamefont {Nyquist}(1928)}]{nyquist1928}%
  \BibitemOpen
  \bibfield  {author} {\bibinfo {author} {\bibfnamefont {H.}~\bibnamefont
  {Nyquist}},\ }\bibfield  {title} {\bibinfo {title} {Thermal {{Agitation}} of
  {{Electric Charge}} in {{Conductors}}},\ }\href
  {https://doi.org/10.1103/PhysRev.32.110} {\bibfield  {journal} {\bibinfo
  {journal} {Physical Review}\ }\textbf {\bibinfo {volume} {32}},\ \bibinfo
  {pages} {110} (\bibinfo {year} {1928})}\BibitemShut {NoStop}%
\bibitem [{\citenamefont {Johnstone}\ and\ \citenamefont
  {Parameswaran}(2004)}]{johnstone2004}%
  \BibitemOpen
  \bibfield  {author} {\bibinfo {author} {\bibfnamefont {R.~W.}\ \bibnamefont
  {Johnstone}}\ and\ \bibinfo {author} {\bibfnamefont {M.}~\bibnamefont
  {Parameswaran}},\ }\href {https://doi.org/10.1007/978-1-4020-8021-0_1} {\emph
  {\bibinfo {title} {An {{Introduction}} to {{Surface Micromachining}}}}}\
  (\bibinfo  {publisher} {{Springer US}},\ \bibinfo {address} {{Boston, MA}},\
  \bibinfo {year} {2004})\BibitemShut {NoStop}%
\bibitem [{\citenamefont {Janschek}(2011)}]{janschek2011mechatronic}%
  \BibitemOpen
  \bibfield  {author} {\bibinfo {author} {\bibfnamefont {K.}~\bibnamefont
  {Janschek}},\ }\href@noop {} {\emph {\bibinfo {title} {Mechatronic systems
  design: methods, models, concepts}}}\ (\bibinfo  {publisher} {Springer
  Science \& Business Media},\ \bibinfo {year} {2011})\BibitemShut {NoStop}%
\bibitem [{\citenamefont {Klimov}\ \emph {et~al.}(2018)\citenamefont {Klimov},
  \citenamefont {Kelly}, \citenamefont {Chen}, \citenamefont {Neeley},
  \citenamefont {Megrant}, \citenamefont {Burkett}, \citenamefont {Barends},
  \citenamefont {Arya}, \citenamefont {Chiaro}, \citenamefont {Chen},
  \citenamefont {Dunsworth}, \citenamefont {Fowler}, \citenamefont {Foxen},
  \citenamefont {Gidney}, \citenamefont {Giustina}, \citenamefont {Graff},
  \citenamefont {Huang}, \citenamefont {Jeffrey}, \citenamefont {Lucero},
  \citenamefont {Mutus}, \citenamefont {Naaman}, \citenamefont {Neill},
  \citenamefont {Quintana}, \citenamefont {Roushan}, \citenamefont {Sank},
  \citenamefont {Vainsencher}, \citenamefont {Wenner}, \citenamefont {White},
  \citenamefont {Boixo}, \citenamefont {Babbush}, \citenamefont {Smelyanskiy},
  \citenamefont {Neven},\ and\ \citenamefont {Martinis}}]{klimov2018}%
  \BibitemOpen
  \bibfield  {author} {\bibinfo {author} {\bibfnamefont {P.~V.}\ \bibnamefont
  {Klimov}}, \bibinfo {author} {\bibfnamefont {J.}~\bibnamefont {Kelly}},
  \bibinfo {author} {\bibfnamefont {Z.}~\bibnamefont {Chen}}, \bibinfo {author}
  {\bibfnamefont {M.}~\bibnamefont {Neeley}}, \bibinfo {author} {\bibfnamefont
  {A.}~\bibnamefont {Megrant}}, \bibinfo {author} {\bibfnamefont
  {B.}~\bibnamefont {Burkett}}, \bibinfo {author} {\bibfnamefont
  {R.}~\bibnamefont {Barends}}, \bibinfo {author} {\bibfnamefont
  {K.}~\bibnamefont {Arya}}, \bibinfo {author} {\bibfnamefont {B.}~\bibnamefont
  {Chiaro}}, \bibinfo {author} {\bibfnamefont {Y.}~\bibnamefont {Chen}},
  \bibinfo {author} {\bibfnamefont {A.}~\bibnamefont {Dunsworth}}, \bibinfo
  {author} {\bibfnamefont {A.}~\bibnamefont {Fowler}}, \bibinfo {author}
  {\bibfnamefont {B.}~\bibnamefont {Foxen}}, \bibinfo {author} {\bibfnamefont
  {C.}~\bibnamefont {Gidney}}, \bibinfo {author} {\bibfnamefont
  {M.}~\bibnamefont {Giustina}}, \bibinfo {author} {\bibfnamefont
  {R.}~\bibnamefont {Graff}}, \bibinfo {author} {\bibfnamefont
  {T.}~\bibnamefont {Huang}}, \bibinfo {author} {\bibfnamefont
  {E.}~\bibnamefont {Jeffrey}}, \bibinfo {author} {\bibfnamefont
  {E.}~\bibnamefont {Lucero}}, \bibinfo {author} {\bibfnamefont {J.~Y.}\
  \bibnamefont {Mutus}}, \bibinfo {author} {\bibfnamefont {O.}~\bibnamefont
  {Naaman}}, \bibinfo {author} {\bibfnamefont {C.}~\bibnamefont {Neill}},
  \bibinfo {author} {\bibfnamefont {C.}~\bibnamefont {Quintana}}, \bibinfo
  {author} {\bibfnamefont {P.}~\bibnamefont {Roushan}}, \bibinfo {author}
  {\bibfnamefont {D.}~\bibnamefont {Sank}}, \bibinfo {author} {\bibfnamefont
  {A.}~\bibnamefont {Vainsencher}}, \bibinfo {author} {\bibfnamefont
  {J.}~\bibnamefont {Wenner}}, \bibinfo {author} {\bibfnamefont {T.~C.}\
  \bibnamefont {White}}, \bibinfo {author} {\bibfnamefont {S.}~\bibnamefont
  {Boixo}}, \bibinfo {author} {\bibfnamefont {R.}~\bibnamefont {Babbush}},
  \bibinfo {author} {\bibfnamefont {V.~N.}\ \bibnamefont {Smelyanskiy}},
  \bibinfo {author} {\bibfnamefont {H.}~\bibnamefont {Neven}},\ and\ \bibinfo
  {author} {\bibfnamefont {J.~M.}\ \bibnamefont {Martinis}},\ }\bibfield
  {title} {\bibinfo {title} {Fluctuations of {{Energy-Relaxation Times}} in
  {{Superconducting Qubits}}},\ }\href
  {https://doi.org/10.1103/PhysRevLett.121.090502} {\bibfield  {journal}
  {\bibinfo  {journal} {Physical Review Letters}\ }\textbf {\bibinfo {volume}
  {121}},\ \bibinfo {pages} {090502} (\bibinfo {year} {2018})}\BibitemShut
  {NoStop}%
\bibitem [{\citenamefont {Schl{\"o}r}\ \emph {et~al.}(2019)\citenamefont
  {Schl{\"o}r}, \citenamefont {Lisenfeld}, \citenamefont {M{\"u}ller},
  \citenamefont {Bilmes}, \citenamefont {Schneider}, \citenamefont {Pappas},
  \citenamefont {Ustinov},\ and\ \citenamefont {Weides}}]{schlor2019}%
  \BibitemOpen
  \bibfield  {author} {\bibinfo {author} {\bibfnamefont {S.}~\bibnamefont
  {Schl{\"o}r}}, \bibinfo {author} {\bibfnamefont {J.}~\bibnamefont
  {Lisenfeld}}, \bibinfo {author} {\bibfnamefont {C.}~\bibnamefont
  {M{\"u}ller}}, \bibinfo {author} {\bibfnamefont {A.}~\bibnamefont {Bilmes}},
  \bibinfo {author} {\bibfnamefont {A.}~\bibnamefont {Schneider}}, \bibinfo
  {author} {\bibfnamefont {D.~P.}\ \bibnamefont {Pappas}}, \bibinfo {author}
  {\bibfnamefont {A.~V.}\ \bibnamefont {Ustinov}},\ and\ \bibinfo {author}
  {\bibfnamefont {M.}~\bibnamefont {Weides}},\ }\bibfield  {title} {\bibinfo
  {title} {Correlating {{Decoherence}} in {{Transmon Qubits}}: {{Low Frequency
  Noise}} by {{Single Fluctuators}}},\ }\href
  {https://doi.org/10.1103/PhysRevLett.123.190502} {\bibfield  {journal}
  {\bibinfo  {journal} {Physical Review Letters}\ }\textbf {\bibinfo {volume}
  {123}},\ \bibinfo {pages} {190502} (\bibinfo {year} {2019})}\BibitemShut
  {NoStop}%
\bibitem [{\citenamefont {Mei{\ss}ner}\ \emph {et~al.}(2018)\citenamefont
  {Mei{\ss}ner}, \citenamefont {Seiler}, \citenamefont {Lisenfeld},
  \citenamefont {Ustinov},\ and\ \citenamefont {Weiss}}]{meissner2018}%
  \BibitemOpen
  \bibfield  {author} {\bibinfo {author} {\bibfnamefont {S.~M.}\ \bibnamefont
  {Mei{\ss}ner}}, \bibinfo {author} {\bibfnamefont {A.}~\bibnamefont {Seiler}},
  \bibinfo {author} {\bibfnamefont {J.}~\bibnamefont {Lisenfeld}}, \bibinfo
  {author} {\bibfnamefont {A.~V.}\ \bibnamefont {Ustinov}},\ and\ \bibinfo
  {author} {\bibfnamefont {G.}~\bibnamefont {Weiss}},\ }\bibfield  {title}
  {\bibinfo {title} {Probing individual tunneling fluctuators with coherently
  controlled tunneling systems},\ }\href
  {https://doi.org/10.1103/PhysRevB.97.180505} {\bibfield  {journal} {\bibinfo
  {journal} {Physical Review B}\ }\textbf {\bibinfo {volume} {97}},\ \bibinfo
  {pages} {180505} (\bibinfo {year} {2018})}\BibitemShut {NoStop}%
\bibitem [{\citenamefont {Sage}\ \emph {et~al.}(2011)\citenamefont {Sage},
  \citenamefont {Bolkhovsky}, \citenamefont {Oliver}, \citenamefont {Turek},\
  and\ \citenamefont {Welander}}]{sage2011}%
  \BibitemOpen
  \bibfield  {author} {\bibinfo {author} {\bibfnamefont {J.~M.}\ \bibnamefont
  {Sage}}, \bibinfo {author} {\bibfnamefont {V.}~\bibnamefont {Bolkhovsky}},
  \bibinfo {author} {\bibfnamefont {W.~D.}\ \bibnamefont {Oliver}}, \bibinfo
  {author} {\bibfnamefont {B.}~\bibnamefont {Turek}},\ and\ \bibinfo {author}
  {\bibfnamefont {P.~B.}\ \bibnamefont {Welander}},\ }\bibfield  {title}
  {\bibinfo {title} {Study of loss in superconducting coplanar waveguide
  resonators},\ }\href {https://doi.org/10.1063/1.3552890} {\bibfield
  {journal} {\bibinfo  {journal} {Journal of Applied Physics}\ }\textbf
  {\bibinfo {volume} {109}},\ \bibinfo {pages} {063915} (\bibinfo {year}
  {2011})}\BibitemShut {NoStop}%
\end{thebibliography}%


%

\end{document}